\documentclass[iop]{emulateapj}
\usepackage{amsmath}
\usepackage{epstopdf}
\usepackage{graphicx}
\usepackage[table]{xcolor}
\usepackage{mathrsfs}
\usepackage{hyperref}

\begin{document}
\title{Synthesizing Exoplanet Demographics: A Single Population of Long-Period Planetary Companions to M Dwarfs Consistent with Microlensing, Radial Velocity, and Direct Imaging Surveys}

\author{Christian Clanton,
B. Scott Gaudi}
\affil{Department of Astronomy, The Ohio State University, 140 W. 18th Ave., Columbus, OH 43210, USA}
\email{clanton@astronomy.ohio-state.edu}

\begin{abstract}
We present the first study to synthesize results from five different exoplanet surveys using three independent detection methods: microlensing, radial velocity, and direct imaging. The constraints derived herein represent the most comprehensive picture of the demographics of large-separation ($\gtrsim 2~$AU) planets orbiting the most common stars in our Galaxy that has been constructed to date. We assume a simple, joint power-law planet distribution function of the form $d^2N_{\rm pl}/(d\log{m_p}~d\log{a})=\mathcal{A}(m_p/M_{\rm Sat})^{\alpha}(a/2.5~{\rm AU})^{\beta}$ with an outer cutoff radius of the separation distribution function of $a_{\rm out}$. Generating populations of planets from these models and mapping them into the relevant observables for each survey, we use actual or estimated detection sensitivities to determine the expected observations for each survey. Comparing with the reported results, we derive constraints on the parameters $\{\alpha, \beta, \mathcal{A}, a_{\rm out}\}$ that describe a single population of planets that is simultaneously consistent with the results of microlensing, RV, and direct imaging surveys. We find median and 68\% confindence intervals of $\alpha=-0.86^{+0.21}_{-0.19}$ ($-0.85^{+0.21}_{-0.19}$), $\beta=1.1^{+1.9}_{-1.4}$ ($1.1^{+1.9}_{-1.3}$), $\mathcal{A}=0.21^{+0.20}_{-0.15}~{\rm dex^{-2}}$ ($0.21^{+0.20}_{-0.15}~{\rm dex^{-2}}$), and $a_{\rm out}=10^{+26}_{-4.7}~$AU ($12^{+50}_{-6.2}~$AU) assuming ``hot-start'' (``cold-start'') planet evolutionary models. These values are consistent with all current knowledge of planets on orbits beyond $\sim 2~$AU around M dwarfs.
\end{abstract}


\section{Introduction}
\label{sec:introduction}

Understanding the demographics of exoplanets is requisite to the development of observationally-constrained formation and migration models. Exoplanet discovery surveys have revealed a large diversity of systems, many of which look nothing like our own Solar system. The overwhelming majority of planets discovered through such surveys have been detected indirectly by monitoring variations in: 1) the centroids of absorption lines in host star spectra (RV), 2) the apparent brightness of the host star (transit), or 3) the magnification of a background source near our line of sight to the host star (microlensing). The physical processes that produce these observables are fundamentally different, and consequently, the properties of the planetary systems they reveal are also different. In other words, each technique is sensitive to planets in a given region of parameter space (e.g. mass-orbital period space), with varying degrees of overlap with regions probed by other techniques.

In general, RV surveys are most sensitive to short-period, massive planets, transit surveys are most sensitive to very short-period planets with large radii, and microlensing surveys are most sensitive to a range of planet masses and orbital separations near the ice line ($\sim$ few AU). Practical limitations currently restrict the sensitivity of direct imaging searches to the most massive ($\gtrsim M_{\rm Jup}$) planets at large separations ($\gtrsim 10~$AU). Coronagraphy and high angular resolution obtained through the use of adaptive optics are required to reveal planetary companions that are several orders-of-magnitude fainter than, and located at small angular separations from, their hosts. Although planet detections from individual methods are constrained to limited regions of planet parameter space, it is possible to synthesize results from multiple techniques to derive a more accurate and complete census of exoplanets \citep[e.g.][]{Howard2012,Gaidos2012,Clanton2014a,Clanton2014b}.

\begin{figure*}[ht]
	\epsscale{0.8}
	\plotone{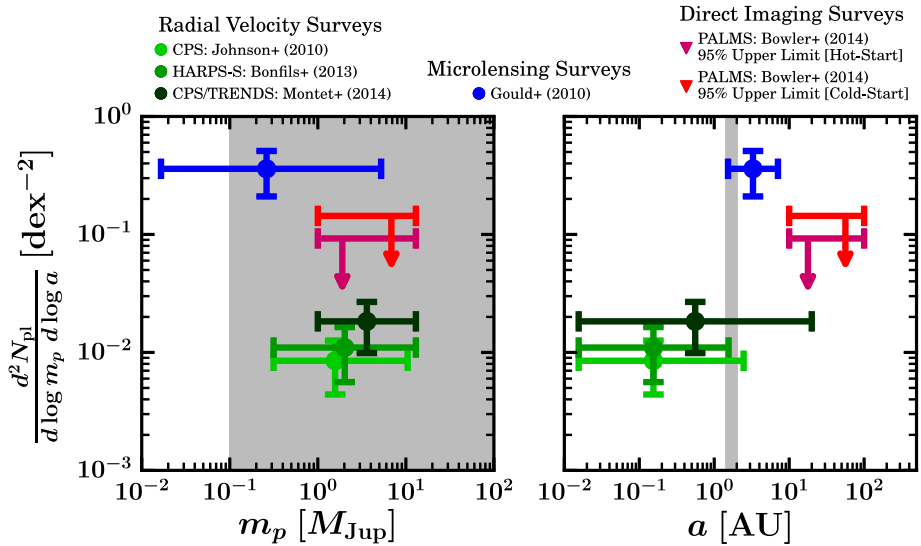}
	\caption{Planet frequency measurements from several exoplanet discovery surveys of M dwarfs as functions of planet mass, $m_p$, and semimajor axis, $a$. The vertical error bars on the points represent the measurement uncertainty on the planet frequency, while the horizontal ``error'' bars represent the approximate ranges of mass and semimajor axis over which these frequencies are measured. The region corresponding to giant planets in the left panel is shaded gray, while the vertical gray line in the right panel represents the approximate location of the ice line around a $0.5~M_{\odot}$ star.
		\label{fig:pf_results_all}}
\end{figure*}

Figure~\ref{fig:pf_results_all} shows several existing constraints on the occurrence rate of planets around M dwarfs for select microlensing, RV, and direct imaging surveys. A direct comparison between these results is complicated by the various selection effects and observational biases intrinsic to each survey. In \citet{Clanton2014b} we perform a careful comparison of the results of the \citet{Gould2010} microlensing survey with the RV studies of HARPS \citep{Bonfils2013} and the California Planet Survey \citep[CPS;][]{Montet2014} (shown in figure~\ref{fig:pf_results_all}) by mapping the observable parameters of the population of planets inferred from microlensing to those of an analogous population orbiting RV-monitored stars. Using the actual detection sensitivities of the HARPS and estimated sensitivities for the CPS, we determine the expected numbers of planet detections and long-term trends for both samples and compare with the reported numbers of detections and trends. We find agreement with the detections, demonstrating that RV and microlensing results are consistent, but found that we could not explain all the long-term trends. These excess trends are either not due to a planetary population, or are due to a population of planets to which microlensing is not sensitive.

A similar approach to comparing the results of microlensing and direct imaging surveys, where we would predict the number of expected planet detections with the actual reported number of detections for a given imaging survey, would not be as informative as it was in the case of microlensing and RV. To zeroth order, the ``sensitivity function'' of direct imaging surveys to detecting planetary companions depends on just two parameters: the planet-to-star contrast ratio and the projected separation. Of course, these parameters are, themselves, dependent on several system-specific properties --- such as age, metallicity, spectral type, distance, planetary atmospheric composition and structure --- that are susceptible to large uncertainties, complicating direct comparisons between results of high-contrast imaging surveys with those of other detection methods. We define the ``typical'' microlensing planet to be one residing in the region of peak sensitivity of microlensing surveys, described by the following parameters: a host star mass of $M_l \sim 0.5~M_{\odot}$, a mass ratio of $q \sim 5\times10^{-4}$, and a projected separation of $r_{\perp} \sim R_E \sim 2.5~$AU, where $R_E \sim 2.5~$AU is the typical Einstein radius for a $0.5~M_{\odot}$ lens \citep{Gould2010}. This corresponds to a typical microlensing planet mass of $m_p = qM_l \sim 0.26~M_{\rm Jup} \sim M_{\rm Sat}$. \citet{Quanz2012b} present the deepest high-contrast image for an M star to date, looking at the $12-50~$Myr old pre-main-sequence star AP Col, which is located $8.4~$pc away, for a planetary companion. The authors present their $5\sigma$ detection limits in their figure 2. It is clear from this figure that a Saturn-mass planet at a projected separation of $\sim 2.5~$AU (i.e. the typical microlensing planet) would be undetectable. This suggests that most, if not all, of the planets to which microlensing is sensitive will be undetectable to imaging surveys, meaning there is probably little overlap between these two techniques.

Thus, we develop new methods in this study to constrain the distributions of the properties of wide-separation ($a\gtrsim 2$~AU) planets by combining the demographics derived from several individual microlensing, direct imaging, and RV surveys. More specifically, we will assume a simple, power-law planet distribution model, map the resultant populations into the relevant observables for each type of survey, and employ the actual detection limits for each survey to determine the expected observations. We will then weight the assumed planet population by its likelihood associated with each given survey, repeating this process many times to derive constraints on the parameters that describe our planet distribution model (i.e. power-law slopes, normalization, and the outer cutoff radius of the separation distribution function).

This is the first study to attempt to perform a joint analysis combining the results from five different exoplanet surveys using three independent detection methods. The constraints derived herein represent the most comprehensive understanding of large-separation planets that orbit the most common stars in our Galaxy and will serve as the standard to which models of planet formation and migration must adhere. In \S~\ref{sec:ml_di_rv_results}, we provide brief descriptions of the surveys and their results included in our analysis. We perform an order-of-magnitude-style analysis in \S~\ref{sec:oom} under several simplifying assumptions, such as a single characteristic host mass, distance, and kinematic parameters, as well as median sensitivity curves to describe each survey. These rough calculations allow us to develop the intuition necessary for understanding the constraining power of both the individual and combined survey results. In \S~\ref{sec:methodology}, relaxing as many of these assumptions as possible, we describe the methodology we develop to perform a detailed analysis that treats uncertainties in the relevant stellar physical and kinematic parameters, orbital parameters, planetary evolutionary models, and variations in detection sensitivities among, and within, the surveys to which we compare. We present and discuss the results of these careful analyses in \S~\ref{sec:results_discussion}, and follow up with a description of all sources of (quantified and unquantified) uncertainty in our analysis in \S~\ref{sec:uncertainties}. We summarize our results and discuss future, related work in \S~\ref{sec:summary_conclusions}.

\section{Microlensing, Direct Imaging, and RV Survey Results}
\label{sec:ml_di_rv_results}
Here we describe the results from several representative microlensing, direct imaging, and RV surveys we use to constrain the demographics of planets around M dwarfs. In this paper, we will make use of the following results: 1) the measured frequency of giant planets from the microlensing survey of \citet{Gould2010}, 2) the slope of the planetary mass-ratio function inferred from the microlensing survey of \citet{Sumi2010}, 3) the detection of four long-term RV trends measured by the CPS TRENDS survey \citep{Montet2014}, 4) the non-detection of any planetary companions in a sample of 72 young, single M stars imaged by the Planets Around Low-Mass Stars \citep[PALMS;][]{Bowler2015} survey, and 5) the non-detection of any planetary companions around the 16 M stars imaged by the Gemini Deep Planet Survey \citep[GDPS;][]{Lafreniere2007}. 

\begin{figure*}[ht]
	\epsscale{0.8}
	\plotone{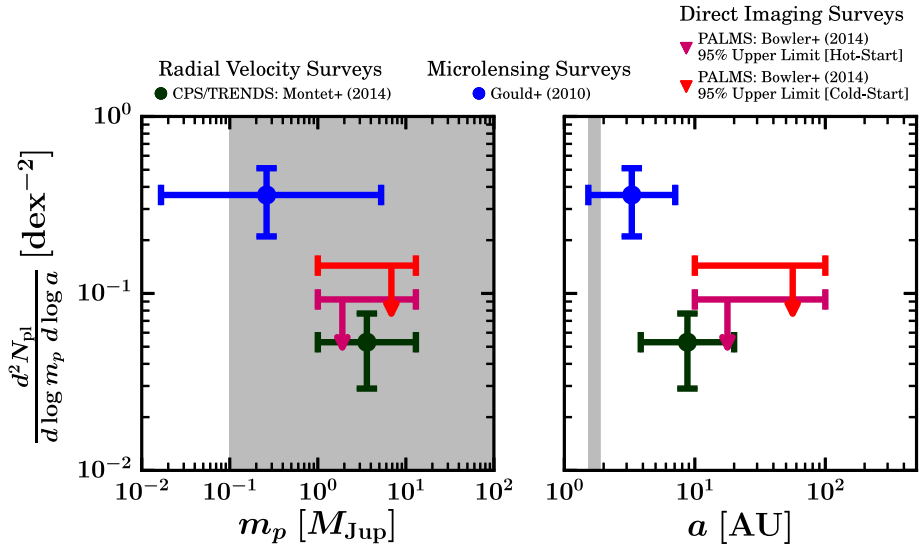}
	\caption{Constraints on planet frequency around M dwarfs for a few of the surveys we consider in this paper as functions of planet mass, $m_p$, and semimajor axis, $a$. The vertical error bars on the points represent the measurement uncertainty on the planet frequency, while the horizontal ``error'' bars represent the approximate ranges of mass and semimajor axis over which these frequencies are measured. The region corresponding to giant planets in the left panel is shaded gray, while the vertical gray line in the right panel represents the approximate location of the ice line around a $0.5~M_{\odot}$ star. The result plotted for the CPS/TRENDS survey of \citet{Montet2014} includes only the constraints from the four significant long-term RV trends they detect (see \S~\ref{subsec:rv_trends_constraints} for description).
		\label{fig:pf_results_used}}
\end{figure*}

Figure~\ref{fig:pf_results_used} graphically displays three of these constraints. Note that the measurement for the CPS/TRENDS survey \citep{Montet2014} differs in this plot relative to that in figure~\ref{fig:pf_results_all}. Since we are concerned with the long-period companions to M dwarfs in this paper, we only use the constraints from the trends that are detected and reported by \citet{Montet2014} (discussed in \S~\ref{subsec:rv_trends_constraints}), and thus the data point plotted here is the result only considering the constraints from their long-term RV drifts. Not shown in figure~{\ref{fig:pf_results_used}} are the two constraints we consider from the GDPS \citep{Lafreniere2007}, as they only report a frequency for all stars in their sample (of which only 16 are M dwarfs; see \S~\ref{subsec:di_constraints}), and that of the \citet{Sumi2010} microlensing survey, which is a measurement of the slope (but not the normalization) of the planetary mass-ratio function (discussed in the following subsection). In the remainder of this section, we describe basic sample properties and results of all five surveys we consider, and place them into the context of the current knowledge of planet occurrences around M dwarfs and solar-type stars.

\subsection{Microlensing}
\label{subsec:mlens_constraints}
{\bf \citet{Gould2010}}: The sample of \citet{Gould2010} is an unbiased selection of 13 high-magnification ($A>200$) microlensing events. Within this sample, there are a total of six planet detections in five systems. They find typical values of the planet-to-star mass ratio, Einstein radius, and lens mass of $q_0\sim 5\times10^{-4}$, $R_{E,0}\sim 3.5~{\rm AU}(M_{\star}/M_{\odot})^{1/2}$, and $M_{l,0}\sim 0.5~M_{\odot}$, respectively. \citet{Gould2010} measure the frequency of ice and gas giant planets with mass ratios in the interval $-4.5\leq \log{q} \leq -2$ to be $\mathcal{G} \equiv d^2N_{\rm pl}/(d\log{q}~d\log{s})=(0.36\pm 0.15)~{\rm dex}^{-2}$ at their mean mass-ratio, $q_0$. This result is sensitive to a wide range of projected separations between $s_{\rm max}^{-1}R_{E,0} \lesssim r_{\perp} \lesssim s_{\rm max}R_{E,0}$, where $s_{\rm max}\sim (q/10^{-4.3})^{1/3}$, corresponding to de-projected separations well beyond the ice lines of these systems. In their analysis, \citet{Gould2010} make the assumption that planets are distributed uniformly in $\log{s}$ near the Einstein radius and demonstrate that this assumption is consistent with the six microlensing detections in their sample.

On its face, this result appears to be in direct conflict with those of recent RV surveys, which have measured frequencies of giant planets around M dwarfs that are smaller by more than an order-of-magnitude \citep{Cumming2008,Johnson2010,Bonfils2013,Montet2014}. Indeed, the publication of \citet{Gould2010} left many uncomfortable with the idea that giant planets may be common around M dwarfs because the theory of giant planet formation by core accretion, supported by the RV measurements, predicts the opposite \citep[e.g.][]{Laughlin2004}. Qualitative arguments invoking the Galactic metallicity gradient or the sensitivity of microlensing surveys to longer periods than RV surveys seemed likely explanations for this vast difference in inferred occurrence rates. However, RV surveys of M dwarfs have already operated long enough to have significant sensitivity, at least in terms of orbital period, to the majority of planets detectable by microlensing. This motivated a study to statistically compare the constraints on exoplanet demographics inferred from these two discovery techniques and develop a thorough understanding of the conflation of the various selection effects intrinsic to them. In \citet{Clanton2014a,Clanton2014b}, we demonstrate that the giant planet frequencies measured by microlensing and RV surveys are actually consistent, and without a need to invoke the Galactic metallicity gradient. Rather, the steep planetary mass-ratio function derived from microlensing discoveries implies that RV surveys are detecting only the high-mass tail of the giant planet population inferred by \citet{Gould2010}. In \citet{Clanton2014a}, we also synthesize constraints from the microlensing surveys of \citet{Gould2010} and \citet{Sumi2010} with those from the HARPS RV survey of \citet{Bonfils2013} and determine the frequency of Jupiters and super-Jupiters with $1\leq m_p\sin{i} / M_{\rm Jup} \leq 13$ and periods $1\leq P/{\rm days}\leq 10^4$ to be $f_J=0.029^{+0.013}_{-0.015}$, a median factor of 4.3 (1.5-14 at 95\% confidence) smaller than the inferred frequency of such planets around FGK stars of $0.11\pm 0.20$ \citep{Cumming2008}. Thus, the combined microlensing and RV constraints are consistent with the generic prediction of core accretion that giant planet frequency is correlated with host mass, although it remains to be seen if these rates are consistent in detail with quantitative predictions of this model.

{\bf \citet{Sumi2010}}: \citet{Sumi2010} perform a likelihood analysis on the 10 microlensing planet discoveries up to their time of publication to determine the planetary mass-ratio distribution function, which they assume has the form $dN_{\rm pl}/d\log{q}=N_0q^n\Theta(q-q_{\rm low})\Theta(q_{\rm up}-q)$, where $q_{\rm low}=10^{-4.5}$ and $q_{\rm up}=10^{-2}$ are the lower and upper limits on the mass ratio, respectively, and where $\Theta$ is the Heaviside step function. The authors report a slope of $n=-0.68\pm 0.20$ but are unable to place constraints on the normalization $N_0$ since they estimate the relative (not absolute) detection efficiencies as a function of $q$. \citet{Sumi2010} infer a slope that is steeper than, but consistent with, that found by RV surveys of solar-type stars of $n_{\rm RV}=-0.31\pm 0.20$ \citep{Cumming2008}.

\subsection{Direct Imaging}
\label{subsec:di_constraints}

{\bf \citet{Bowler2015}}: The PALMS survey is a direct imaging campaign targeting a sample of 122 young M stars that have been selected for their youth and proximity, with the goals of discovering sub-stellar companions and determining the frequency of giant planets orbiting M dwarfs at separations beyond $\sim 10~$AU. The median spectral type and  distance of the full PALMS sample is approximately M3.5 and 21~pc, respectively. The median age of the sample is 135~Myr and 90\% of the targets are younger than 620~Myr. Imaging of these 122 M dwarfs reveal that 44 are close stellar binaries. \citet{Bowler2015} do not include these close binaries in their statistical analysis in order to construct a sample that most closely matches those of RV surveys, which are routinely vetted of such binaries. Their final statistical sample is thus composed of 78 single M stars and represents the largest imaging search for planets around low-mass stars ($0.1-0.6~M_{\odot}$) to date. In our analysis, we further refine this sample to construct a sample to which we can fairly compare the results of other studies (in particular, the CPS RV sample of M dwarfs which have been explicitly vetted for binaries), eliminating three stars with spectral types earlier than M0: TYC 523-573-1 (K7.5), NLTT 26359 (K5), and TYC 1752-63-1 (K7). We also eliminate three stars which are close visual binaries with angular separations $<1''$, where there are no stable S-type orbits: 1RXS J034231.8+121622 AB, GJ 3629AB, and GJ 4185 Aab. This leaves us with a sample of 72 single M stars. Some of these stars are actually in wide binaries, but following the analysis of \citet{Bowler2015}, we only examine separations out to which there is a stable S-type orbit and assume no sensitivity to separations beyond this maximum stable separation.

\citet{Bowler2015} report typical contrasts of $12-14~$mag with Keck/NIRC2 \citep{Wizinowich2000} and $9-13~$mag with Subaru/HiCIAO \citep{Hayano2010,Hodapp2008,Suzuki2010} at an angular separation of $1''$, which correspond to limiting planet masses between roughly $0.5-10~M_{\rm Jup}$ at projected separations between $5-33~$AU for 85\% of their statistical sample. Overall, the PALMS survey is most sensitive to planets with projected separations between $10-100~$AU, demonstrating an ability to detect $\sim 5\%$ of $1~M_{\rm Jup}$ companions, $\sim 20\%$ of $2~M_{\rm Jup}$ companions, and $\sim 50\%$ of $5~M_{\rm Jup}$ companions in this range (see figure~24 of \citealt{Bowler2015} for the full PALMS survey sensitivity map).

A total of 167 faint point sources (planet candidates) are identified around 45 stars from first epoch images. \citet{Bowler2015} were able to obtain follow-up imaging and recover 56\% of these candidates, showing that all are consistent with background stars. The nature of the remaining candidates with just a single epoch of astrometry is uncertain, however only 8 of these reside at projected separations within $100~$AU. \citet{Bowler2015} do not confirm any planets orbiting the single M dwarfs in their sample. These non-detections provide 95\% confidence upper limits on the frequency of giant planets ($1-13~M_{\rm Jup}$) orbiting single M dwarfs between $10-100~$AU of $<10.3\%$ ($<16.0\%$), assuming logarithmically-uniform distributions in planet mass and semimajor axis and ``hot-start'' (``cold-start'') planet evolutionary models of \citet{Baraffe2003} \citep{Fortney2008}.

\citet{Bowler2015} point out that the results of the PALMS survey, when compared with similarly large surveys for the direct detection of exoplanets around samples of A \citep[e.g.][]{Nielsen2013,Vigan2012} and FGK stars \citep[e.g.][]{Nielsen2010,Chauvin2015}, provide no statistical evidence for a dependency of the frequency of long-period, giant planets on host star mass. Larger sample sizes in all mass stellar mass regimes will be required of future surveys in order to distinguish small differences in the relative occurrence rates of long-period ($>10~$AU) giant planets around low- and high-mass stars.

Finally, \citet{Bowler2015} report the detection of four brown dwarfs, deriving a frequency of brown dwarf companions to single M dwarfs with masses between $13-75~M_{\rm Jup}$ and separations between $10-100~$AU of $2.8^{+2.4}_{-1.5}\%$, consistent with previous estimates for BD companions to M dwarfs \citet{Dieterich2012} and to solar-type primaries \citet{Metchev2009}.

In this study, we will employ the sensitivity limits for the 72 single M stars targeted by the PALMS survey to identify classes of planet populations that are consistent with non-detections and compare these with the populations consistent with the results of the other exoplanet surveys from microlensing, RV, and the Gemini Deep Planet Survey \citep[GDPS;][]{Lafreniere2007}. The contrast curves provided in table~5 of \citet{Bowler2015} are the deepest contrast curves for each star, but are not necessarily the ones employed in their statistical analysis (see their \S~6.1 for a description of how they select the appropriate contrast curves for their analysis). In order to most accurately determine the expected results of the PALMS survey for generic (power-law) planet populations and compare with the actual reported results, we instead use the same set of contrast curves used in their statistical analysis (B.P. Bowler, private communication).

{\bf \citet{Lafreniere2007}}: The GDPS is a direct imaging survey of young, nearby stars, with the primary science driver of constraining the occurrence rates of massive, planetary companions at orbital separations ranging between $10-300~$AU. The GDPS sample includes 85 FGKM stars within 35~pc at ages ranging between $\sim 10-6700~$Myr. The median values for several parameters of the full sample are a distance of $22~$pc, spectral type of K0, age of $\sim 200~$Myr, $H$ magnitude of 5.75, proper-motion amplitude of $240~{\rm mas~yr^{-1}}$, and metallicity of $[{\rm Fe/H}]=0.00~$dex (with a standard deviation of $0.21~$dex). \citet{Lafreniere2007} do not detect any planetary companions, but report a typical $5\sigma$ detection limit sufficient to detect planets with masses $>2~M_{\rm Jup}$ at projected separations between $40-200~$AU. These authors analyze the detection limits of each star in their sample, deriving upper limits on the fraction of stars hosting at least one planet with $m_p>2~M_{\rm Jup}$ to be 0.23 for companions with separations between $25-420~$AU and 0.12 for companions with separations between $50-295~$AU at 95\% confidence, assuming the ``hot-start'' planet evolutionary models of \citet{Baraffe2003}.

\citet{Lafreniere2007} provide the $5\sigma$ detection limits they derive for each star in the GDPS sample in tabular form, which we obtain and use to determine expected detections from generic planet populations. We perform analyses on the subset of M stars in their sample, which should more closely represent the population of microlensing surveys, at least in terms of stellar mass, than the more massive stars. There are a total of 16 M dwarfs among their full sample.

While the GDPS is an overall deeper survey than PALMS in terms of contrast, the sample sizes differ by a factor of $\sim 4$. Thus, the reported upper limits on giant planet frequency at large separations from the PALMS survey seem to be more constraining than those reported by the GDPS. However, it is difficult to quantify the actual differences in completeness, as the GDPS results are quoted over a larger range of semimajor axis ($25-420~$AU) than that from PALMS ($10-100$~AU). Furthermore, the GDPS result includes the non-detections from the higher-mass stars, and it is not clear if the planet evolutionary models were incorporated in a consistent fashion between the two surveys. This highlights the importance of performing a self-consistent analysis between surveys using the same detection technique in order to perform a fair comparison of their results. To this end, we perform our own analysis of both the GDPS and PALMS sample within the same framework, using the exact same planet evolutionary models that have been interpolated to a given time resolution and extrapolated to a larger range of planet masses and to later ages. We describe in detail how we incorporate the evolutionary models of \citet{Baraffe2003} and \citet{Fortney2008} in \S~\ref{subsec:imaging_observables}.

\subsection{Radial Velocity Drifts}
\label{subsec:rv_trends_constraints}
{\bf \citet{Montet2014}}: The CPS TRENDS survey \citep{Johnson2010,Montet2014} targets a sample of 111 M dwarfs (which they define as having $B-V>1.44$) brighter than $V=13.5$, closer than 16~pc, and vetted for known binaries. The sample has a median number of observations of 29 over a median time baseline of 11.8~years, with typical Doppler precisions of a couple meters per second. With spectral types from M0 to M5.5 and masses ranging between $0.10-0.64~M_{\odot}$ (median of $0.41~M_{\odot}$), the CPS M dwarfs cover a wide range in metallicity, $-0.81\leq [{\rm Fe/H}]\leq 0.52$, with a slightly sub-Solar median value of $-0.1~$dex.

\citet{Montet2014} identify four stars within their sample that exhibit long-term RV trends, signaling the existence of wide-separation companions. In one of the cases, curvature is detected in the RV curve, allowing \citet{Montet2014} to constrain the companion mass (and to a lesser extent, the orbital period) and show the long-term RV variation must be due to a giant planet. \citet{Montet2014} obtain images with AO to rule out stellar companions and most brown dwarf masses as the cause of their observed RV trends. Thus, considering the low inferred frequency of brown dwarf companions to M dwarfs \citep{Metchev2009,Dieterich2012}, \citet{Montet2014} argue these trends are most likely due to giant planets rather than lower-mass brown dwarfs that are not quite ruled out by their AO observations. Assuming all their trends are indeed due to planetary-mass companions, and including the planets with full orbits that have been detected within their sample, \citet{Montet2014} infer a frequency of Jupiters and super-Jupiters with $1\leq m_p/M_{\odot}\leq 13$ and orbital separations $a<20~$AU around M dwarfs to be $0.065\pm0.030$. This is consistent with the constraint we derive in \citet{Clanton2014b} from synthesizing microlensing and RV results of $0.035^{+0.017}_{-0.019}$ in the same mass and semimajor axis intervals, and is inconsistent with the result for FGK stars of $0.11\pm 0.02$ \citep{Cumming2008}.

Although \citet{Montet2014} infer a frequency of Juipters and super-Jupiters that is consistent with \citet{Clanton2014b}, it is nevertheless a median factor of 1.9 ($0.11-7.9$ at 95\% confidence) times larger. This is potentially due to a population of very long-period super-Jupiters to which microlensing is not sensitive but is being inferred by \citet{Montet2014} from these long-term RV trends. It could be that this population is responsible for some or all of the short-timescale events observed in the microlensing survey of \citet{Sumi2011}.

\section{Order of Magnitude Evaluations}
\label{sec:oom}
The primary objective of this study is to constrain the properties of wide ($a\gtrsim 2~$AU) planetary companions to M dwarfs by testing if a simple parameterized planet distribution model can explain the microlensing, RV, and direct imaging data. In order to do this correctly, we will need to map the parameters of an assumed model of the distributions of planet masses and semimajor axes into the appropriate observables for each discovery technique and use the actual detection sensitivities for each specific survey to determine their expected number of detections for a given planet population. The full procedure will require marginalizing over several stellar, Galactic, kinematic, and orbital parameters, but first we will forgo several of these steps to estimate --- at an order-of-magnitude level of precision --- the expected results for microlensing, direct imaging, and RV surveys. We perform this rough calculation first to develop intuition on the potential constraining power of each individual detection technique, as well as that from their combination. These evaluations are also a way to validate our more careful---but also more complicated and harder to interpret---methodology.

We begin by assuming planet populations that are described by four parameters: the slope of the mass distribution function, $\alpha$, the slope of the separation distribution function, $\beta$, their normalization, $\mathcal{A}$, and the outer cutoff radius of the separation distribution function, $a_{\rm out}$. The populations are thus distributed in planet mass and semimajor axis as
\begin{equation}
    \frac{d^2N_{\rm pl}}{d\log{m_p}~d\log{a}} = \mathcal{A}\left(\frac{m_p}{M_{\rm Sat}}\right)^{\alpha}\left(\frac{a}{2.5~{\rm AU}}\right)^{\beta} \label{eqn:planet_dist_function}\; ,
\end{equation}
with a maximum semimajor axis of $a_{\rm out}$. We choose to normalize our distribution functions at a Saturn mass and at a semimajor axis of 2.5~AU, which correspond to the mass ratio and projected separation in units of the Einstein radius in the region of peak sensitivity for microlensing surveys, $q_0=5\times10^{-4}$ and $s_0=1$, for a lens mass of $M_l\sim 0.5~M_{\odot}$. Note that $2.5~$AU is the median Einstein radius of an ensemble of simulated microlensing events with parameters consistent with those observed for actual samples of events (see \citealt{Clanton2014a,Clanton2014b} for more information).

We generate random planet populations, enumerated by the index $i$, each of which is described by a random set of parameters: $\left\{\alpha_i, \beta_i, \mathcal{A}_i, a_{\rm out, i}\right\}$. We randomly (and independently) draw values of $\alpha$ and $\beta$ uniformly within the range $\left[-4,4\right]$, values of $\log{(\mathcal{A}/{\rm dex^{-2}})}$ uniformly within the range $\left[-3,3\right]$, and values of $\log{(a_{\rm out}/{\rm AU})}$ uniformly within the range $\left[-1,3\right]$. We then map the properties of these populations into the observables relevant to microlensing, RV, and direct imaging surveys, and apply the appropriate detection criteria to determine the expected observations for each survey. Each planet population is then assigned a statistical weight according to its likelihood associated with all the surveys to which we compare. We thus derive maximum likelihood distributions of each of our population parameters ($\alpha$, $\beta$, $\mathcal{A}$, and $a_{\rm out}$) which describe planet populations consistent with the results of all surveys. This process is described in greater detail in the following sections.

\subsection{Individual Constraints}
\label{subsec:oom_individual_constraints}

\subsubsection{Microlensing}
\label{subsubsec:oom_microlensing}
For the following order of magnitude comparisons with microlensing surveys, we transform between $(q,s)\leftrightarrow (m_p,a)$ assuming a characteristic lens mass of $M_l\sim 0.5~M_{\odot}$ and the median projection angle of a randomly oriented, circular orbit where necessary. These assumptions will be relaxed in later sections when we perform more detailed calculations.

{\bf \citet{Gould2010}:} We analytically compute the integrated planet frequency over the mass and semimajor axis ranges of $5~M_{\oplus}\leq m_p\leq 5~M_{\rm Jup}$ and $1.4\leq a/{\rm AU}\leq 7.2$, respectively, for each population. These correspond roughly to the ranges in mass ratio and projected separation to which the \citet{Gould2010} survey is sensitive ($-4.5\leq \log{q} \leq -2$ and $0.5\leq s \leq 2.5$, respectively). The integrated frequency is given by the equation
\begin{align}
\mathcal{G}_i = & {} \mathcal{A}_i\int^{\log{a_{{\rm max},i}}}_{\log{(1.4~{\rm AU})}}\int^{\log (5~M_{\rm Jup})}_{\log (5~M_{\oplus})}\left(\frac{m_p}{M_{\rm Sat}}\right)^{\alpha_i} \nonumber \\
    & {} \times \left(\frac{a}{2.5~{\rm AU}}\right)^{\beta_i}d\log{m_p}~d\log{a}\; , \label{eqn:g_freq}
\end{align}
where 
\begin{equation}
    a_{{\rm max},i} = 
    \begin{cases}
    1.4~{\rm AU} & {\rm for}~a_{{\rm out},i}\leq 1.4~{\rm AU} \\
    \min(7.2~{\rm AU}, a_{{\rm out},i}) & {\rm for}~a_{{\rm out},i}> 1.4~{\rm AU}.
    \end{cases}
\end{equation}
We then weight the parameters of each planet population by the likelihoods associated with our calculated values of $\mathcal{G}_i$. These likelihoods are assigned according to the likelihood function measured by \citet{Gould2010}, to which we assign the variable $\mathscr{L}_{\mathcal{G}}$.

\begin{figure*}[!ht]
	\epsscale{0.7}
	\plotone{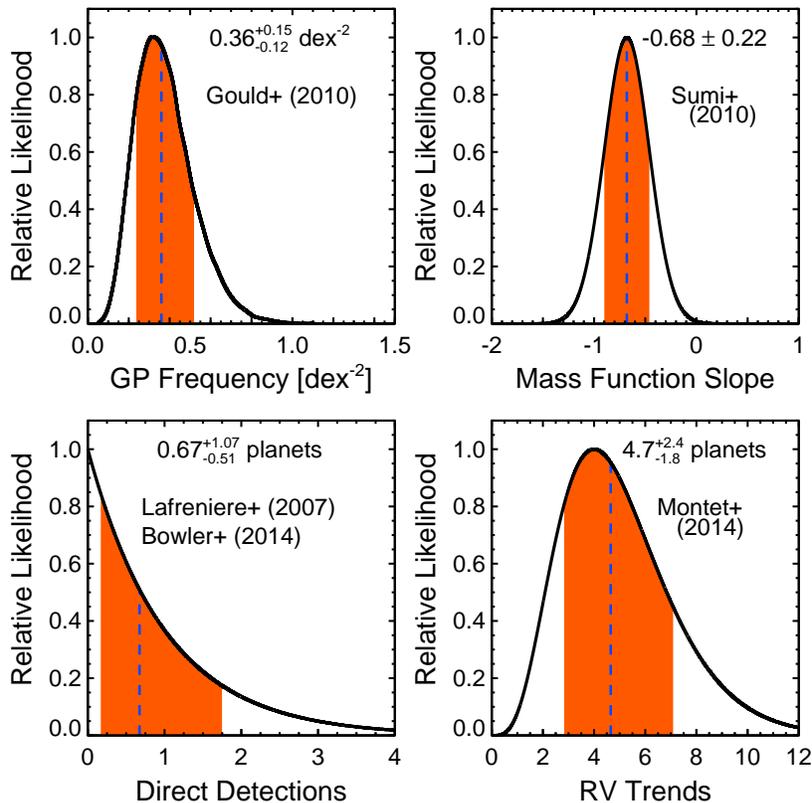}
	\caption{Likelihood functions on the measurements made by the various surveys to which we compare. The upper left panel shows our approximation to the likelihood function of giant planet frequency measured by \citet{Gould2010}, $\mathscr{L}_{\mathcal{G}}$, while the upper right panel shows the likelihood function of the slope of the mass-ratio distribution function, $\mathscr{L}_{\rm S10}$, measured by \citet{Sumi2010}. The lower left panel shows the likelihood functions of the number of planets existing around stars in both the GDPS \citep{Lafreniere2007}, $\mathscr{L}_{\rm L07}$, and PALMS \citep{Bowler2015}, $\mathscr{L}_{\rm B15}$, samples, derived from upper limits due to non-detections of planets by these surveys. The lower right panel plots the likelihood function of the number of RV trend detections, $\mathscr{L}_{\rm tr}$, measured by the CPS TRENDS survey \citep{Montet2014}. The median values and $68\%$ confidence intervals are indicated in each plot by a blue, vertical, dotted line and a red, shaded region, respectively, and printed at the top of each panel.
		\label{fig:lhood_functions}}
\end{figure*}

Similar to \citet{Clanton2014b}, we approximate $\mathscr{L}_{\mathcal{G}}$ by first drawing a non-integer number from a Poisson distribution with a mean value equal to the number of detections in the \citet{Gould2010} sample (six) to represent the average number of planets per star. We then use this number to compute the implied planet frequency, $\mathcal{G}$, in a similar manner as described in \citet{Gould2010}. After repeating this process many times, we normalize the resultant distribution of planet frequencies such that the median value is equal to the maximum likelihood value reported by \citet{Gould2010} of $0.36~{\rm dex^{-2}}$. The likelihood function we derive from this process is plotted in figure~\ref{fig:lhood_functions} and has a median value and $68\%$ uncertainties of $0.36^{+0.15}_{-0.12}~{\rm dex^{-2}}$. While this is an approximation of the actual likelihood function derived by \citet{Gould2010}, who report $68\%$ uncertainties of $\pm 0.15~{\rm dex^{-2}}$, we show in \S~\ref{sec:uncertainties} that other sources of uncertainty will dominate over that introduced by our approximation of $\mathscr{L}_{\mathcal{G}}$.

\begin{figure}[!ht]
	\epsscale{1.1}
	\plotone{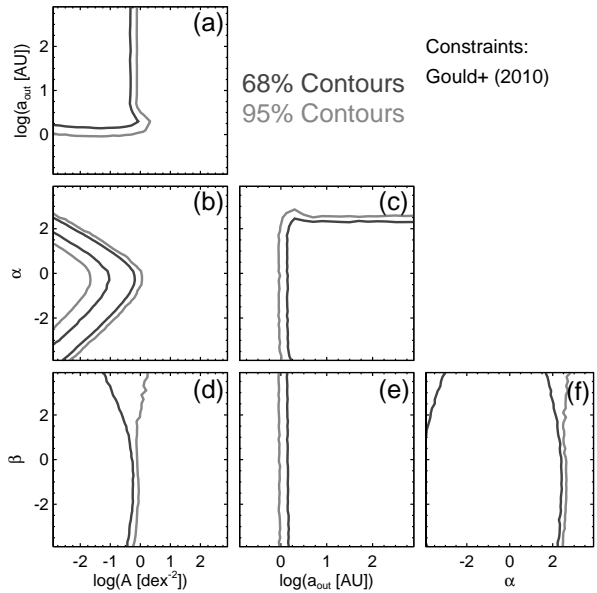}
	\caption{Likelihood contours as a function of pairs of parameters describing planet distribution functions that are found to be consistent (in our order of magnitude evaluations) with the measurement of the integrated planet frequency, $\mathcal{G}$, by the \citet{Gould2010} microlensing survey. Contours are drawn at levels of $68\%$ and $95\%$ of the peak likelihood and are marginalized over all other parameters except those being plotted.
		\label{fig:oom_constraints_g10}}
\end{figure}

\begin{figure}[!ht]
	\epsscale{1.1}
	\plotone{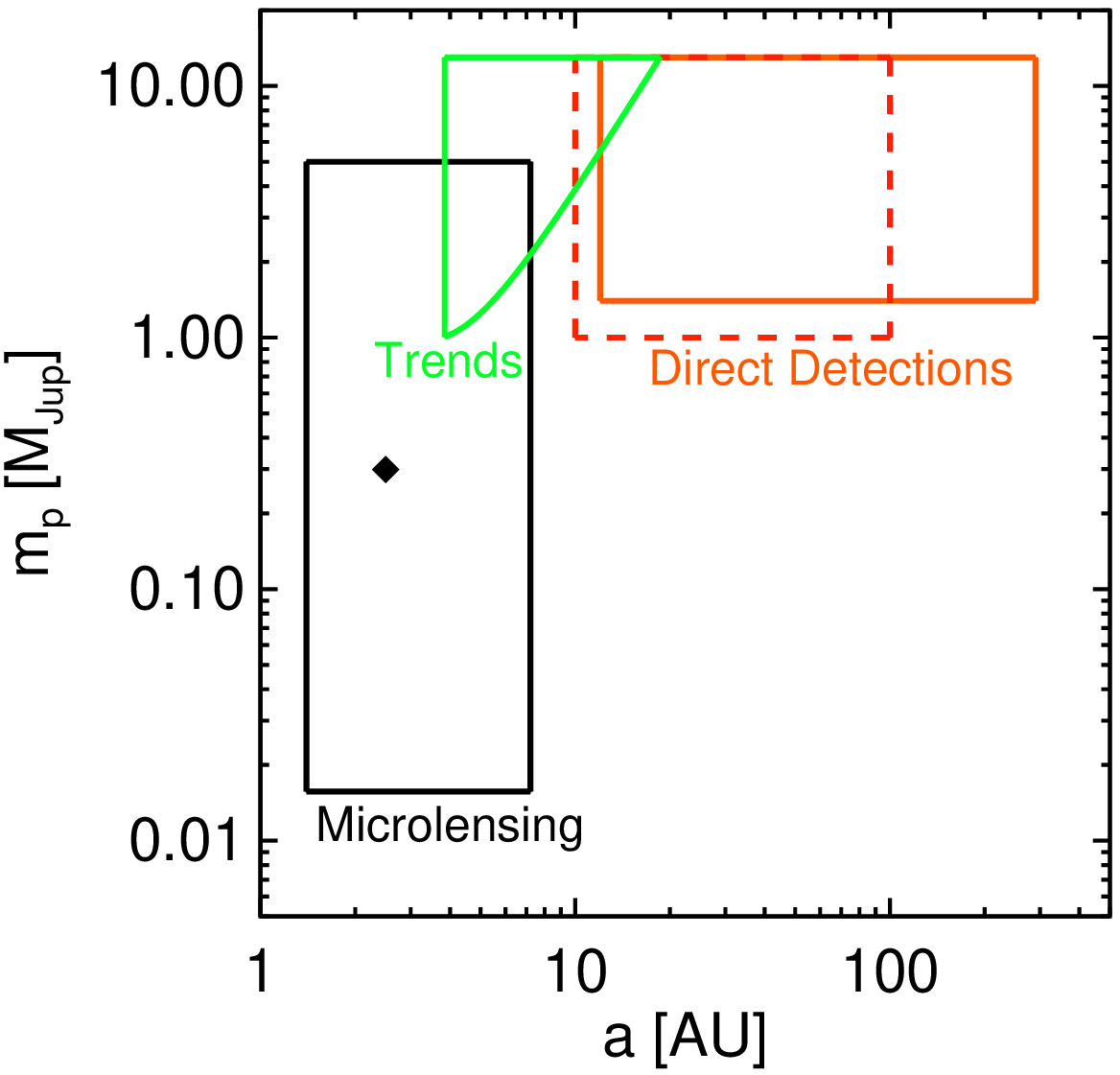}
	\caption{Approximate regions of sensitivity for the various surveys we consider. The solid, black lines bound the region for the microlensing surveys of \citet{Gould2010} and \citet{Sumi2010}. The green region shows the area within which the CPS TRENDS survey \citep{Montet2014} has demonstrated an ability to detect long-term RV trends, while the red regions show the areas within which the GDPS \citep[solid; ][]{Lafreniere2007} and the PALMS \citep[dashed; ][]{Bowler2015} surveys are sensitive to direct detections. The solid, black diamond shows the ``pivot point'' of our planet distribution function, i.e. the point at which we choose to normalize our distribution function ($m_p=M_{\rm Sat}$, $a=2.5~$AU). Note that this pivot point is located inside the region of sensitivity for the microlensing surveys of \citet{Gould2010} and \citet{Sumi2010} and outside the regions of sensitivity for the remaining surveys. This has consequences that are reflected in the constraints we derive on the parameters of our planet distribution models for the individual surveys (see text, figure~\ref{fig:oom_constraints_g10}, and figures~\ref{fig:oom_constraints_s10}--\ref{fig:oom_constraints_mb14}).
		\label{fig:regions_plot}}
\end{figure}

Figure~\ref{fig:oom_constraints_g10} shows the resultant constraints on the parameters of populations consistent with this survey. These can be understood by considering how changes in the parameters of our models affect the integrated planet frequency given by equation~(\ref{eqn:g_freq}). We have assumed the measurement of \citet{Gould2010} uniformly covers a given range of $q$ and $s$, and thus effectively, a given range of $m_p$ and $a$ (although in reality, the range in $m_p$ and $a$ varies for each system depending on the orbital phase and primary mass -- this will be addressed in the more detailed calculations presented in later sections). The black box in figure~\ref{fig:regions_plot} illustrates this area, over which the \citet{Gould2010} survey is sensitive. Another important detail that must be kept in mind when thinking about these constraints is the point at which our planet distribution models are normalized. We refer to this as the ``pivot point'' of our models. The black diamond in figure~\ref{fig:regions_plot} shows the location of the pivot point, at a mass of $m_p=M_{\rm Sat}$ and a semimajor axis of $a=2.5~$AU, relative to the regions of sensitivity for the various surveys. Note that the pivot point lies interior to the region of sensitivity of the \citet{Gould2010} survey. As we will show, this has consequences that are reflected in the constraints we derive on the parameters of our planet models from comparison with this survey.

Comparison of our planet models with the results of the \citet{Gould2010} survey essentially does three things: 1) puts a lower bound on $a_{\rm out}$, 2) puts an upper bound on $\mathcal{A}$, and 3) constrains a combination of $\alpha$ and $\mathcal{A}$. Panels~(a), (c), and (e) in figure~\ref{fig:oom_constraints_g10} show that values of $a_{\rm out}\leq 1.4$~AU are not allowed (although the bin size makes it appear as values down to $\approx 1~$AU are allowed), as that would imply there are no planets beyond $1.4~$AU, which is clearly discrepant with the observations. Panels~(a), (b), and (d) demonstrate that there is a maximum value of $\mathcal{A}$ allowed. This is best understood mathematically. First consider cases where $a_{\rm out}>7.2~$AU. At a fixed value of $\mathcal{A}$, there are specific values of $\alpha$ and $\beta$ (that are completely independent of $a_{\rm out}$) that minimize the integrals in equation~(\ref{eqn:g_freq}). At these minimizing values, $\mathcal{A}$ becomes the only way to adjust $\mathcal{G}$, meaning that there is a maximum value of $\mathcal{A}$ beyond which the implied planet frequency becomes too large to be consistent with the measurement of \citet{Gould2010}.

To understand the constraint on the combination of $\alpha$ and $\mathcal{A}$, consider what happens to the integrated planet frequency at fixed values of $\alpha$, $\beta$, and $a_{\rm out}$. In this case, we can see directly from equation~(\ref{eqn:g_freq}) that $\mathcal{G}\propto \mathcal{A}$. Next, consider fixed values of $\alpha$, $\beta$, and $\mathcal{A}$. Larger values of $a_{\rm out}$ will result in larger values of $\mathcal{G}$, while smaller $a_{\rm out}$ result in smaller $\mathcal{G}$. However, at fixed $\beta$, $a_{\rm out}$, and $\mathcal{A}$, increasing (decreasing) $\alpha$ will {\it not always} lead to increased (decreased) values of $\mathcal{G}$. This is due to the fact that the pivot point lies interior to the region of sensitivity of the \citet{Gould2010} survey. For fixed values of $\mathcal{G}$, $\beta$, $a_{\rm out}$, and $\mathcal{A}$, there are two values of $\alpha$ that will satisfy equation~(\ref{eqn:g_freq}). Qualitatively, what is happening in this case, is a redistribution of planets from greater numbers of more (less) massive planets to greater numbers of less (more) massive planets for decreasing (increasing) values of $\alpha$, while conserving the overall planet frequency. This is illustrated in panel~(b) of figure~\ref{fig:oom_constraints_g10}. Taking a slice through these contours at a fixed value of $\mathcal{A}=0.01~{\rm dex^{-2}}$ results in a bimodal distribution in $\alpha$. If we had a perfect measurement of $\mathcal{G}$ and knew the exact values of $\beta$ and $a_{\rm out}$, this distribution would consist of two ``spikes,'' or delta functions, at specific values of $\alpha$. However, uncertainty in the measurement of $\mathcal{G}$ and mutual degeneracies between $\alpha$, $\beta$, and $a_{\rm out}$ (see panels~c, e, and f) give these modes finite width. As $\mathcal{A}$ increases to a maximum allowed value, these modes move closer together and eventually merge into a single peak at a value of $\alpha$ that minimizes the contribution of the mass integral in equation~(\ref{eqn:g_freq}). Panel~(d) displays a similar behavior in the point constraints of $\beta$ and $\mathcal{A}$, although it is less pronounced than that between $\alpha$ and $\mathcal{A}$ because the range of $a$ covered by the \citet{Gould2010} survey is smaller relative to that of $m_p$.

Note that even though panels~(b), (c), and (f) seem to indicate that there is a maximum value of $\alpha$ allowed, this is not a real constraint. Instead, it is a consequence of the limited range of $\mathcal{A}$ we consider, and if we had considered smaller values of $\mathcal{A}$, the contours would expand to include larger values of $\alpha$. However, some values of $\alpha$ would not be consistent with the distribution of mass ratios seen by \citet{Gould2010} (and similarly, some values of $\beta$ would not be consistent with their observed distribution of projected separations). We address this point in later sections.

\begin{figure}[!ht]
	\epsscale{1.1}
	\plotone{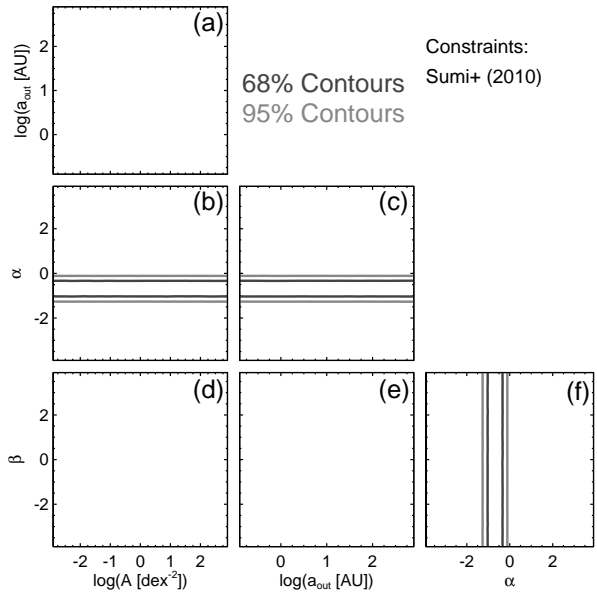}
	\caption{Likelihood contours as a function of pairs of parameters describing planet distribution functions that are found to be consistent (in our order of magnitude evaluations) with the measurement of the slope of the mass-ratio distribution function by the \citet{Sumi2010} microlensing survey. Contours are drawn at levels of $68\%$ and $95\%$ of the peak likelihood and are marginalized over all other parameters except those being plotted.
		\label{fig:oom_constraints_s10}}
\end{figure}

{\bf \citet{Sumi2010}:} This survey measures the slope of the mass-ratio distribution function, $dN_{\rm pl}/d\log{q}\propto q^{p}$, reporting a maximum likelihood value and 68\% uncertainties of $p=-0.68\pm 0.22$. Roughly, the area in $m_p-a$ space over which this measurement is made is illustrated by the black box in figure~\ref{fig:regions_plot}. \citet{Sumi2010} report their likelihood function of $p$ in their figure~10, which is well described by a Gaussian with mean $\mu_p=-0.68$ and standard deviation $\sigma_p=0.22$. Since $q=m_p/M_l$ and we are assuming a characteristic lens mass of $M_l\sim 0.5~M_{\odot}$ for this order of magnitude comparison, we have that $dN_{\rm pl}/d\log{q}\propto dN_{\rm pl}/dlog{a}$, or equivalently, $p=\alpha$. Thus, \citet{Sumi2010} provide a direct constraint on $\alpha$, and the likelihood function we adopt has the simple analytic form $\mathscr{L}_{\rm S10}=\exp{[-(\alpha_i-\mu_p)^2/2\sigma^2_p]}$. We show a plot of $\mathscr{L}_{\rm S10}$ in figure~\ref{fig:lhood_functions}. Figure~\ref{fig:oom_constraints_s10} shows the resultant constraints on $\alpha$. In reality, conflation of the planet mass and lens mass distribution functions complicate the relationship between $\alpha$ and $p$, which we consider in later sections when we perform more detailed calculations.

\subsubsection{Direct Imaging}
\label{subsubsec:oom_di}
The following order of magnitude comparisons with the results of direct imaging surveys use mean detection limits in terms of planet mass and semimajor axis derived using the ``hot-start'' and ``cold-start'' planet evolutionary models of \citet{Baraffe2003} and \citet{Fortney2008}, respectively. In more detailed calculations described later, we make use of each individual contrast curve reported by the surveys, along with the stellar ages and distances (and our predicted planet masses) to determine the expected numbers of detections for direct imaging surveys, again using both ``hot-'' and ``cold-start'' models.

{\bf \citet{Bowler2015}:} We compute the expected number of planet detections per star for the PALMS survey for a given planet population as
\begin{align}
N_{{\rm det, img}, i} = & {} \mathcal{A}_i\int^{\log{a_{{\rm max}, i}}}_{\log{a_{\rm min}}}\int^{\log{m_{p,{\rm max}}}}_{\log{m_{p,{\rm min}}}}\left(\frac{m_p}{M_{\rm Sat}}\right)^{\alpha_i} \nonumber \\
    & {} \times \left(\frac{a}{2.5~{\rm AU}}\right)^{\beta_i}\Phi_{\rm det}(\log{m_p},\log{a}) \nonumber \\
    & {} \times d\log{m_p}~d\log{a}\; , \label{eqn:ndet_img_oom}
\end{align}
where $a_{{\rm max},i}=\max{(a_{\rm min}, a_{{\rm out},i})}$ and where the function $\Phi_{\rm det}(\log{m_p},\log{a})$ represents the detection limits of the PALMS survey in terms of $\log{m_p}$ and $\log{a}$. The integration limits of equation~(\ref{eqn:ndet_img_oom}) are set by the region over which we assume our planet distribution function holds. We adopt the values $a_{\rm min} =1~$AU, $m_{p,{\rm min}}=1~M_{\oplus}$, and $m_{p,{\rm max}}=13~M_{\rm Jup}$.

We assume a very simple form for the detection sensitivity function of the PALMS survey of $\Phi_{\rm det}(\log{m_p},\log{a})=P_{\rm det}(\log{m_p})\Theta(\log{a}-\log{a_{\rm min, det, img}})\Theta(\log{a_{\rm max, det, img}}-\log{a})$, where $\Theta$ is the Heaviside step function, $P_{\rm det}(\log{m_p})$ is the probability of detecting a planet of a given mass (which we assume to be independent of $a$ in the range of semimajor axes over which we integrate) and where $a_{\rm min, det, img}$ and $a_{\rm max, det, img}$ are the minimum- and maximum-detectable semimajor axes, respectively, for the survey. Since the main result of the PALMS survey to which we will compare this order-of-magnitude estimate is the upper limit on the occurrence rate of giant planets with semimajor axes between $10-100~$AU, we set $a_{\rm min, det, img}=10~$AU and $a_{\rm max, det, img}=100~$AU.

We determine $P_{\rm det}(\log{m_p})$ by examining Figure~24 of \citet{Bowler2015}, which displays the mean survey sensitivities for both hot- and cold-start planet evolutionary models. Specifically, we fit (by eye) the contour levels, which correspond to the detection probability of a given planet mass, at a constant semimajor axis ($50~$AU) as a function of $\log{m_p}$, finding 
\begin{equation}
    P_{\rm det}(\log{m_p})\approx
    \begin{cases}
    0.7\log{\left(m_p/M_{\rm Jup}\right)}+0.07 & \text{for ``hot-start,''} \\
    0.5\log{\left(m_p/M_{\rm Jup}\right)}+0.01 & \text{for ``cold-start,''} \label{eqn:pdet_b15_oom}
    \end{cases}
\end{equation}
for planet masses $1\leq m_p/M_{\rm Jup}\leq 13$, and assume that this probability is independent of $a$. We set $P_{\rm det}(\log{m_p})=0$ for $m_p < M_{\rm Jup}$ (since \citealt{Bowler2015} measure the frequency of Jupiter- and super-Jupiter-mass planets) and further assume this form of $P_{\rm det}(\log{m_p})$ is the same for each of the 72 single M stars we consider from the PALMS sample.

Under these assumptions, for a given planet population, the number of expected direct imaging detections per star, $N_{{\rm B15}, i}$, for the PALMS survey is thus determined analytically as
\begin{align}
    N_{{\rm B15}, i} = & {} \mathcal{A}_i\int^{\log{a_{{\rm max}, i}}}_{\log{(10~{\rm AU})}}\int^{\log{(13~M_{\rm Jup})}}_{\log{(1~M_{\rm Jup})}}\left(\frac{m_p}{M_{\rm Sat}}\right)^{\alpha_i} \nonumber \\
    & {} \times \left(\frac{a}{2.5~{\rm AU}}\right)^{\beta_i}P_{\rm det}(\log{m_p})d\log{m_p}~d\log{a}\; , \label{eqn:ndet_b15_oom}
\end{align}
where we now have
\begin{equation}
    a_{{\rm max}, i} =
    \begin{cases}
        \max(10~{\rm AU}, a_{{\rm out}, i}) & {\rm for}~a_{{\rm out}, i} \leq 100~{\rm AU}, \\
        100~{\rm AU} & {\rm for}~a_{{\rm out}, i} > 100~{\rm AU},
    \end{cases}
\end{equation}
and where $P_{\rm det}(\log{m_p})$ is given by equation~(\ref{eqn:pdet_b15_oom}). The total number of expected detections for the full sample is then $N_{{\rm tot, B15},i}=72N_{{\rm B15}, i}$. We assign a statistical weight to each planet population according to the likelihood function we define to be $\mathscr{L}_{\rm B15}\equiv P(0|N_{{\rm tot, B15}, i})$, where $P(0|N_{{\rm tot, B15}, i})=\exp(-N_{{\rm tot, B15}, i})$ is the Poisson probability of detecting zero planets among the 72 single M stars in the PALMS sample if the true number of detectable planets around such stars is actually $N_{{\rm tot, B15}, i}$. We plot this likelihood function in figure~\ref{fig:lhood_functions}, where the abscissa in this case corresponds to $N_{{\rm tot, B15}, i}$.

\begin{figure}[!t]
	\epsscale{1.1}
	\plotone{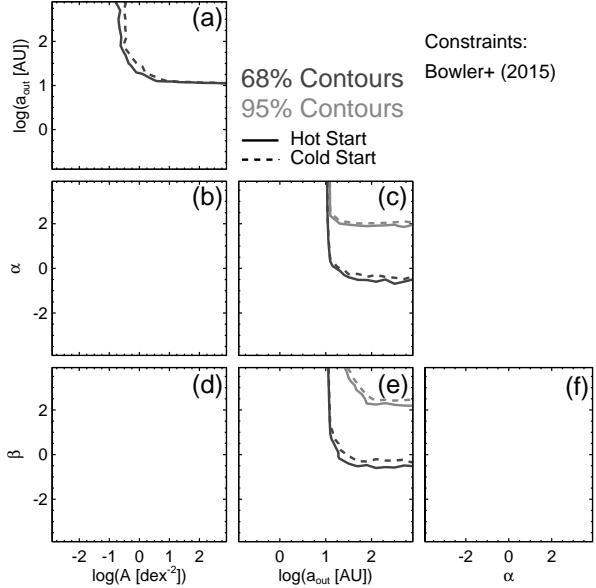}
	\caption{Likelihood contours as a function of pairs of parameters describing planet distribution functions that are found to be consistent (in our order of magnitude evaluations) with the non-detection of any planetary companions in the PALMS direct imaging survey \citep{Bowler2015}. Contours are drawn at levels of $68\%$ and $95\%$ of the peak likelihood and are marginalized over all other parameters except those being plotted. Note that the contours enclose the bottom left regions of the panels.
		\label{fig:oom_constraints_bb15}}
\end{figure}

Figure~\ref{fig:oom_constraints_bb15} shows the resultant constraints on the parameters of populations we find to be consistent with this survey. As can be seen in equation~(\ref{eqn:ndet_img_oom}), there are degeneracies between $\alpha$, $\beta$, and $\mathcal{A}$ which result in no constraining power on combinations of these parameters from direct imaging results. The only real constraints (although weak) allowed by the PALMS survey individually are on $a_{\rm out}$. Panels~(a), (c), and (e) demonstrate that there is a maximum allowed value of $a_{\rm out}$ for a given value of $\alpha$, $\beta$, and $\mathcal{A}$. The shapes of these contour lines are straightforward to understand. For $a_{\rm out}\leq 10~$AU, which is the minimum separation at which PALMS is sensitive to planets, there is no constraint on $a_{\rm out}$ because all planets, regardless of their distribution (and overall occurrence rate) in mass and semimajor axis, would be undetectable by PALMS and thus consistent with the reported non-detections.

Panel~(a) illustrates that for a small enough normalization, $a_{\rm out}$ can be arbitrarily large (within the range of $a_{\rm out}$ we consider). Similarly, there are values of $\alpha$ and $\beta$ below which there is no constraint on $a_{\rm out}$, where the planets are distributed in such a way that they are undetectable. Since in the above-described order of magnitude calculation, we have assumed the sensitivity of the PALMS survey is the same for each star in their sample, the transition from having a constraint on $a_{\rm out}$ to having no constraint is quite sharp. As we will demonstrate, this will not be the case when we perform more careful calculations that make use of each individual sensitivity curve reported by \citet{Bowler2015}. Variations in the limiting contrast (and thus, planet mass) at a given separation among the stars in the PALMS survey will provide more sensitivity to these parameters, particularly $\alpha$, as these detection limits will sample our planet distribution functions at a range of masses and semimajor axes that is not probed by the mean sensitivity curve we employ for this order of magnitude comparison.

Finally, we note that there is not much difference between the constraints obtained from the ``hot-start'' models versus those obtained using the ``cold-start'' models. We explain the reasons for this in \S~\ref{subsec:results_syn_constraints}.

{\bf \citet{Lafreniere2007}:} Our comparison with the GDPS is similar to that described above for the PALMS survey, where the number of expected planet detections per star for a given planet population is given by equation~(\ref{eqn:ndet_img_oom}). Here, we assume a different form for $\Phi_{\rm det}(\log{m_p},\log{a})$ to represent the average detection efficiency of the GDPS in terms of $\log{m_p}$ and $\log{a}$. In their figure~4, \citet{Lafreniere2007} present a median detection limit for a star at the median distance ($\sim 22~$pc) of their sample. We assume the relation $\Phi_{\rm det}(\log{m_p},\log{a})=\Theta(\log{m_p}-\log{m_{p,{\rm min, L07}}})\Theta(\log{a}-\log{a_{\rm min, L07}})\Theta(\log{a_{\rm max, L07}}-\log{a})$, where $m_{p,{\rm min, L07}}$, $a_{\rm min, L07}$, and $a_{\rm max, L07}$ are the minimum-detectable planet mass, and the minimum- and maximum-detectable semimajor axes, respectively, for the survey. There is a maximum-detectable semimajor axis that is set by the field of view (FOV) of the GDPS observations of $\sim 11''$ in radius, which corresponds a projected separation in physical units of roughly $r_{\perp}\sim 250~$AU at their median target distance, and thus we adopt $a_{\rm max, L07}\sim 290$, implicitly assuming a randomly-oriented, circular orbit.

We infer values of $m_{p,{\rm min, L07}}$ by taking the maximum contrast to which the GDPS is sensitive of $\Delta {\rm mag}=16.3$ in the CH4-short filter (given by the representative median sensitivity curve plotted in their figure~4) and converting this to a planet mass using their median distance, an age of 100~Myr, and a host absolute H-band magnitude of 4.0 (assumed to be equivalent to its magnitude in the CH4-short filter---see \citealt{Lafreniere2007}), along with the predictions of planet evolutionary models. For the ``hot-start'' models of \citet{Baraffe2003} we derive a minimum-detectable mass of $m_{p,{\rm min, LO7,hot}}\approx 1.4~M_{\rm Jup}$, and for the ''cold-start'' models of \citet{Fortney2008} we derive a minimum-detectable mass of $m_{p,{\rm min, L07,cold}}\approx 3.2~M_{\rm Jup}$. We adopt a minimum-detectable semimajor axis of $a_{\rm min, L07}\approx 12~$AU (which corresponds to a median value of $r_{\perp, {\rm min, L07}}\approx 10~$AU for a randomly-oriented, circular orbit) directly from figure~4 of \citet{Lafreniere2007}. We again make the assumption that the resultant form of $\Phi_{\rm det}(\log{m_p},\log{a})$ is representative of all 16 M dwarfs in the GDPS sample.

Under these assumptions, equation~(\ref{eqn:ndet_img_oom}) takes the form
\begin{align}
N_{{\rm L07}, i} = & {} \mathcal{A}_i\int^{\log{(a_{{\rm max}, i})}}_{\log{(12~{\rm AU})}}\int^{\log{(13~M_{\rm Jup})}}_{\log{m_{p,{\rm min, L07}}}}\left(\frac{m_p}{M_{\rm Sat}}\right)^{\alpha_i} \nonumber \\
    & {} \times \left(\frac{a}{2.5~{\rm AU}}\right)^{\beta_i}d\log{m_p}~d\log{a}\; , \label{eqn:ndet_l07_oom}
\end{align}
where we have
\begin{equation}
    a_{{\rm max}, i} =
    \begin{cases}
        \max(12~{\rm AU}, a_{{\rm out}, i}) & {\rm for}~a_{{\rm out}, i} \leq 290~{\rm AU}, \\
        290~{\rm AU} & {\rm for}~a_{{\rm out}, i} > 290~{\rm AU},
    \end{cases}
\end{equation}
and
\begin{equation}
    m_{p,{\rm min, L07}} =
    \begin{cases}
        1.4~M_{\rm Jup} & \text{for ``hot-start''}, \\
        3.2~M_{\rm Jup} & \text{for ``cold-start''}.
    \end{cases}
\end{equation}

For a given planet population, the number of expected planet detections per star, $N_{{\rm L07}, i}$, for the GDPS M stars can then be determined analytically via equation~(\ref{eqn:ndet_l07_oom}). The total number of detections for the 16 M stars in the GDPS is simply $N_{{\rm tot, L07},i}=16N_{{\rm L07}, i}$. As before, we assign a statistical weight to each planet population according to the likelihood function we define to be $\mathscr{L}_{\rm L07}\equiv P(0|N_{{\rm tot, L07}, i})$, where $P(0|N_{{\rm tot, L07}, i})=\exp(-N_{{\rm tot, L07}, i})$ is the Poisson probability of detecting zero planets among the 16 M stars in the GDPS sample if the true number of detectable planets around such stars is actually $N_{{\rm tot, L07}, i}$. We plot this likelihood function in figure~\ref{fig:lhood_functions}, where the abscissa in this case corresponds to $N_{{\rm tot, L07}, i}$.

\begin{figure}[!t]
	\epsscale{1.1}
	\plotone{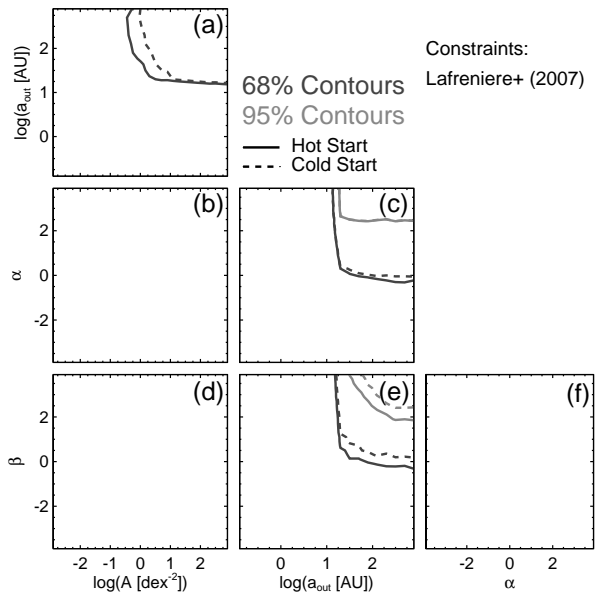}
	\caption{Likelihood contours as a function of pairs of parameters describing planet distribution functions that are found to be consistent (in our order of magnitude evaluations) with the non-detection of any planetary companions in the GDPS \citep{Lafreniere2007}. Contours are drawn at levels of $68\%$ and $95\%$ of the peak likelihood and are marginalized over all other parameters except those being plotted. Note that the contours enclose the bottom left regions of the panels.
		\label{fig:oom_constraints_l07}}
\end{figure}

Figure~\ref{fig:oom_constraints_l07} shows the resultant constraints on the parameters of populations we find to be consistent with the GDPS. The interpretation of these constraints (and non-constraints) is identical to that described for the PALMS survey above. Note that although the GDPS is, in general, deeper than the PALMS survey, their constraints are somewhat weaker due to the fact that the PALMS sample size is larger than that of the GDPS by a factor of 4.5.

\subsubsection{Radial Velocities}
\label{subsubsec:oom_rv}

\citet{Montet2014} report the measurement of four significant RV trends in the CPS M dwarf sample due (most probably) to planetary companions, demonstrating an ability to measure accelerations of $\dot v\sim 1~{\rm m~s^{-1}~yr^{-1}}$ over long time baselines ($\sim 10~$yr). In order to estimate the number of long-term RV trends, $N_{{\rm tr}, i}$, expected for the CPS M dwarfs for a given planet population, we use the median sensitivity we derive in \citet{Clanton2014b} for the CPS sample to determine the minimum-detectable acceleration as a function of semimajor axis, $\dot v_{\rm thr}(a)$. Mathematically, the number of expected detectable trends is quantified by equation~(\ref{eqn:ndet_img_oom}) with $N_{{\rm det, img}, i}\rightarrow N_{{\rm tr}, i}$ and $\Phi_{\rm det}(\log{m_p},\log{a})\rightarrow \Theta(P-T)\Theta(\dot v-\dot v_{\rm thr})$, where $T$ is the time baseline of observations for the CPS TRENDS survey. This integral cannot be evaluated analytically, so we instead evaluate it numerically with some assumptions.

First, we assume all planets described by our distribution function are on randomly oriented, circular orbits. In order to compute the magnitude of the RV trends these planets would induce, $\dot v$, we employ Kepler's third law (in the limit $M_l\gg m_p$), which yields the period
\begin{equation}
P(M_l, a) = 2\pi\left(\frac{a^3}{GM_l}\right)^{1/2}\; . \label{eqn:p_kep}
\end{equation}
For our typical parameter values, the period described by equation~(\ref{eqn:p_kep}) evaluates to $P\approx 5.6~{\rm yr}(M_l/0.5~M_{\odot})^{-1/2}(a/2.5~{\rm AU})^{3/2}$.

Equation~(\ref{eqn:p_kep}) in conjunction with the standard radial velocity equation yield the velocity semi-amplitude for a circular orbit as:
\begin{equation}
K(M_l, a, m_p, i) = \left(\frac{G}{aM_l}\right)^{1/2}m_p\sin{i}\; , \label{eqn:k_std}
\end{equation}
where $i$ is the inclination of the orbit. We adopt a convention wherein the inclination angle is measured relative to the plane of the sky, such that such that $i=0^\circ$ corresponds to an orbit viewed face-on, while $i=90^\circ$ corresponds to an orbit viewed edge-on. Equation~(\ref{eqn:k_std}) evaluates to $K\approx 7.6~{\rm m~s^{-1}}(M_l/0.5~{M_{\odot}})^{-1/2}(a/2.5~{\rm AU})^{-1/2}(m_p/M_{\rm Sat})\sin{i}$ for our typical parameter values.

The magnitude of the acceleration of a host star due to a companion on a circular orbit is $\dot v = (2\pi K\cos{M_0})/P$ (where $M_0$ is the mean anomaly, which describes the orbital phase at a given epoch), which combined with equations~(\ref{eqn:p_kep}--\ref{eqn:k_std}) takes the form
\begin{equation}
\dot v(a, m_p, i, M_0) = \left(\frac{Gm_p}{a^2}\right)\sin{i}\cos{M_0}\; . \label{eqn:dotv_std}
\end{equation}
Typically, this corresponds to an acceleration of $\dot v\approx 8.6~{\rm m~s^{-1}~yr^{-1}}(a/2.5~{\rm AU})^{-2}(m_p/M_{\rm Sat})\sin{i}\cos{M_0}$. When computing these accelerations, rather than marginalizing over unknown orbital parameters, we will assume that the inclination and mean anomaly of each system are at their median respective values of $\left.\sin{i}\right|_{\rm med}\approx 0.867$ and $\left.\cos{M_0}\right|_{\rm med}=0.5$ for the purposes of this order of magnitude calculation.

To determine which planets produce accelerations which are actually detectable by the CPS, we estimate the minimum-detectable acceleration as a function of their orbital period (and thus, implicity, $a$) in the following manner. In \citet{Clanton2014b}, we estimate the sensitivity of the CPS to each of their M dwarfs in terms of a signal-to-noise (S/N) ratio. We adopt these same sensitivities for this order of magnitude evaluation to infer $\dot v_{\rm thr}(P)$. In \citet{Clanton2014a}, we also derive an expression for the S/N at which an RV survey can detect planetary companions as a function of the magnitudes of the signal $K$ and measurement uncertainties $\sigma$, the time baseline of observations $T$, and total number of observations $N$. We found the ``phase-averaged'' S/N, which we designate as $\mathcal{Q}$, to be
\begin{equation}
\mathcal{Q} = \left(\frac{N}{2}\right)^{1/2}\left(\frac{K}{\sigma}\right)\left\{1-\frac{1}{\pi^2}\left(\frac{P}{T}\right)^2\sin^2{\left(\frac{\pi T}{P}\right)}\right\}^{1/2}\; , \label{eqn:q_snr}
\end{equation}
for which we find the CPS sample to have a median value of $\mathcal{Q}_{\rm med}=8.26$ \citep{Clanton2014b}.

\begin{figure}[!t]
	\epsscale{1.1}
	\plotone{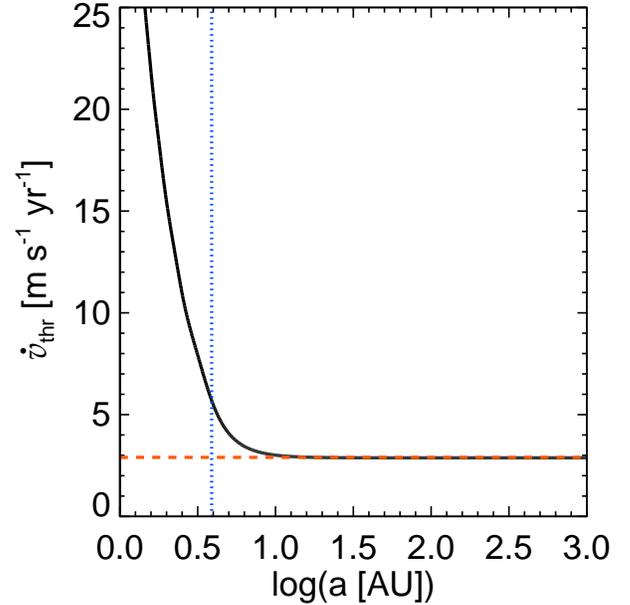}
	\caption{The minimum-detectable acceleration as a function of semimajor axis at a S/N of $\mathcal{Q}=8.26$ for an RV-monitored target with $N=29$ observations over a time baseline of $T=11.8~$yr, measurement uncertainties of $\sigma=4.5~{\rm m~s^{-1}}$, and a stellar mass of $M_{\star}=0.41~M_{\odot}$. These values are the median parameters of the CPS M dwarf sample. The dotted blue, vertical line marks the log semimajor axis that roughly corresponds to $T=11.8~$yr, $\log{a_T}\sim 0.59$, assuming $M_{\star}= 0.41~M_{\odot}$. In the limit $a\gg a_T$, $\dot v_{\rm thr}$ becomes a very weak function of the orbital period (and thus $a$), and is approximately $2.9~{\rm m~s^{-1}~yr^{-1}}$ for the chosen parameter values (shown by the red, dashed line).
		\label{fig:dotv_thr}}
\end{figure}

Solving equation~(\ref{eqn:q_snr}) for $K$ yields the minimum-detectable velocity semi-amplitude, $K_{\rm min}$, for a given value of $\mathcal{Q}$, which we then use to determine the minimum-detectable acceleration as $\dot v_{\rm thr}(P) = (2\pi K_{\rm min})/P$. We derive this to be
\begin{align}
\dot v_{\rm thr} (P) = & {} \left(\frac{2\pi}{P}\right)\left(\frac{2}{N}\right)^{1/2}\sigma\mathcal{Q} \nonumber \\
    & {} \times \left\{1-\frac{1}{\pi^2}\left(\frac{P}{T}\right)^2\sin^2{\left(\frac{\pi T}{P}\right)}\right\}^{-1/2}\; , \label{eqn:dotv_thr}
\end{align}
which is implicity a function of $a$ via Kepler's third law (equation~\ref{eqn:p_kep}). Since we are aiming for a rough estimate, we take the median parameters of the CPS M dwarfs to determine $\dot v_{\rm thr}$ and assume the resulting threshold as a function of $a$ is the same for each of the 111 M dwarfs in their sample. The median parameters for the CPS sample are $N_{\rm med}=29$, $\sigma_{\rm med}=4.5~{\rm m~s^{-1}}$ (inclues instrumental noise and stellar jitter), and $T_{\rm med}=11.8~$yr. We plot this minimum-detectable acceleration as a function of log semimajor axis in figure~\ref{fig:dotv_thr}. The dotted blue, vertical line marks the log semimajor axis that roughly corresponds to the median time baseline of observations, $T_{\rm med}$, for the CPS M dwarfs, $\log{a_T}\sim 0.59$, assuming the median stellar mass of the CPS M dwarfs of $M_{\star}\sim 0.41~M_{\odot}$. As $a$ increases beyond $a_T$, $\dot v_{\rm thr}$ becomes an extremely weak function of $P$ (and thus $a$). Expanding the Maclaurin series of $\sin^2(x)=x^2-x^4/3+\mathcal{O}(x^6)$, where $x\equiv \pi T/P$, and substituting this expression into equation~(\ref{eqn:dotv_thr}), we find
\begin{equation}
\dot v_{\rm thr}\simeq \left(\frac{24}{N}\right)^{1/2}\left(\frac{\sigma}{T}\right)\mathcal{Q}\; , \label{eqn:dotv_thr_pggt}
\end{equation}
in the limit of $T \ll P$. This evaluates to $\dot v_{\rm thr}\approx 2.9~{\rm m~s^{-1}~yr^{-1}}$ at the median parameter values for the CPS M dwarfs listed above. We plot this in figure~\ref{fig:dotv_thr} as a red, dashed, horizontal line, verifying that it is a good approximation for equation~(\ref{eqn:dotv_thr}), shown by the solid, black curve, for $a\gg a_T$.

For each planet population, we numerically evaluate equation~(\ref{eqn:ndet_img_oom}), making use of equations~(\ref{eqn:p_kep}), (\ref{eqn:dotv_std}), and (\ref{eqn:dotv_thr}), and adopting $\Phi_{\rm det}(\log{m_p},\log{a})\rightarrow \Theta(P-T)\Theta(\dot v-\dot v_{\rm thr})$, $a_{\rm min}=0.1~$AU, $a_{{\rm max}, i}=\max(a_{\rm min}, a_{{\rm out}, i})$, $m_{p, {\rm min}}=1~M_{\oplus}$, and $m_{p, {\rm max}}=13~M_{\rm Jup}$. This yields the expected number of long-term RV trend detections per star, $N_{{\rm tr}, i}$, for the CPS TRENDS survey, assuming their median sample parameters. The total number of trends we expect the CPS to detect for a given planet population is then $N_{{\rm tot, tr}, i} = 111N_{{\rm tr}, i}$. Since \citet{Montet2014} report the detection of four significant long-term RV trends, we adopt a likelihood function of the form $\mathscr{L}_{\rm tr}\equiv P(4|N_{{\rm tot, tr}, i})$, where $P(4|N_{{\rm tot,tr}, i})=(N_{{\rm tot, tr}, i})^4\exp(-N_{{\rm tot,tr}, i})/4!$ is the Poisson probability of detecting exactly four trends from the 111 M stars in the CPS sample if $N_{{\rm tot,tr}, i}$ is the true mean number of detectable trends within the sample. We plot this likelihood function in figure~\ref{fig:lhood_functions}, where the abscissa in this case corresponds to $N_{{\rm tot, tr}, i}$.

\begin{figure}[!t]
	\epsscale{1.1}
	\plotone{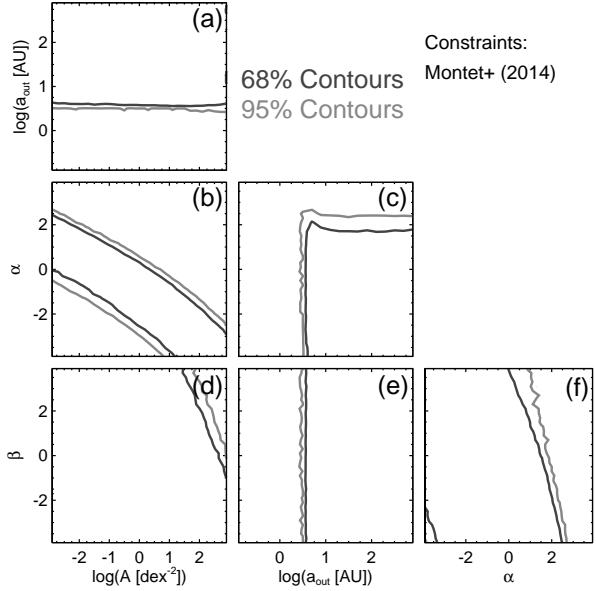}
	\caption{Likelihood contours as a function of pairs of parameters describing planet distribution functions that are found to be consistent (in our order of magnitude evaluations) with the detection of four significant, long-term RV trends in the CPS M dwarf sample \citep{Montet2014}. Contours are drawn at levels of $68\%$ and $95\%$ of the peak likelihood and are marginalized over all other parameters except those being plotted.
		\label{fig:oom_constraints_mb14}}
\end{figure}

Figure~\ref{fig:oom_constraints_mb14} shows the resultant constraints on the parameters of populations consistent with the results of \citet{Montet2014}. There is not much orthogonality between these constraints and those supplied by the \citet{Gould2010} microlensing survey; the RV trends also put a lower bound on $a_{\rm out}$ and constrain a combination of $\alpha$ and $\mathcal{A}$, but do not put an upper bound on $\mathcal{A}$. The lower bound on $a_{\rm out}$ (panels~a, c, and e) comes from the fact that we require the period of planets that we expect to be detected as a trend to be longer than the time baseline of observations, i.e. $P>T$. Since we use only a median sensitivity to be representative of the full CPS M dwarf sample, this lower bound on $a_{\rm out}$ is sharp at $a=a_T=3.85~$AU. Later in this paper, we will make use of each individual sensitivity limit and the lower bound placed on $a_{\rm out}$ will not be as sharp, as the stars in the CPS sample have been monitored for a range of $T$. The observed anti-correlation between $\alpha$ and $\mathcal{A}$ in panel~(b) is a reflection of the fact that the pivot point of our planet distribution function is outside the region of sensitivity for the CPS trends, as illustrated in figure~\ref{fig:regions_plot}. At fixed $a_{\rm out}$ and $\beta$, increases (decreases) in $\mathcal{A}$ must be compensated by decreases (increases) in $\alpha$ to preserve the number of observed trends.

The apparent constraints on the combination of $\alpha$ and $\beta$ in panel~(f) are not actual constraints; if we were to consider a larger range of $\mathcal{A}$, this full space would be allowed. Similarly, there would not appear to be a maximum allowed value of $\alpha$, as implied by panel~(c).

\subsection{Combined Order of Magnitude Constraints}
\label{subsec:combined_oom_constraints}
Individually, the results of any single exoplanet discovery survey are unable to place precise constraints on each of the four parameters of our assumed planet distribution function. The results of single surveys are limited to small regions of planet parameter space (see figure~\ref{fig:regions_plot}), but we can obtain real constraining power by simultaneously comparing with all surveys.

\begin{figure}[!t]
	\epsscale{1.1}
	\plotone{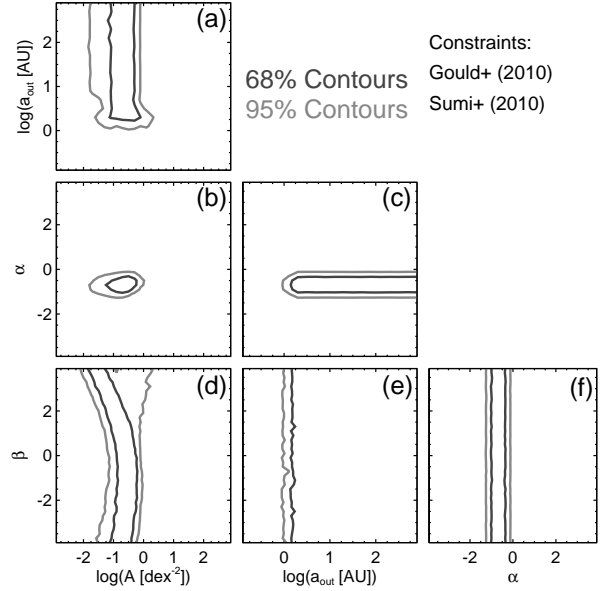}
	\caption{Likelihood contours as a function of pairs of parameters describing planet distribution functions that are found to be consistent (in our order of magnitude evaluations) with the results of the microlensing surveys of \citet{Gould2010} and \citet{Sumi2010}. Contours are drawn at levels of $68\%$ and $95\%$ of the peak likelihood and are marginalized over all other parameters except those being plotted.
		\label{fig:oom_constraints_g10s10}}
\end{figure}

As shown in the preceeding sections, the \citet{Gould2010} microlensing survey constrains the combination of $\alpha$ and $\mathcal{A}$. 
Since the \citet{Sumi2010} survey basically fixes $\alpha$, the combination of this constraint with the results of the \citet{Gould2010} survey then fixes $\mathcal{A}$. This is illustrated in panel~(b) of figure~\ref{fig:oom_constraints_g10s10}, which displays the combined constraints on the parameters of our planet distribution function from both the \citet{Gould2010} and \citet{Sumi2010} microlensing surveys. However, we find that a degeneracy between $\beta$ and $\mathcal{A}$ remains (panel~d), and there is still no constraining power on $a_{\rm out}$, other than the previously described lower limit of $a_{\rm out}>1.4~$AU.

With such tight constraints on $\alpha$ and $\mathcal{A}$, the results of direct imaging surveys \citep{Lafreniere2007,Bowler2015} and those of the CPS TRENDS survey \citep{Montet2014} become sensitive to the values of $\beta$ and $a_{\rm out}$ in complementary ways. As we will demonstrate, the RV trends exclude small values of $\beta$ and put a lower bound on $a_{\rm out}$, while direct imaging surveys tend to exclude large values of $\beta$ and limit large values of $a_{\rm out}$ (to a certain point that is dictated by the FOV of the direct imaging surveys).

\begin{figure}[!b]
	\epsscale{1.1}
	\plotone{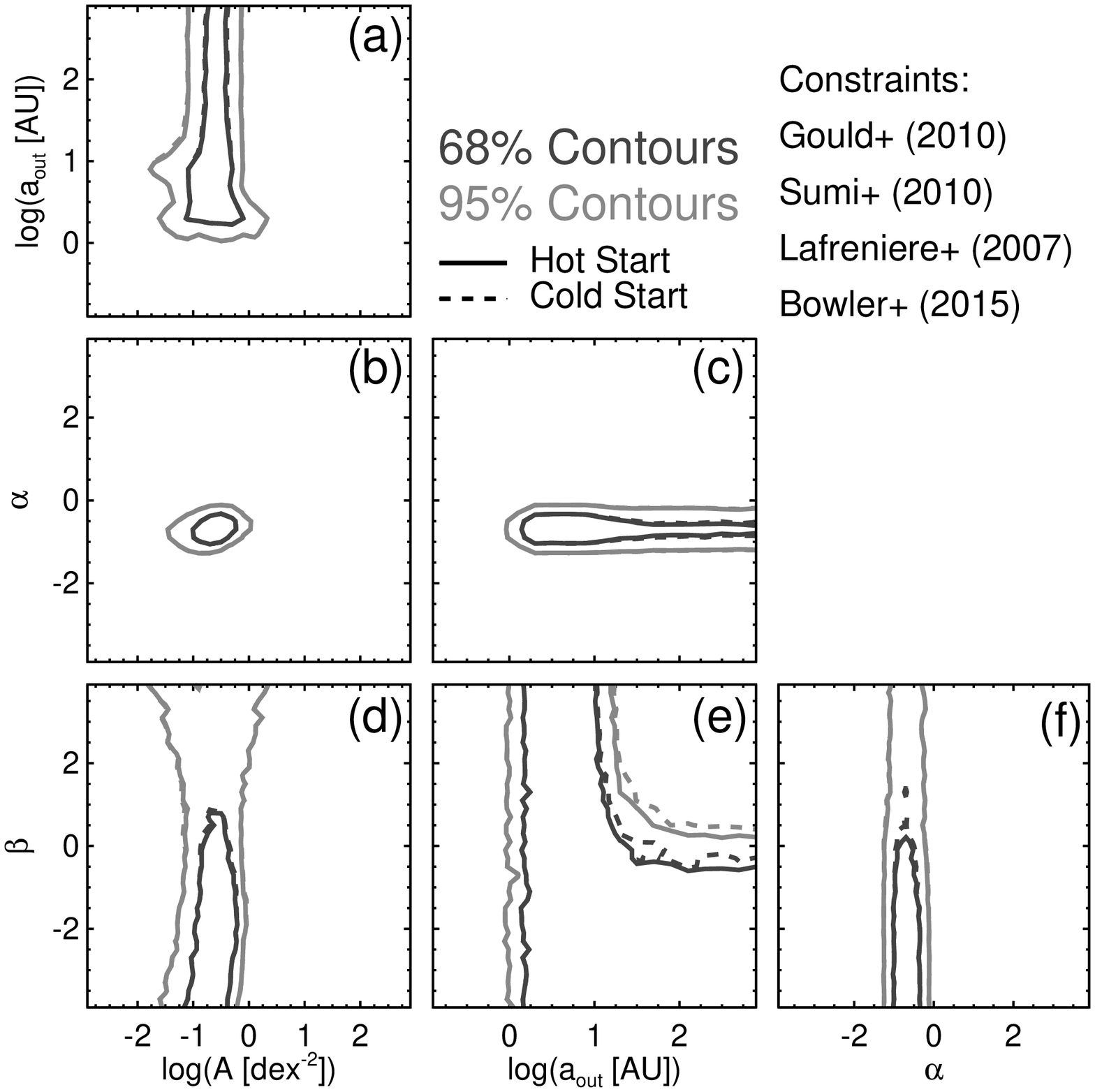}
	\caption{Likelihood contours as a function of pairs of parameters describing planet distribution functions that are found to be consistent (in our order of magnitude evaluations) with the results of microlensing \citep{Gould2010,Sumi2010} and direct imaging \citep{Lafreniere2007,Bowler2015} surveys. Contours are drawn at levels of $68\%$ and $95\%$ of the peak likelihood and are marginalized over all other parameters except those being plotted.
		\label{fig:oom_constraints_g10s10l07bb15}}
\end{figure}

Figure~\ref{fig:oom_constraints_g10s10l07bb15} shows our constraints from the combination of microlensing and direct imaging surveys. Panels~(d), (e), and (f) demonstrate that inclusion of direct imaging results excludes large values of $\beta$ (for $a_{\rm out}\gtrsim 10~$AU where there is no sensitivity from direct imaging) which would

\begin{figure}[!b]
	\epsscale{1.1}
	\plotone{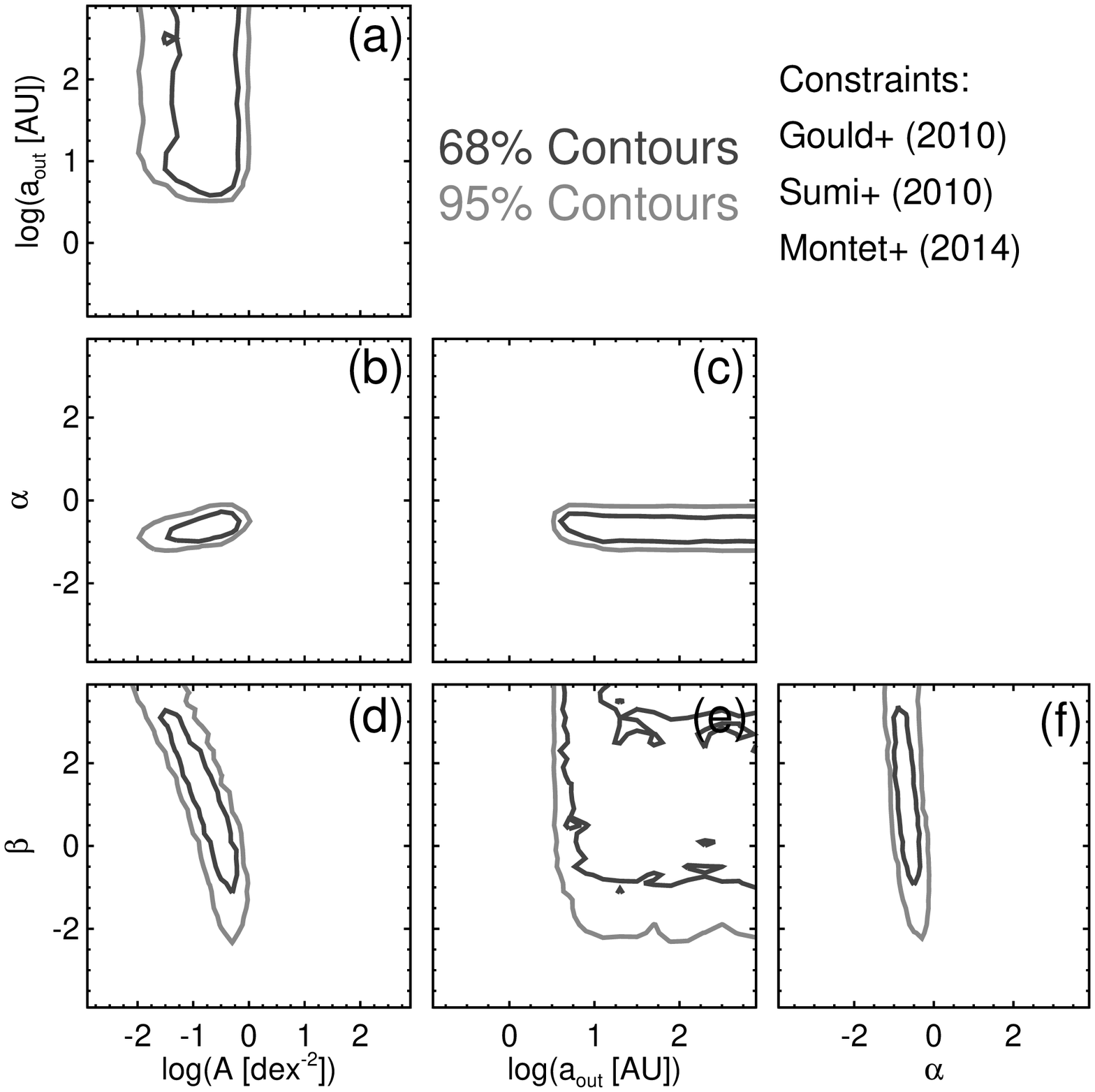}
	\caption{Likelihood contours as a function of pairs of parameters describing planet distribution functions that are found to be consistent (in our order of magnitude evaluations) with the results of microlensing surveys \citep{Gould2010,Sumi2010} and the RV trends detected by the CPS TRENDS survey \citep{Montet2014}. Contours are drawn at levels of $68\%$ and $95\%$ of the peak likelihood and are marginalized over all other parameters except those being plotted.
		\label{fig:oom_constraints_g10s10mb14}}
\end{figure}

\begin{figure*}[!ht]
	\epsscale{1.0}
	\plotone{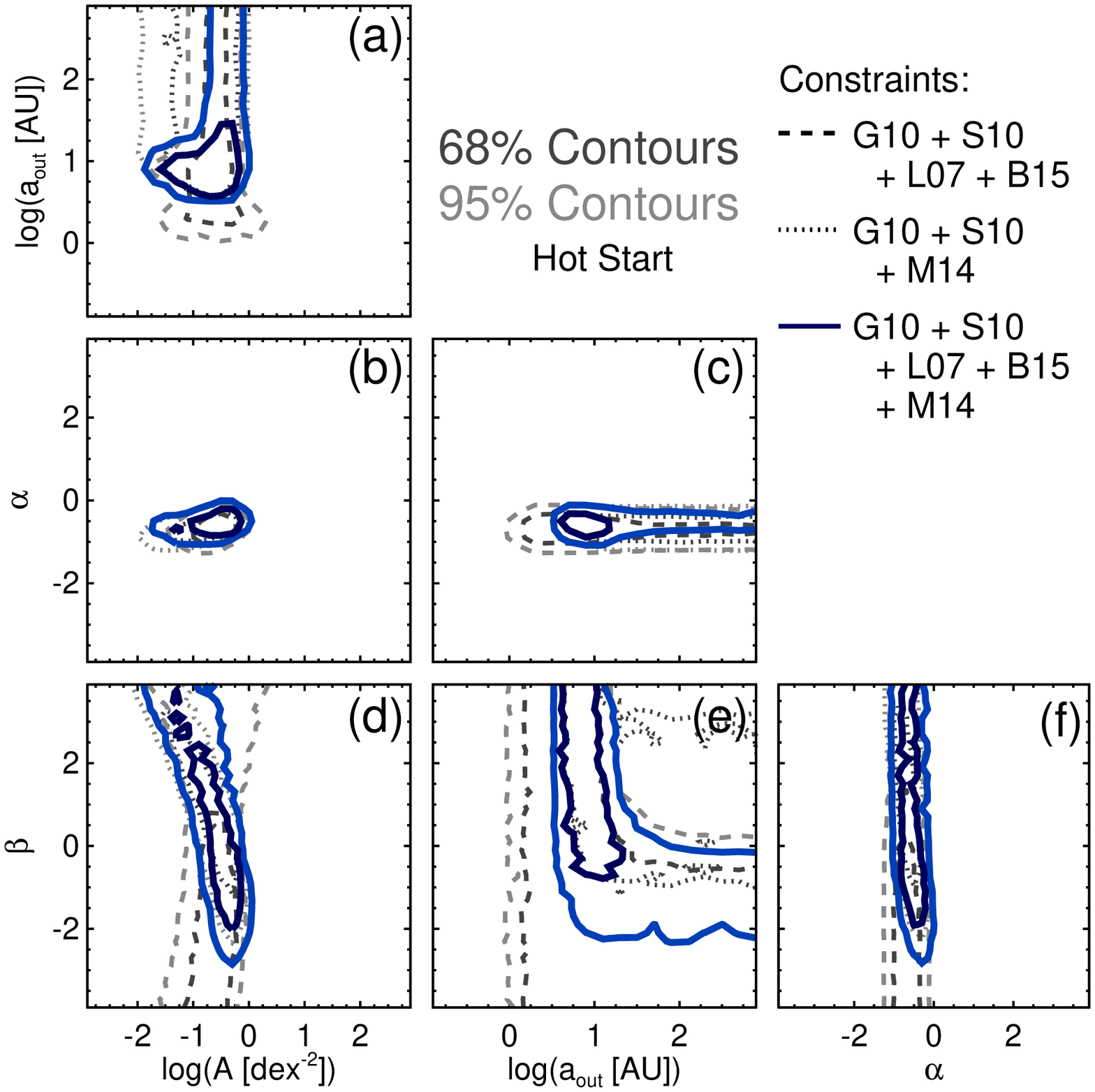}
	\caption{Likelihood contours as a function of pairs of parameters describing planet distribution functions that are found to be consistent (in our order of magnitude evaluations) with the results of microlensing (G10--\citealt{Gould2010}; S10--\citealt{Sumi2010}), direct imaging (L07--\citealt{Lafreniere2007}; B15--\citealt{Bowler2015}), and RV (M14--\citealt{Montet2014}) surveys. Here, we have utilized the predictions of the ``hot-start'' planet evolutionary models of \citet{Baraffe2003}. Contours are drawn at levels of $68\%$ and $95\%$ of the peak likelihood and are marginalized over all other parameters except those being plotted.
		\label{fig:oom_constraints_main5_combos_hot}}
\end{figure*}

\noindent predict (at the essentially fixed values of $\alpha$ and $\mathcal{A}$) large numbers of direct detections that are inconsistent with the observations. Panel~(e) shows that as $\beta$ decreases, $a_{\rm out}$ must decrease, up to the maximum separation at which these surveys are able to detect companions ($\sim 300$~AU for the GDPS), set by their FOV.

Figure~\ref{fig:oom_constraints_g10s10mb14} shows our constraints from the combination of results from microlensing surveys and the trends detected by the CPS TRENDS survey. Panels~(d), (e), and (f) demonstrate that inclusion of RV trend detections excludes small values of $\beta$ that would predict (at the essentially fixed values of $\alpha$ and $\mathcal{A}$) too few long-term trend detections to be consistent with the observations of \citet{Montet2014}. Panels~(a), (c), and (e) show that there is a lower bound on $a_{\rm out}$, interior to which we would predict no planets to be detected as trends. Since the acceleration of a host star induced by the presence of a planetary companion depends strongly on orbital separation ($\dot v\propto a^{-2}$), the sensitivity of RV surveys to trend detections rapidly declines with $a$. Thus, there is no constraining power on the upper limit for $a_{\rm out}$ by inclusion of the RV results of \citet{Montet2014}.

\begin{figure*}[!t]
	\epsscale{1.0}
	\plotone{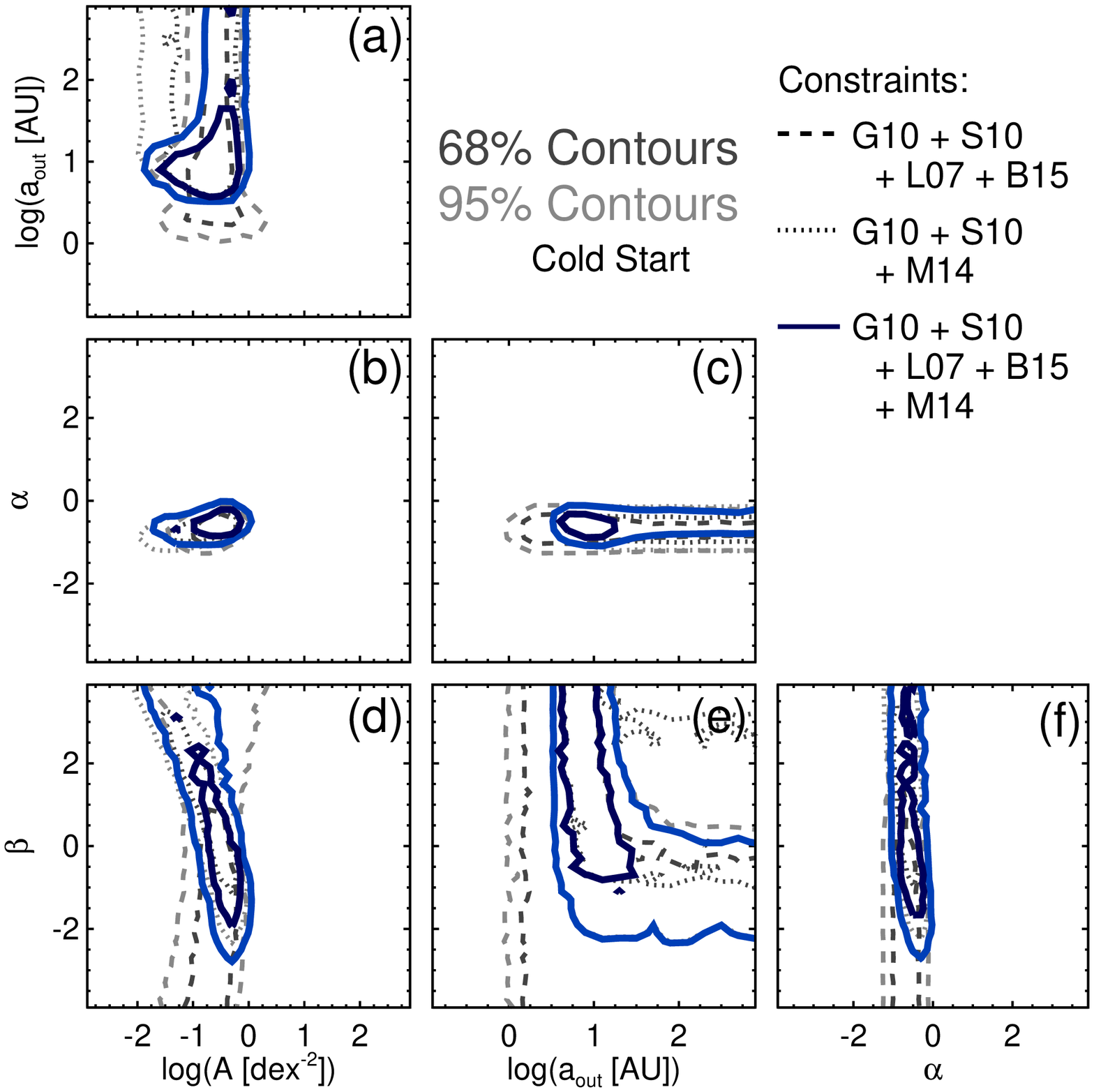}
	\caption{Identical to figure~\ref{fig:oom_constraints_main5_combos_hot}, but utilizing the predictions of the ``cold-start'' planet evolutionary models of \citet{Fortney2008}.
		\label{fig:oom_constraints_main5_combos_cold}}
\end{figure*}

Finally, combining the results of all microlensing, direct imaging, and RV surveys we consider, we derive the constraints displayed in figures~\ref{fig:oom_constraints_main5_combos_hot} (``hot-start models'') and \ref{fig:oom_constraints_main5_combos_cold} (``cold-start models''). For reference, we have also included the constraints obtained by combining the above-described combinations of survey results. These constraints are derived from order of magnitude-level comparisons, but suggest that our simple, power-law planet distribution function can simultaneously explain the results of multiple exoplanet surveys, warranting more detailed and careful calculations (described in \S~\ref{sec:methodology}).

\begin{figure*}[!t]
	\epsscale{0.8}
	\plotone{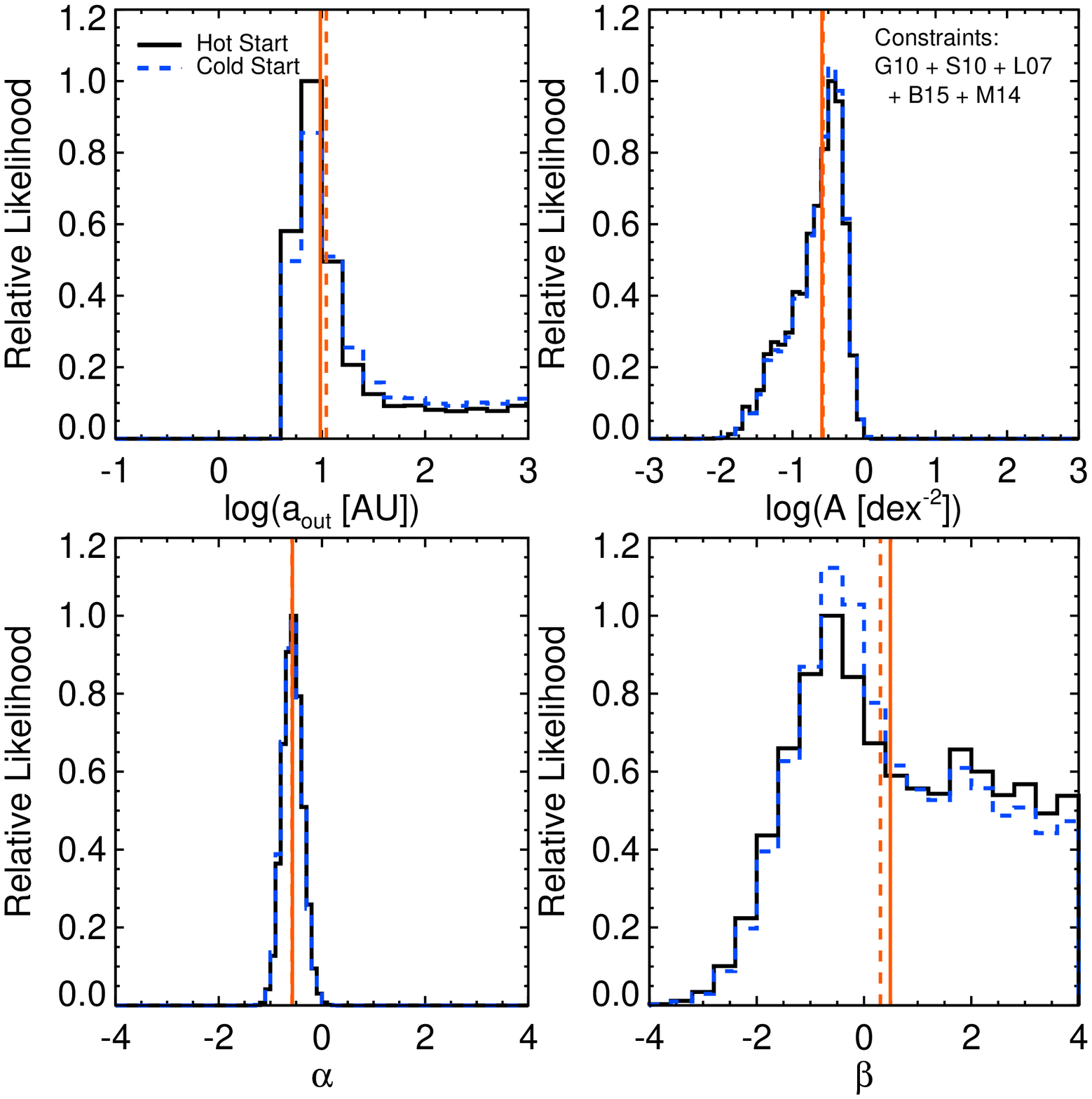}
	\caption{Relative likelihoods for each parameter, marginalized over the other three parameters, derived from our order of magnitude comparison with microlensing (G10--\citealt{Gould2010}; S10--\citealt{Sumi2010}), direct imaging (L07--\citealt{Lafreniere2007}; B15--\citealt{Bowler2015}), and RV (M14--\citealt{Montet2014}) surveys. The vertical, red lines in each plot indicate the median values for the ``hot-start'' models (solid) and ``cold-start'' models (dashed). The median values and their 68\% confidence intervals are listed in table~\ref{tab:1d_oom_medians}.
		\label{fig:oom_constraints_1d_main5}}
\end{figure*}

We plot the corresponding one-dimensional likelihood functions on each parameter, marginalized over the other three parameters, in figure~\ref{fig:oom_constraints_1d_main5}, and summarize the median values (and their 68\% confidence intervals) in table~\ref{tab:1d_oom_medians} for both ``hot-'' and ``cold-start'' planet evolutionary models.

\begin{table*}[!t]
	\caption{\label{tab:1d_oom_medians} Median values and 68\% uncertainties on the parameters of our planet distribution models for our order of magnitude comparisons with the results of the microlensing surveys by \citet{Gould2010} and \citet{Sumi2010}, the GDPS \citep{Lafreniere2007} and PALMS \citep{Bowler2015} direct imaging surveys, and the CPS TRENDS survey of long-term RV trends \citep{Montet2014}.}
	\centering
	\begin{tabular}{c||c|c|c|c}
		\hline \hline
		Planet Evolutionary  &  \multicolumn{4}{c}{Median Values and 68\% Uncertainties} \\
		Models &  $\alpha$  & $\beta$  & $\mathcal{A}~[{\rm dex^{-2}}]$  & $a_{\rm out}~[{\rm AU}]$ \\
		\hline
		\begin{tabular}{@{}c@{}} ``Hot-Start'' \\ (\citealt{Baraffe2003})\end{tabular} & $-0.57^{+0.20}_{-0.19}$& $0.49^{+2.32}_{-1.63}$ & $0.26^{+0.21}_{-0.18}$ & $9.7^{+62}_{-3.6}$ \\
		\hline
		\begin{tabular}{@{}c@{}} ``Cold-Start'' \\ (\citealt{Fortney2008})\end{tabular} & $-0.58^{+0.20}_{-0.19}$& $0.31^{+2.39}_{-1.39}$ & $0.27^{+0.20}_{-0.18}$ & $11^{+105}_{-5}$ \\
		\hline\hline
	\end{tabular}
\end{table*}

\begin{figure}[!t]
	\epsscale{1.1}
	\plotone{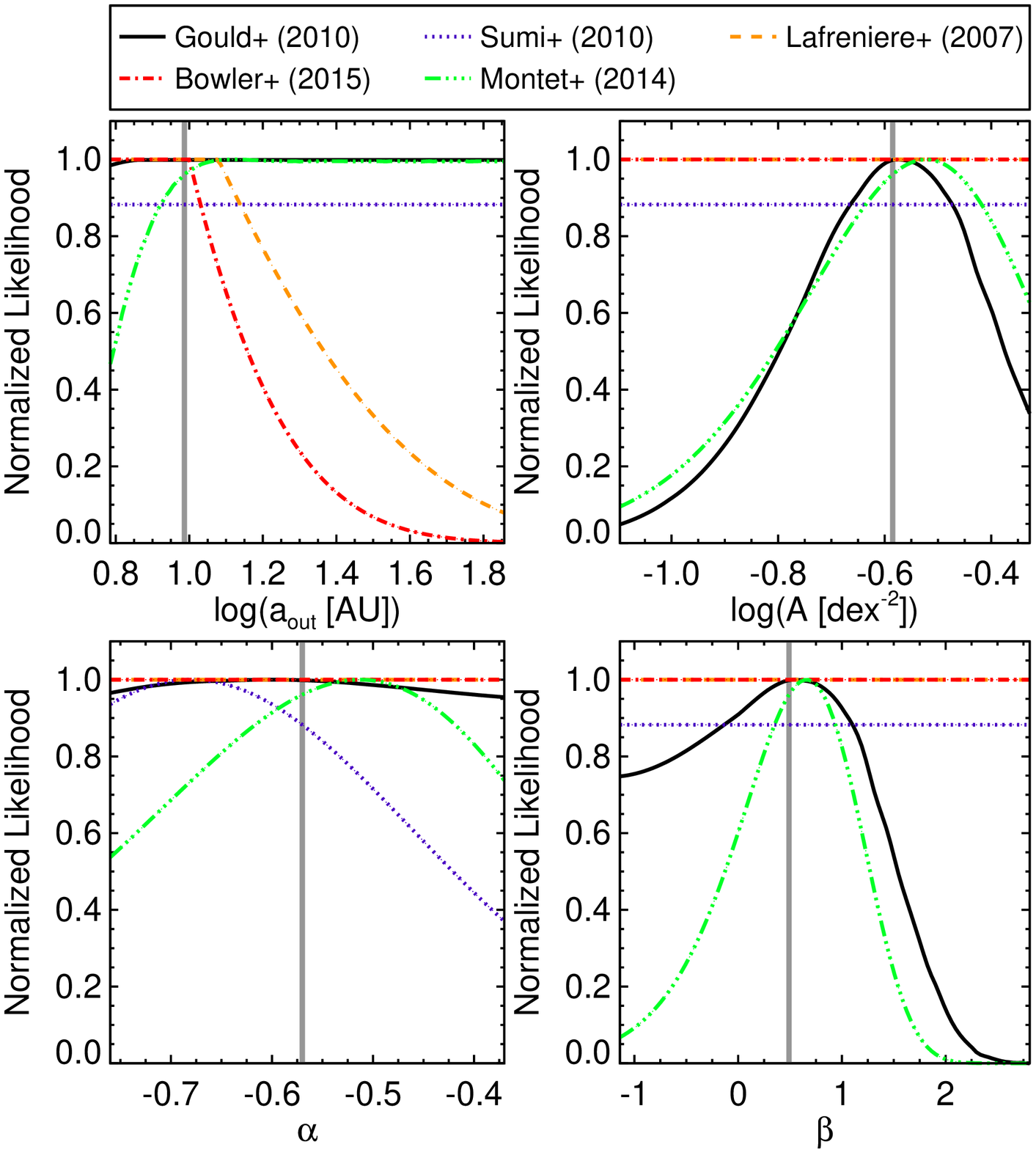}
	\caption{Likelihood functions of the parameters of our planet population model derived from individual exoplanet surveys across the 68\% confidence intervals we infer from our order of magnitude comparison for each parameter from the combination of all surveys, assuming the remaining three parameters are at their inferred median values. The vertical, grey lines indicate the median values of each parameter listed in table~\ref{tab:1d_oom_medians} for the ``Hot-Start'' results. Note that the orange lines (representing the likelihood according to the results of \citealt{Lafreniere2007}) in the upper right and bottom panels are included, but are not visible, as they are behind the red lines at a likelihood of one, since in each of these panels, $a_{\rm out}$ is held fixed at its median value, which lies interior to the inner working angle of the GDPS.
		\label{fig:range_lhoods_plot_main5_hot}}
\end{figure}

\begin{figure}[!t]
	\epsscale{1.1}
	\plotone{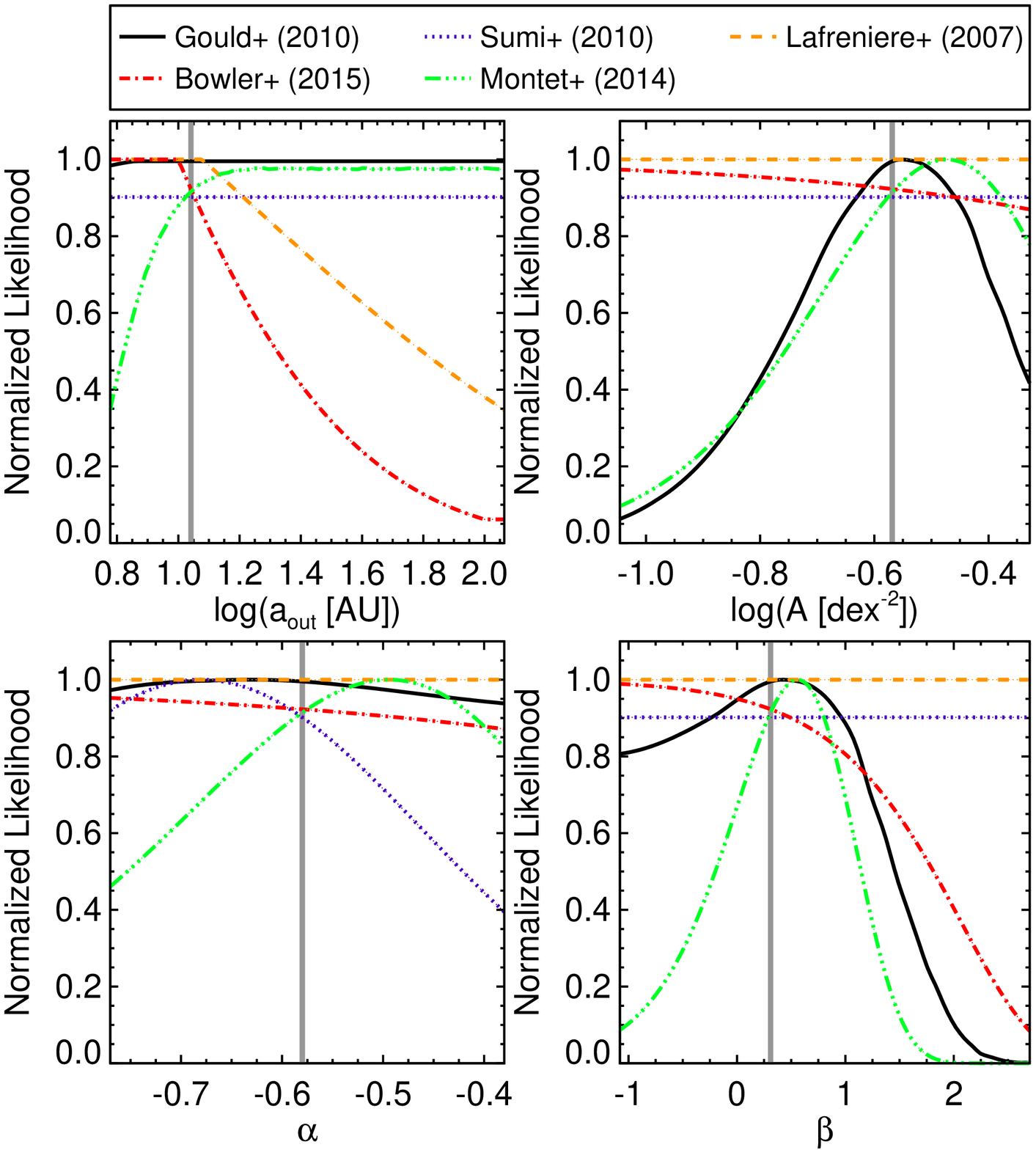}
	\caption{Identical to figure~\ref{fig:range_lhoods_plot_main5_hot}, but utilizing the inferred median values and 68\% confidence intervals inferred for the ``Cold-Start'' comparison results listed in table~\ref{tab:1d_oom_medians}.
		\label{fig:range_lhoods_plot_main5_cold}}
\end{figure}

We have shown that our order of magnitude comparisons reveal that there is overlap in parameter space from the results of microlensing, RV, and direct imaging surveys. We are also interested in determining the probability of the resultant parameter distributions given by each survey individually. That is, how much does the likelihood function from a given survey vary over the 68\% confidence interval determined from the combination of all surveys? In figures~\ref{fig:range_lhoods_plot_main5_hot} (``hot-start'') and \ref{fig:range_lhoods_plot_main5_cold} (``cold-start''), we plot the normalized likelihood functions (i.e. the likelihoods normalized by the maximum likelihood value for each individual survey) of given parameters across their 68\% confidence intervals, assuming the remaining three parameters are at their median values (see table~\ref{tab:1d_oom_medians}). These figures show that the likelihoods of these parameters from each individual survey are consistent to within $1\sigma$ (i.e. their normalized likelihood is greater than $\approx 0.32$) across much of their inferred 68\% confidence intervals, and with normalized likelihoods $\gtrsim 0.85$ at the median value for all surveys (indicated by the vertical, grey lines). This demonstrates that this overlap in parameter space we identify is not improbable, but has a rather good chance of being real. In other words, these order of magnitude evaluations indicate that the results of microlensing, RV, and direct imaging surveys can be plausibly explained by a single planet population modeled with power-law distributions in mass and semimajor axis.

Note that while the direct imaging results, particularly those of the PALMS survey, in the ``cold-start'' case (figure~\ref{fig:range_lhoods_plot_main5_cold}) seem to be more constraining than those in the ``hot-start'' case (figure~\ref{fig:range_lhoods_plot_main5_hot}), this is not actually the case. The median value of $a_{\rm out}$ in the ``cold-start'' case is larger than the minimum-detectable semimajor axis for the PALMS survey ($10~$AU), while this is not so for the ``hot-start'' case (refer to table~\ref{tab:1d_oom_medians}). This is not surprising, as one would expect a larger median value on $a_{\rm out}$ for the ``cold-start'' models relative to the ``hot-start'' models. However, it does make the interpretation of figures~\ref{fig:range_lhoods_plot_main5_hot} and \ref{fig:range_lhoods_plot_main5_cold} a bit complicated. Since the values of the parameters over which we marginalize in each panel of these figures are set to the their respective median values, direct imaging surveys are fully consistent (normalized likelihood of one) over the full 68\% confidence interval of the parameter plotted in each panel, as zero detections are predicted for $a_{\rm out}<10~$AU, with the exception of the top-left panel displaying the likelihood of $a_{\rm out}$ itself. Thus, since the median value of $a_{\rm out}$ is larger than $10~$AU for the ``cold-start'' models, the number of expected direct imaging detections is non-zero, giving rise to the constraints within the 68\% confidence intervals each parameter displayed in figure~\ref{fig:range_lhoods_plot_main5_cold}.

\section{Methodology}
\label{sec:methodology}
The order of magnitude-style calculations presented in the last section are useful for developing intuition about the ways in which individual surveys and their combination can constrain the distribution of properties of planets. In order to obtain more robust constraints, we now relax as many assumptions from those calculations as possible. In this section, we account for variations in stellar properties (mass, distance, age), Galactic properties (velocity dispersion, stellar number density along a given sight line), planetary orbital parameters, planet evolutionary models (``hot-start'', ``cold-start''), and detection sensitivities for each star/target in each survey.

The general procedure we adopt is as follows. We begin by generating a population of planets from our simple, power-law distribution function, assigning randomly oriented orbits and host star properties. Next, we derive the expected observables for each planet in the population for microlensing, RV, and direct imaging observations. Applying the sensitivity limits of the various surveys, we predict the results of each survey given the specific planet population under consideration, comparing these predictions with the actual results to assign a likelihood to the population. This process is repeated many different planet populations until we have adequately sampled all the relevant parameter space. We describe this methodology in detail in the remainder of this section. 

\subsection{Generating Planet Populations}
\label{subsec:gen_planet_pops}

\subsubsection{Masses and Separations}
\label{subsubsec:masses_seps}
In a given realization, $i$, corresponding to a single planet population described by $\{\alpha_i, \beta_i, \mathcal{A}_i, a_{{\rm out}, i}\}$, we generate an ensemble of $N_{\rm ens}=10^6$ planets enumerated by the index $j$. Each planet is described by a mass, $m_{p, i, j}$, and a semimajor axis, $a_{i, j}$, drawn from the distribution function
\begin{equation}
    \left.\frac{d^2N_{\rm pl}}{d\log{m_p}~d\log{a}}\right|_{i} = \mathcal{A}_i\left(\frac{m_p}{M_{\rm Sat}}\right)^{\alpha_i}\left(\frac{a}{2.5~{\rm AU}}\right)^{\beta_i}\; .
\end{equation}
The average number of planets per star in each realization within the range of planetary masses and separations we consider is then given by
\begin{align}
    \mathcal{F}_i = & {} \displaystyle \int_{\log{(a_{\rm in})}}^{\log{(a_{{\rm out},i})}} \int_{\log{(m_{p,{\rm min}})}}^{\log{(m_{p,{\rm max}})}}\left.\frac{d^2N_{\rm pl}}{d\log{m_p}~d\log{a}}\right|_{i} \nonumber \\
    & {} \times d\log{m_p}~d\log{a}\; . \label{eqn:f_val}
\end{align}
For all realizations, we fix the minimum semimajor axis at $a_{\rm in}=0.1~$AU, and the minimum and maximum planet masses are set to $m_{p,{\rm min}}=1~M_{\oplus}$ and $m_{p,{\rm max}}=13~M_{\rm Jup}$.

\subsubsection{Orbits}
\label{subsubsec:orbits}
Mapping the masses and semimajor axes of planets belonging to such populations into the observables for various types of surveys requires generating orbits. We assign a random orbit to each planet by drawing values from prior distributions of eccentricity, $e_{i,j}$, inclination, $i_{i,j}$, argument of pericenter, $\omega_{i,j}$, and mean anomaly, $M_{0,i,j}$. Note that we do not need to assign a longitude of ascending node, $\Omega_{i,j}$, to our planets since the projected separation (required for microlensing and direct imaging) is independent of this angle, as are the relevant RV observables. Eccentricities are drawn from the distribution determined by \citet{Zakamska2011}, which consists of two components. The first, representing 38\% of the population, is characterized by circular orbits, while the second, representing ``dynamically active'' planets, follows the form of \citet{Juric2008} given by
\begin{equation}
\frac{dN_{\rm pl}}{de} \propto e \exp{\left[-\frac{1}{2}\left(\frac{e}{0.3}\right)^2\right]} \; . \label{eqn:jt_ecc_prior}
\end{equation}
The priors on $\omega_{i,j}$ and $M_{0,i,j}$ are uniform between $[0,2\pi)$, and that on $\left.\cos{i}\right|_{i,j}$ is uniform between $[0,1]$. At this point, each planet is described by the set of parameters $\left\{m_{p, i, j}, a_{i, j}, e_{i, j}, i_{i, j}, \omega_{i, j}, M_{0, i, j}\right\}$. These parameters, along with properties of the host stars (which we describe in the following sections), will be used to determine the appropriate observables for each survey.

\subsection{Microlensing Observables}
\label{subsec:mlens_observables}
In order to compare with microlensing surveys, we need to map the planet masses and semimajor axes of populations of planets into planet-to-star mass ratios and projected separations in units of the Einstein radius, i.e. $(m_{p, i, j}, a_{i, j})\rightarrow(q_{i,j},s_{i,j})$. This transformation requires that we associate each planet with a host star acting as a lens in a microlensing event. We generate a ensemble of microlensing events for each planet population by drawing event parameters from mostly naive prior distributions. We then calculate posterior distributions of parameters by weighting each event by its associated event rate---a statistical probability based on stellar, kinematic, and Galactic properties.

The priors from which we draw event parameters are as follows. We choose to fix the source in each microlensing event in the bulge at a distance of $D_{s,i,j} =10~$kpc. In reality, there is some distribution of source distances, however there exists a strong prior on $D_s$. This prior is based on two facts: 1) the vast majority of sources are in the Galactic bulge, and 2) a majority of lenses are also located in the bulge. Since the lens must be located in front of the source, the range of $D_s$ is constrained to be a few kpc, placing most sources behind the Galactic center and in front of the back edge of the bulge. If we were to allow the source distance to vary, it would not significantly change our results, so we choose to fix it to simplify our calculations. The lens distances, $D_{l,i,j}$, are drawn from a uniform distribution ranging between $[0~{\rm kpc}, D_s]=[0~{\rm kpc},10~{\rm kpc}]$. We draw lens masses, $M_{l,i,j}$, from a mass distribution function we describe later in this section that includes brown dwarf, main sequence, and white dwarf lenses, with (initial) masses ranging between $[0.01~M_{\odot},8~M_{\odot}]$. We choose a random line-of-sight for each event with Galatic longitude, $l_{i,j}$, and latitude, $b_{i,j}$, drawn uniformly between $-8^{\circ}\leq l \leq0^{\circ}$ and $-5^{\circ}\leq b \leq 8^{\circ}$, respectively. These ranges in $l$ and $b$ are chosen because they cover the most densely monitored fields by the Optical Gravitational Lensing Experiment \citep[OGLE;][]{Udalski2003}. Finally, we determine the distribution of lens-source relative proper motions, $\boldsymbol{\mu}_{i,j}$, drawing two components in the Galactic north, $\mu_{N,i,j}$, and east, $\mu_{E,i,j}$, directions uniformly between $[0~{\rm mas~yr^{-1}},20~{\rm mas~yr^{-1}}]$. Each planetary microlensing event is uniquely described by the set of parameters $\left\{m_{p, i, j}, a_{i, j}, e_{i, j}, i_{i, j}, \omega_{i, j}, M_{0, i, j}, D_{l,i,j}, M_{l,i,j}, l_{i,j}, b_{i,j}, \boldsymbol{\mu}_{i,j}\right\}$.

The planet-to-star mass ratios for each system are then $q_{i,j} = m_{p,i,j}/M_{l,i,j}$ and we compute the projected separation in physical units, $r_{\perp, i, j}$, by projecting each orbit onto the plane of the sky (see equations~13--20 in \S~4 of \citealt{Clanton2014a}). We then normalize these projected separations by the Einstein radius, $R_{E,i,j}$, of the associated lensing system, $s_{i,j} = r_{\perp, i, j} / R_{E,i,j}$. The Einstein radius is a function of the lens mass and distance, as well as the source distance, and is given by
\begin{align}
R_E\left(M_l, D_l, D_s\right) = & {}~ \left[\frac{4GM_l}{c^2}\frac{D_l(D_s-D_l)}{D_s}\right]^{1/2} \label{eqn:einstein_radius} \\ 
\approx & {}~ 3.2~{\rm AU}\left(\frac{M_l}{0.5~M_{\odot}}\right)^{1/2}\left(\frac{D_s}{10~{\rm kpc}}\right)^{1/2} \nonumber \\
    & {} \times \left[\frac{x\left(1-x\right)}{0.25}\right]^{1/2}\; , \nonumber
\end{align}
where $x\equiv D_l/D_s$.

\subsubsection{The Rate Equation}
\label{subsubsec:rate_eqn}
Microlensing events, are rare, transient phenomena resulting from chance (and very precise) alignments of a source star and a foreground lens. The rate of a microlensing event with given lens mass, $M_l$, lens distance, $D_l$, and lens-source relative proper motion, $\boldsymbol{\mu}$, along a given sight line $(l,b)$ is
\begin{equation}
\frac{d^4\Gamma}{dD_ldM_ld\boldsymbol{\mu}} = 2R_Ev_{\rm rel}\nu\frac{d^2\Gamma}{d\boldsymbol{\mu}}\frac{d\Gamma}{dM_l}\; , \label{eqn:event_rate}
\end{equation}
where $R_E$ is the Einstein radius in physical units, $v_{\rm rel}$ is the lens-source relative (linear) velocity, $\nu$ is the local number density of lenses at the event location determined by $D_l$ and $(l,b)$, $d^2\Gamma/d\boldsymbol{\mu}$ is the two-dimensional probability density of a given lens-source relative proper motion, $\boldsymbol{\mu}$, and $d\Gamma/dM_l$ is the lens mass function (see \citealt{Griest1991} and \citealt{Kiraga1994} for more information on robust calculations of event rates). The lens-source relative velocity in physical units is simply $v_{\rm rel} = \left|\boldsymbol{\mu}\right|D_l$.

We provide a brief description of these terms to explain how relative event rates for an ensemble of microlensing events are calculated. For a complete explanation, refer to \S~5 of \citet{Clanton2014a} (specifically equations 33-54). There are a few differences in our assumptions in this paper relative to \citet{Clanton2014a,Clanton2014b}, which we note below, such as the lens mass function and the velocity dispersion of bulge stars we adopt.

\subsubsection{Lens Mass Function}
\label{subsubsec:lens_mass_func}
\begin{figure*}[!ht]
	\epsscale{0.7}
	\plotone{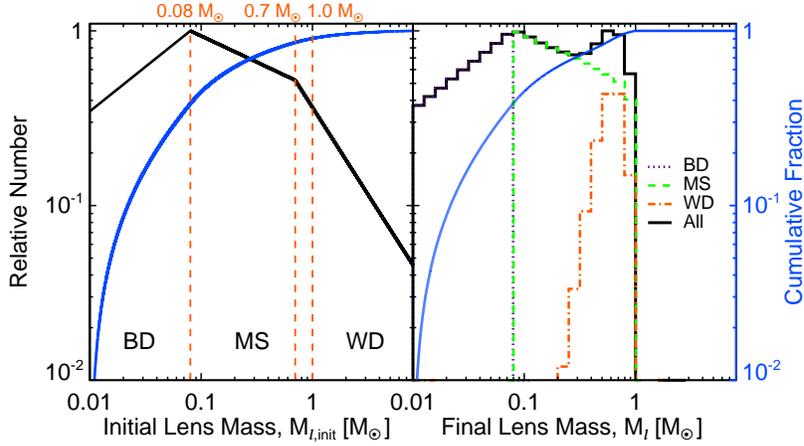}
	\caption{The left panel plots the distribution of ``initial lens masses'', $M_{l,{\rm init}}$, while the right panel plots the distribution of ``final lens masses'', $M_l$. The need to show both distributions arises from the fact that white dwarf lenses will have a mass distribution that is different from their initial mass distribution. The relative number of white dwarf lenses is drawn from the initial mass distribution, each of which is then given a final mass according to a Gaussian centered at $0.6~M_{\odot}$ with a standard deviation of $0.16~M_{\odot}$. See text for details.
		\label{fig:lmf}}
\end{figure*}

We adopt a lens mass function consisting of brown dwarf, main sequence, and white dwarf lenses similar to those of \citet{Gould2000} and model 1 of \citet{Sumi2011}. We choose not to include contributions from black hole and neutron star lenses, as these lenses only constitute a few percent of all microlensing events \citep{Gould2000}, so their exclusion will not significantly affect our results. Our distribution of ``initial lens masses'', $M_{l,{\rm init}}$, is thus
\begin{equation}
\frac{dN}{d\log{M_{l,{\rm init}}}} \propto 
\begin{cases}
M_{l,{\rm init}}^{0.51} & {\rm for}~0.01 \leq M_{l,{\rm init}}/M_{\odot} < 0.08 \\
M_{l,{\rm init}}^{-0.3} & {\rm for}~0.08 \leq M_{l,{\rm init}}/M_{\odot} < 0.70 \\
M_{l,{\rm init}}^{-1} & {\rm for}~0.70 \leq M_{l,{\rm init}}/M_{\odot} < 8.0.
\end{cases}
\end{equation}
Our distribution of ``final lens masses'', $M_l$, is identical up to $M_l < 1.0~M_{\odot}$, but the white dwarf lenses will have final masses described by a Gaussian, $\left.dN/dM_l\right|_{\rm WD}\propto \exp{[(M_l-\mu_{\rm WD})^2/(2\sigma_{\rm WD}^2)]}$, with a mean mass of $\mu_{\rm WD}=0.6~M_{\odot}$ and a standard deviation of $\sigma_{\rm WD}=0.16~M_{\odot}$ \citep{Sumi2011}. Figure~\ref{fig:lmf} plots the initial and final lens mass functions we adopt.

Our final mass distribution yields ratios of the numbers of brown dwarf, main sequence, and white dwarf lenses of (38:52:10). The fractions of total mass contained in each class of objects is (6:69:25), and the fractions of microlensing events (weighted by event rate) due to each object class is (17:65:17). These fractions are consistent with those reported by \citet{Gould2000}, but note that we do not include black hole and neutron star lenses, and the numbers reported by \citet{Gould2000} group together the brown dwarf and main sequence lenses.

\subsubsection{Local Number Density of Lenses}
\label{subsubsec:local_number_density_lenses}
Since we only need to compute the relative event rates for a given ensemble of microlensing events, we can use the local mass density, $\rho(x,y,z)$, rather than number density, $\nu(x,y,z)$, to compute the event rate for any particular lens. Here, $(x,y,z)$ are Galactocentric coordinates. We adopt a two-component Galactic model, comprised of an axisymmetric (``double-exponential'') disk with a 1~kpc hole and a barred, anisotropic bulge. This Galactic model is described by equations~39--46 of \citet{Clanton2014a}, including the coordinate transformations between $(D_l,l,b)\rightarrow(x,y,z)$ and the normalizations of the bulge and disk mass density functions. We demonstrate in \citet{Clanton2014a} that this Galactic model produces optical depths to microlensing events that are in rough agreement with those reported by \citet{Han2003}.

\subsubsection{Lens-Source Relative Proper Motion Distribution}
\label{subsubsec:lens_source_rel_mu_dist}
We assume that the velocity dispersion of lenses and sources are Gaussian. For lenses in the disk\footnote{Sources in the disk would have the same mean velocity and dispersion, but we assume all sources are located in the bulge.}, we assume a mean velocity in Galactic north and east directions of $\mathbf{v}_{l}=(v_{l,N},v_{l,E})=(0,v_{\rm rot})-(0,10)~{\rm km~s^{-1}}$ with a dispersion of $\sigma_{\mathbf{v}}=(\sigma_{v,N},\sigma_{v,E})=(20,30)~{\rm km~s^{-1}}$, where $v_{\rm rot}=220~{\rm km~s^{-1}}$. Lenses/sources that are located in the bulge have mean velocities of $\mathbf{v}_{l/s}=(v_{l/s,N},v_{l/s,E})=(0,0)~{\rm km~s^{-1}}$ with a dispersion of $\sigma_{\mathbf{v}}=(\sigma_{v,N},\sigma_{v,E})=(130,130)~{\rm km~s^{-1}}$. Dividing these velocities and their dispersions by either $D_l$ or $D_s$ converts them into proper motions. Note that we have adopted a slightly larger velocity dispersion for the objects in the bulge than in \citet{Clanton2014a,Clanton2014b}---$130~{\rm km~s^{-1}}$ versus $100~{\rm km~s^{-1}}$---in order to better reproduce the observed distribution of event timescales measured by \citet{Sumi2011}. We further justify this choice later in this section when we describe our comparison with the results of \citet{Sumi2011}.

\subsubsection{Statistical Weights for Each Event}
\label{subsubsec:statistical_weights_per_event}
The statistical weight we assign to each microlensing event according to the event rate is simply the product
\begin{equation}
    \Gamma_{i,j} = 2R_{E,i,j}v_{{\rm rel},i,j}\rho_{i,j}(x,y,z)\frac{d\Gamma}{d\boldsymbol{\mu}_{i,j}} \; . \label{eqn:Gamma_weight}
\end{equation}
These weights are applied to (probabilistically) correct for the fact that many of the stellar, kinematic, and Galactic properties are drawn from uninformed priors and the fact that we do not know with certainty whether or not a given lens is a bulge star or a disk star. These weights do not indicate the likelihood of a planet population given the results of microlensing (or other) studies. Note that since we draw our lens masses from an informed prior, the events will already be weighted according to $d\Gamma/dM_l$ and we do not need to explicitly include this factor in $\Gamma_{i,j}$.

\subsection{Application to Microlensing Surveys}
\label{subsec:microlensing_methods_application}

\subsubsection{\citet{Gould2010}}
\label{subsubsec:g10_methods_application}
We apply the actual detection sensitivities (``triangle diagrams'') of \citet{Gould2010} to our derived distributions of $(q_{i,j}, s_{i,j})$ to determine the number of planet detections their survey should detect, $N_{{\rm det, G10},i}$ and thus the planet occurrence rate they would infer, $\mathcal{G}_i$, for each population. A likelihood is then assigned to each planet population according to the likelihood function we derive for the \citet{Gould2010} study, $\mathscr{L}_{\mathcal{G}}$ (plotted in figure~\ref{fig:lhood_functions}), as described in \S~\ref{subsubsec:oom_microlensing}.

The mathematical formalism behind this calculation, developed in \S~4 of \citet{Clanton2014a}, is as follows. We implicitly assume a planet distribution function of the form
\begin{align}
    \left.\frac{d^8N_{\rm pl}}{d\left\{\alpha\right\}}\right|_{{\rm \mu lens},i,j} = & {} \left.\frac{dN_{\rm pl}}{di}\right|_{i,j}\left.\frac{dN_{\rm pl}}{de}\right|_{i,j}\left.\frac{dN_{\rm pl}}{d\omega}\right|_{i,j}\left.\frac{dN_{\rm pl}}{dM_0}\right|_{i,j} \nonumber \\
    & {} \times \left.\frac{d^2N_{\rm pl}}{d\log{m_p}~d\log{a}}\right|_{i,j}\left.\frac{dN_{\rm pl}}{dM_l}\right|_{i,j} \nonumber \\
    & {} \times \left.\frac{dN_{\rm pl}}{dD_l}\right|_{i,j}\; ,
\end{align}
such that the joint distribution of $\log{q}$ and $\log{s}$ for a given realization (i.e. planet population) is
\begin{align}
    & {} \left.\frac{d^4N_{\rm pl}}{d\log{q}~d\log{s}~d\log{m_p}~d\log{a}}\right|_i = \frac{\mathcal{F}_i}{N_{\rm ens}}\displaystyle \int_{\left\{\alpha\right\}}d\left\{\alpha\right\} \nonumber \\
    & {} ~~~~~~~~~\times \left.\frac{d^8N_{\rm pl}}{d\left\{\alpha\right\}}\right|_{{\rm \mu lens},i,j}\delta(\log{q}-\log{q_{i,j}})\delta(\log{s}-\log{s_{i,j}}) \nonumber \\
    & {} ~~~~~~~~~\times \delta(\log{m_p}-\log{m_{p,i,j}})\delta(\log{a}-\log{a_{i,j}})\; , \label{eqn:logq_logs_joint_dist}
\end{align}
where $N_{\rm ens}$ is the total ensemble number of planets of the $i$th population. The term $\mathcal{F}_i/N_{\rm ens}$ in equation~(\ref{eqn:logq_logs_joint_dist}) normalizes the joint distribution of $\log{q}$ and $\log{s}$ such that there are a total of $\mathcal{F}_i$ planets per star in the region of $\log{q}$--$\log{s}$ space we consider. If each of the $N_{\rm G10}=13$ microlensing events of \citet{Gould2010} is indexed by $k$, then the number of detections we expect for a single event in a given realization is then
\begin{align}
    N_{{\rm det, G10},i,k} = & {} \frac{1}{\sum_j \Gamma_{i,j}}\int d\log{q} \int d\log{s} \int d\log{m_p} \int d\log{a} \nonumber \\
    & {} \times \left.\frac{d^4N_{\rm pl}}{d\log{q}~d\log{s}~d\log{m_p}~d\log{a}}\right|_i \nonumber \\
    & {} \times \Gamma_{i,j}\Phi(\log{q},\log{s})_{{\rm det},k}\; ,
\end{align}
where $\Phi(\log{q},\log{s})_{{\rm det},k}$ is the detection limit for the $k$th microlensing event, having the form of the ``triangle diagrams'' of \citet{Gould2010}. There is no analytic form of $\Phi(\log{q},\log{s})_{{\rm det},k}$, so these calculations are performed numerically. The total number of expected detections for the \citet{Gould2010} survey for a specific planet population is then just the sum over all events, $N_{{\rm det, G10},i} = \displaystyle\sum_k N_{{\rm det, G10},i,k}$, and the expected inferred planet occurrence rate is thus $\mathcal{G}_i=N_{{\rm det, G10},i}/N_{\rm G10}$. The likelihood of each population according to the results of the \citet{Gould2010} survey, $\mathscr{L}_{\mathcal{G}}(\mathcal{G}_i)$, is given by the function plotted in figure~\ref{fig:lhood_functions}. We present and discuss these results in \S~\ref{sec:results_discussion}.

\subsubsection{\citet{Sumi2010}}
\label{subsubsec:s10_methods_application}
We construct the distribution $\left.dN_{\rm pl}/\log{q}\right|_{i}$ for each planet population and measure its slope over the range of mass ratios to which the \citet{Sumi2010} survey is sensitive. Specifically, we bin values of $\log{q_{i,j}}$, each weighted by the associated event rate, $\Gamma_{i,j}$, and use a least-squares fit to determine the slope of the resultant distribution in the range $-4.5\leq \log{q} \leq -2$. A likelihood is then assigned to each planet population according to the likelihood function inferred by the \citet{Sumi2010} study, $\mathscr{L}_{\rm S10}$ (plotted in figure~\ref{fig:lhood_functions}), as described in \S~\ref{subsubsec:oom_microlensing}. The results of this comparison are reported in \S~\ref{sec:results_discussion}.

\subsection{Direct Imaging Observables}
\label{subsec:imaging_observables}
To determine whether or not a given planet in a given population will be detectable by a direct imaging survey, we must derive the mapping $(m_{p, i, j}, a_{i, j})\rightarrow(\Delta {\rm mag}_{i,j},r_{\perp,i,j})$, where $\Delta {\rm mag}_{i,j}$ is the difference in magnitude in a given band (e.g. $H$ or $K_s$) between the planet and its host star and $r_{\perp,i,j}$ is the planet/star projected separation in physical units. Since we are testing whether or not a single planet population is consistent with all surveys, we use the same orbital parameters generated for each planet in the comparison with microlensing (see \S~\ref{subsubsec:orbits}) to obtain the projection $a_{i, j}\rightarrow r_{\perp,i,j}$ following equations~13--20 in \S~4 of \citealt{Clanton2014a}.

We require the use of planet evolutionary models to determine an absolute (e.g. $H-$band) magnitude for the planet given its mass and age, $H_{p,i,j}(m_{p,i,j},{\rm age}_{p,i,j})$. This introduces some amount of unquantified uncertainty in our analysis, but is a necessary step (see \S~\ref{subsec:uncertainties_PEM} for an in-depth description and discussion). We perform two separate analyses, one for each class of evolutionary models. We use the ``cold-start'' models at $5\times$ Solar (planet) atmospheric metallicity reported in Table~2 of \citet{Fortney2008}. To fill in this sparse data, we interpolate and extrapolate such that we end up with cooling curves with ages $\log{\rm (age/yr)} \in [6,10]$ at uniform sampling of $\Delta \log{\rm (age/yr)} = 0.01$, for masses $(m_p/M_{\rm Jup}) \in [0.5,13]$ at a uniform sampling of $\Delta (m_p/M_{\rm Jup})=0.01$ (Fortney, J. J., private communication). For the ``hot-start'' models of \citet{Baraffe2003}, we downloaded isochrones\footnote{\url{http://perso.ens-lyon.fr/isabelle.baraffe/COND03_models}} and extrapolated and interpolated in an identical manner as we did for the \citet{Fortney2008} models.

We then convert the absolute magnitude $H_{p,i,j}(m_{p,i,j},{\rm age}_{p,i,j})$ derived from the evolutionary models to an apparent magnitude, $h_{p,i,j}(H_{p,i,j},d_{i,j})$, given the distance to the system, $D_{\star,i,j}$. The difference in (apparent) magnitude between the planet and host star is then simply
\begin{equation}
    \Delta h_{i,j} = H_{p,i,j} - 5\left[1-\log{\left(\frac{D_{\star,i,j}}{\rm pc}\right)}\right] - h_{\star,i,j}\; , \label{eqn:app2abs_mag}
\end{equation}
where $h_{\star,i,j}$ is the apparent $H-$band magnitude of the host.

\subsection{Application to Direct Imaging Surveys}
\label{subsec:imaging_methods_application}

\subsubsection{\citet{Bowler2015}}
\label{subsubsec:b15_methods_application}
We apply the exact same detection limits used by \citet{Bowler2015} in their statistical analysis of the PALMS sample to calculate the number of expected detections given a particular planet population (Bowler, B. P. private communication). These have the same form as the contrast curves presented in Table~5 of \citet{Bowler2015}, but are different sets of limits. For each of the 72 single M stars to which we compare (see \S~\ref{subsec:di_constraints} for a description of this sample), we were provided information on which filter (either $H$ or $K_s$) was used for observations of the target, an apparent $H$ or $K_s$ magnitude, $7\sigma$ contrast limits as a function of angular separation (in arcseconds), and FOV coverage fraction as a function of angular separation. We obtain distances and ages (complete with uncertainties) for each star from Table~2 of \citet{Bowler2015}.

For a given planet population, $i$, we determine the number of expected detections around each of the 72 single M stars in the PALMS sample, which we will enumerate by the subscript $k$, in the following manner. For each planet, $j$, in this particular population we draw an age from the estimates provided by \citet{Bowler2015}. For stars in young moving groups, we assume Gaussian uncertainties, while those with an age range listed, we assume a distribution that is uniform in log-space. Using the age for each planet around a given star we draw from one of these distributions, ${\rm age}_{i,j,k}$, the distance to that system, $D_{\star,k}$, and the apparent magnitude of the star in the appropriate band, we determine the planet-to-star contrast $\Delta {\rm mag}_{i,j,k}$ as described above. Combining this contrast with the projected separation we calculate for each system, $r_{\perp, i, j}$, the FOV coverage fraction at this projected separation, and the stability limits for the host, $a_{{\rm stab},k}$, (if applicable, see below for description), we determine whether or not such a planet is expected to be detected by the PALMS survey.

We developed a generalized mathematical formalism for such calculations in \S~4 of \citet{Clanton2014a}, which we apply here. We implicitly assume a planet distribution function of the form
\begin{align}
\left.\frac{d^9N_{\rm pl}}{d\left\{\alpha\right\}}\right|_{{\rm img},i,j} = & {} \left.\frac{dN_{\rm pl}}{di}\right|_{i,j}\left.\frac{dN_{\rm pl}}{de}\right|_{i,j}\left.\frac{dN_{\rm pl}}{d\omega}\right|_{i,j}\left.\frac{dN_{\rm pl}}{dM_0}\right|_{i,j} \nonumber \\
    & {} \times \left.\frac{d^2N_{\rm pl}}{d\log{m_p}~d\log{a}}\right|_{i,j}\left.\frac{dN_{\rm pl}}{d({\rm mag}_{\star})}\right|_{i,j}\left.\frac{dN_{\rm pl}}{dD_{\star}}\right|_{i,j} \nonumber \\
    & {} \times \left.\frac{dN_{\rm pl}}{d{\rm (age)}}\right|_{i,j}\; ,
\end{align}
where ${\rm mag_{\star}}$ is the apparent magnitude of the hosts in either $H-$ or $K_s-$band, such that the joint distribution of $\Delta {\rm mag}$ and $r_{\perp}$ for a given realization (i.e. planet population) is
\begin{align}
& {} \left.\frac{d^4N_{\rm pl}}{d(\Delta {\rm mag})~dr_{\perp}~d\log{m_p}~d\log{a}}\right|_i = \frac{\mathcal{F}_i}{N_{\rm ens}}\displaystyle \int_{\left\{\alpha\right\}}d\left\{\alpha\right\} \nonumber \\
& {} ~~~~~~~~~\times \left.\frac{d^9N_{\rm pl}}{d\left\{\alpha\right\}}\right|_{{\rm img},i,j}\delta(\Delta {\rm mag}-\Delta {\rm mag}_{i,j})\delta(r_{\perp}-r_{\perp, i, j}) \nonumber \\
& {} ~~~~~~~~~\times \delta(\log{m_p}-\log{m_{p,i,j}})\delta(\log{a}-\log{a_{i,j}})\; , \label{eqn:dmag_rperp_joint_dist}
\end{align}
where $N_{\rm ens}$ is the total ensemble number of planets of the $i$th population. The term $\mathcal{F}_i/N_{\rm ens}$ in equation~(\ref{eqn:dmag_rperp_joint_dist}) normalizes the joint distribution of $\Delta {\rm mag}$ and $r_{\perp}$ such that there are a total of $\mathcal{F}_i$ planets per star in the region of $\Delta {\rm mag}$--$r_{\perp}$ space we consider. The number of detections we expect for a single target in the PALMS sample (and in a given realization) is then
\begin{align}
N_{{\rm det, B15},i,k} = & {} \int d(\Delta{\rm mag}) \int dr_{\perp} \int d\log{m_p} \int d\log{a} \nonumber \\
& {} \times \left.\frac{d^4N_{\rm pl}}{d(\Delta {\rm mag})~dr_{\perp}~d\log{m_p}~d\log{a}}\right|_i \nonumber \\
& {} \times \Phi(\Delta{\rm mag},r_{\perp})_{{\rm det},k}\Phi(r_{\perp})_{{\rm FOV},k}\Phi(a)_{{\rm stab},k}\; , \label{eqn:ndet_imaging}
\end{align}
where $\Phi(\Delta {\rm mag},r_{\perp})_{{\rm det},k}$ is the contrast limit, $\Phi(r_{\perp})_{{\rm FOV},k}$ is the FOV coverage fraction as a function of $r_{\perp}$ (a number between zero and one), and $\Phi(a)_{{\rm stab},k}=\Theta(a_{{\rm stab},k}-a)$ is the planet stability criterion as a function of semi-major axis for the $k$th star in the PALMS sample provided to us (Bowler, B. P. private communication). Some of the stars in the PALMS sample have known (wide-separation) binary companions that are expected to threaten the long-term stability of some planetary orbits about the primary \citep{Holman1999}. In their statistical analysis, \citet{Bowler2015} assume no sensitivity to planets with separations that would place them in unstable orbits according to the stability criterion developed in \citet{Holman1999}. We adopt these same stability limits and assume there is no sensitivity to planets on orbits beyond the maximum stable semimajor axis, $a_{{\rm stab},k}$, calculated by \citet{Bowler2015}.

There are no analytic forms of $\Phi(\log{q},\log{s})_{{\rm det},k}$ and $\Phi(r_{\perp})_{{\rm FOV},k}$, so these calculations are performed numerically. The total number of expected detections for the PALMS survey given a specific planet population is then just the sum over all targets, $N_{{\rm det, B15},i} = \displaystyle\sum_k N_{{\rm det, B15},i,k}$. The likelihood of each population according to the results of the PALMS survey, which reported zero planetary companion detections over all their targets, is then $\mathscr{L}_{B15}(N_{{\rm det, B15},i})=\exp(-N_{{\rm det, B15},i})$. This process is repeated for many different planet populations to build up constraints on the parameters of our assumed planet distribution function, which we present and discuss in \S~\ref{sec:results_discussion}.

\subsubsection{\citet{Lafreniere2007}}
\label{subsubsec:l07_methods_application}
We determine the number of expected detections for the GDPS given a particular planet population in a similar fashion as we describe above for the PALMS survey, with a few minor differences. For each of the 16 M stars in the GDPS, we obtain $H-$band magnitudes, distances, and age ranges from Table~1 of \citet{Lafreniere2007}, and we obtain $5\sigma$ detection limits from their Table~4. The detection limits provided by \citet{Lafreniere2007} are magnitude differences in the NIRI CH4-short filter (1.54--1.65$\mu$m) as a function of projected separation (in arcseconds). For planets in a particular population, we compute the expected contrast in this CH4 filter, $\Delta m_{{\rm CH4},i,j}$, as
\begin{equation}
    \Delta m_{{\rm CH4},i,j} = \Delta h_{i,j} - 2.5\log{\left(\frac{f_{\rm CH4}}{f_{H}}\right)}\; , \label{eqn:ch4mag_diff}
\end{equation}
where $\Delta h_{i,j}$ is the magnitude difference in $H-$band given by equation~(\ref{eqn:app2abs_mag}), $f_{\rm CH4}$ and $f_{H}$ are the mean flux densities of the planet in the NIRI CH4-short and $H$ filters, respectively, and the host star magnitudes in these two filters are assumed to be equal (see equation~2 and surrounding text in \S~4.1 of \citealt{Lafreniere2007}). Since the ratio of flux densities in these filters typically lies in the range 1.5--2.5 for giant planets, we adopt a value of $f_{\rm CH4}/f_{H}\equiv 2$ for all planets \citep{Allard2001,Baraffe2003,Lafreniere2007}.

The mathematical formalism behind this calculation is identical to that presented for the PALMS survey above, but with $\Phi(r_{\perp})_{{\rm FOV},k}\rightarrow \Theta[r_{\perp} - (\rho_{\rm FOV,L07}D_{\star,i,j})]$ and $\Phi(a)_{{\rm stab},k}\rightarrow 1$ in equation~(\ref{eqn:ndet_imaging}), where $\rho_{\rm FOV,L07}=10''$ is the (radial) FOV of the GDPS, out to which we assume they have a uniform coverage fraction of one. We follow \citet{Lafreniere2007} and do not adopt a stability criterion for binaries in our comparison (note that \citealt{Lafreniere2007} only report one of their M stars as having a known companion, GJ 234AB).

Thus, for each planet population, we numerically compute a number of expected detections for the GDPS, $N_{{\rm det,L07},i,k}$, where the subscript $k$ enumerates the 16 M stars in the sample. The total expected detections for their survey (for a given population) is then $N_{{\rm det, L07},i} = \displaystyle\sum_k N_{{\rm det, L07},i,k}$. We assign a likelihood to each population according to the likelihood function $\mathscr{L}_{L07}(N_{{\rm det, L07},i})=\exp(-N_{{\rm det, L07},i})$ since the GDPS discovered no planetary companions to any of their targets. Repeating this process for many different planet populations, we derive the constraints presented and discussed in \S~\ref{sec:results_discussion}.

\subsection{RV-Trend Observables and Application to the CPS TRENDS Survey}
\label{subsec:rv_trend_observables}
We compute the number of expected long-term RV trend detections for the CPS TRENDS survey given a specific planet population by calculating the magnitude of acceleration of the host star due to a planet on a given orbit and comparing with the detection sensitivities of the sample. Again, since we are testing whether or not a single planet population is consistent with the results of all the surveys we consider, we use the same orbital parameters we generated for the comparison with microlensing and direct imaging (see \S~\ref{subsubsec:orbits}). This allows us to immediately compute the expected acceleration for each planet in a given population, $\dot v_{i,j}(a_{i,j},m_{p,i,j},i_{i,j},M_{0,i,j})$, using equation~(\ref{eqn:dotv_std}). We also need to know the periods of the planets' orbits, $P_{i,j}$, to determine their detectability as a long-term trend, rather than as ``secure'' detections. For a planet to be detected as a long-term RV trend, the period of the planet must be larger than the baseline of observations, such that the RVs of the host are monitored for only a fraction of an orbital period (see \citealt{Clanton2014a,Clanton2014b}). The orbital period of each planet is easily computed via Kepler's Third Law (equation~\ref{eqn:p_kep}) given the masses of the CPS TRENDS M dwarfs and the planet's semimajor axis.

\subsubsection{\citet{Montet2014}}
\label{subsubsec:m14_methods_application}

For each star in the CPS TRENDS survey, which we will enumerate with the index $k$, we determine the ``threshold'' acceleration as a function of orbital period, above which planets are expected to be detected as long-term RV trends, $\dot v_{{\rm thr},k}(P)$, directly from equation~(\ref{eqn:dotv_thr}). We describe the reasoning behind this minimum-detectable acceleration in \S~\ref{subsubsec:oom_rv} and in \citet{Clanton2014a,Clanton2014b}. Previously, in the order of magnitude comparison presented in \S~\ref{subsubsec:oom_rv}, we used the median sample parameters of the CPS TRENDS survey to derive rough constraints on the properties of planets consistent with the survey results. Now, we will perform a more accurate calculation, including a different detection limit for each star in the sample that is determined from the specifics of the observations of each target. We obtain masses, $M_{\star, k}$, for each of the 111 M dwarfs monitored by the CPS TRENDS survey from Table~1 of \citet{Montet2014}, and from their Table~2, we obtain the total number of observations, $N_k$, the baseline of observations, $T_k$, and the total RV measurement uncertainty (the quadrature sum of the estimated photon noise and stellar jitter), $\sigma_k$, for each target. Thus, for each star, we compute the threshold acceleration (i.e. trend detection limit) as a function of orbital period using equation~(\ref{eqn:q_snr}) to obtain the ``phase-averaged'' S/N, $\mathcal{Q}_k$, and substituting this into equation~(\ref{eqn:dotv_thr}). An example of such a detection limit is shown in figure~\ref{fig:dotv_thr}.

We implicitly assume a planet distribution function of the form
\begin{align}
\left.\frac{d^7N_{\rm pl}}{d\left\{\alpha\right\}}\right|_{{\rm RV},i,j} = & {} \left.\frac{dN_{\rm pl}}{di}\right|_{i,j}\left.\frac{dN_{\rm pl}}{de}\right|_{i,j}\left.\frac{dN_{\rm pl}}{d\omega}\right|_{i,j}\left.\frac{dN_{\rm pl}}{dM_0}\right|_{i,j} \nonumber \\
    & {} \times \left.\frac{d^2N_{\rm pl}}{d\log{m_p}~d\log{a}}\right|_{i,j}\left.\frac{dN_{\rm pl}}{dM_{\star}}\right|_{i,j}\; ,
\end{align}
such that the joint distribution of $\dot{v}$ and $\log{P}$ for a given realization (i.e. planet population) is
\begin{align}
& {} \left.\frac{d^4N_{\rm pl}}{d\dot{v}~d\log{P}~d\log{m_p}~d\log{a}}\right|_i = \frac{\mathcal{F}_i}{N_{\rm ens}}\displaystyle \int_{\left\{\alpha\right\}}d\left\{\alpha\right\} \nonumber \\
& {} ~~~~~~~~~\times \left.\frac{d^7N_{\rm pl}}{d\left\{\alpha\right\}}\right|_{{\rm RV},i,j} \delta(\dot{v}-\dot{v}_{i,j})\delta(\log{P}-\log{P_{i,j}}) \nonumber \\
& {} ~~~~~~~~~\times \delta(\log{m_p}-\log{m_{p,i,j}})\delta(\log{a}-\log{a_{i,j}})\; , \label{eqn:vdot_p_joint_dist}
\end{align}
where $N_{\rm ens}$ is the total ensemble number of planets of the $i$th planet population. The term $\mathcal{F}_i/N_{\rm ens}$ in equation~(\ref{eqn:vdot_p_joint_dist}) normalizes the joint distribution of $\dot{v}$ and $\log{P}$ such that there are a total of $\mathcal{F}_i$ planets per star in the region of $\dot{v}$--$\log{P}$ space we consider. The number of long-term RV trend detections we expect for a single target in the CPS TRENDS sample for a given population is then
\begin{align}
N_{{\rm tr},i,k} = & {} \int d\dot{v} \int d\log{P} \int d\log{m_p} \int d\log{a} \nonumber \\
& {} \times \left.\frac{d^4N_{\rm pl}}{d\dot{v}~d\log{P}~d\log{m_p}~d\log{a}}\right|_i \nonumber \\
& {} \times \Phi(\dot{v})_{{\rm tr},k}\Phi(\log{P})_{{\rm tr},k}\; ,
\end{align}
where $\Phi(\dot{v})_{{\rm tr},k}=\Theta[\dot{v}(\log{P})-\dot{v}_{{\rm thr},k}(\log{P})]$ is the acceleration detection limit for the $k$th target at the orbital period of a given planet and where $\Phi(\log{P})_{{\rm tr},k}=\Theta(\log{P}-\log{T}_k)$. The first detection criterion simply requires that the acceleration of the host star due to a planet has large enough magnitude to be detectable, while the second criterion ensures that the period of the planetary orbit is longer than the baseline of observations. The total number of expected long-term RV trend detections for the CPS TRENDS survey for a specific planet population is then just the sum over all targets in the sample, $N_{{\rm tr},i} = \displaystyle\sum_k N_{{\rm tr},i,k}$, and the likelihood of each planet population is given by the likelihood function $\mathscr{L}_{tr}(N_{{\rm tr},i})=(N_{{\rm tr},i})^4\exp{(-N_{{\rm tr},i})}/4!$, which we plot in figure~\ref{fig:lhood_functions}. We present and discuss the results of our comparison with the CPS TRENDS survey in \S~\ref{sec:results_discussion}.

\section{Results and Discussion}
\label{sec:results_discussion}
We find that the results of microlensing, RV, and direct imaging surveys are consistent with a single planet population described by a joint power-law distribution function in planet mass and semimajor axis. We present our final results in figures~\ref{fig:2d_constraints_combined_hot} (``hot-start''), \ref{fig:2d_constraints_combined_cold} (``cold-start'') and \ref{fig:1d_constraints_combined}, and summarize the median values and 68\% confidence intervals for each of the four parameters of our planetary distribution function, which are consistent with all surveys we consider, in table~\ref{tab:1d_medians}. We plot the associated likelihoods of the individual surveys across these 68\% confidence intervals in figures~\ref{fig:range_lhoods_plot_hot} and \ref{fig:range_lhoods_plot_cold}, demonstrating that the resultant population of planets has a reasonably high likelihood of producing the observations reported by each survey.

There are several subtleties embedded in the constraints obtained from the synthesis of all survey results, so in order to demonstrate that we understand the results of our calculations, we first present the constraints obtained from each individual survey and compare with those we found in the order of magnitude evaluations of \S~\ref{sec:oom} before reporting our final results.

\subsection{Constraints from Individual Surveys}
\label{subsec:results_ind_constraints}

\subsubsection{Microlensing}
\label{subsubsec:results_mlens_only}

\begin{figure}[!t]
	\epsscale{1.1}
	\plotone{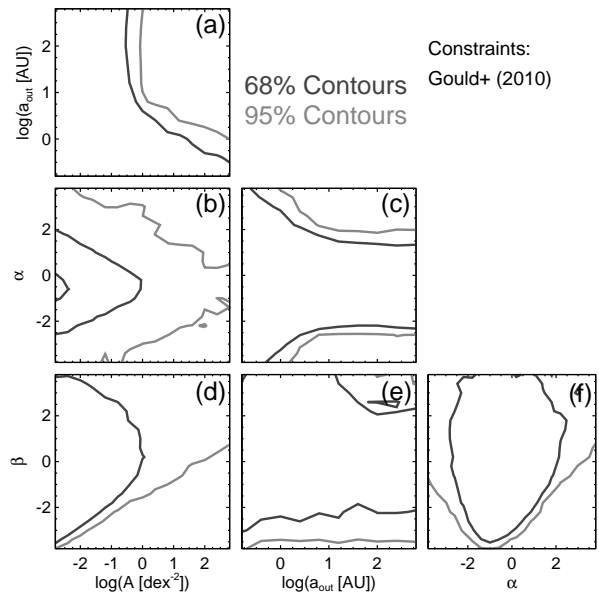}
	\caption{Likelihood contours as a function of pairs of parameters describing planet distribution functions that are found to be consistent with the measurement of the integrated planet frequency, $\mathcal{G}$, by the \citet{Gould2010} microlensing survey. Contours are drawn at levels of $68\%$ and $95\%$ of the peak likelihood and are marginalized over all other parameters except those being plotted.
		\label{fig:2d_constraints_g10}}
\end{figure}

{\bf \citet{Gould2010}:} Figure~\ref{fig:2d_constraints_g10} shows the likelihood contours on the parameters of our planet distribution function that are consistent with the results of this survey. These constraints have some differences with those we obtained in the order of magnitude evaluations, but the general intuition we develop in \S~\ref{subsubsec:oom_microlensing} for understanding these constraints still holds.

As we found in \S~\ref{subsubsec:oom_microlensing}, the most important constraint here is on the combination of $\alpha$ and $\mathcal{A}$ shown in panel~(b) for the same reasons. We find that the constraints on the combination of $\beta$ and $\mathcal{A}$ shown in panel~(d) are more pronounced because, by applying different detection limits for each of the 13 microlensing events in the \citet{Gould2010} sample, we are effectively sampling the planet distribution function over various ranges of $s$ (and thus $a$), thus providing more sensitivity to $\beta$. As a result, the constraint on the combination of $\alpha$ and $\beta$ is slightly stronger in this calculation relative to those of \S~\ref{subsubsec:oom_microlensing}.

Another difference here is that there is no minimum value of $a_{\rm out}$ (within the range we consider) as we found for the order of magnitude calculations. There, we had a sharp cutoff at 1.4~AU due to the sensitivity criteria we applied, which is not apparent here. However, panel~(a) shows that small values of $a_{\rm out}$ require very large normalizations to be consistent with the \citet{Gould2010} results. Finally, in panel~(c), we see that very large and very small values of $\alpha$ are excluded unless $a_{\rm out}$ is small.

\begin{figure}[!t]
	\epsscale{1.1}
	\plotone{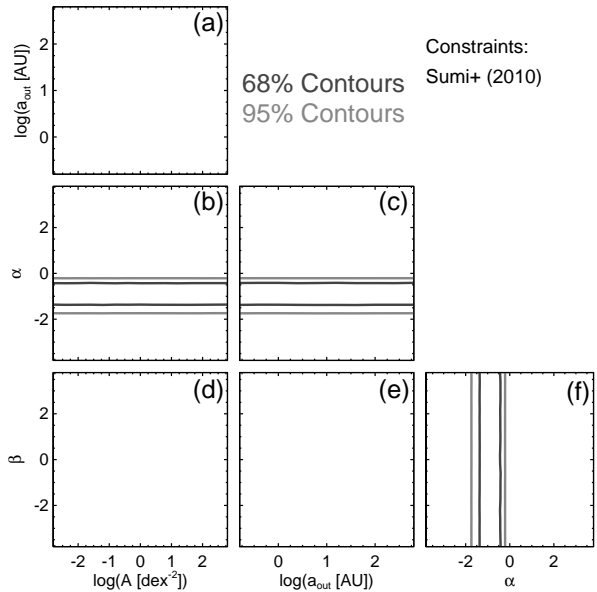}
	\caption{Likelihood contours as a function of pairs of parameters describing planet distribution functions that are found to be consistent with the measurement of the slope of the mass-ratio distribution function by the \citet{Sumi2010} microlensing survey. Contours are drawn at levels of $68\%$ and $95\%$ of the peak likelihood and are marginalized over all other parameters except those being plotted.
		\label{fig:2d_constraints_s10}}
\end{figure}

{\bf \citet{Sumi2010}:} Figure~\ref{fig:2d_constraints_s10} shows the likelihood contours on the parameters of our planet distribution function that are consistent with the measurement of the slope of the mass ratio distribution function ($p=-0.68\pm 0.20$) by the \citet{Sumi2010} study. It is obvious why the only constraining power offered by this survey is on $\alpha$, but we note that the contours are not centered at $\alpha=-0.68$ and the 68\% contours spread larger than $0.20$. Since $q\equiv m_p/M_l$, the conflation of $dN_{\rm pl}/d\log{m_p}$ with the lens mass function (see \S~\ref{subsubsec:lens_mass_func}) means that $dN_{\rm pl}/d\log{m_p}\not\propto dN_{\rm pl}/d\log{q}$. This is in contrast to the order of magnitude constraints presented in \S~\ref{subsubsec:oom_microlensing}, where we had assumed a characteristic lens (host) mass rather than marginalizing over the population.

\subsubsection{Direct Imaging}
\label{subsubsec:results_imaging_only}
\begin{figure}[!t]
	\epsscale{1.1}
	\plotone{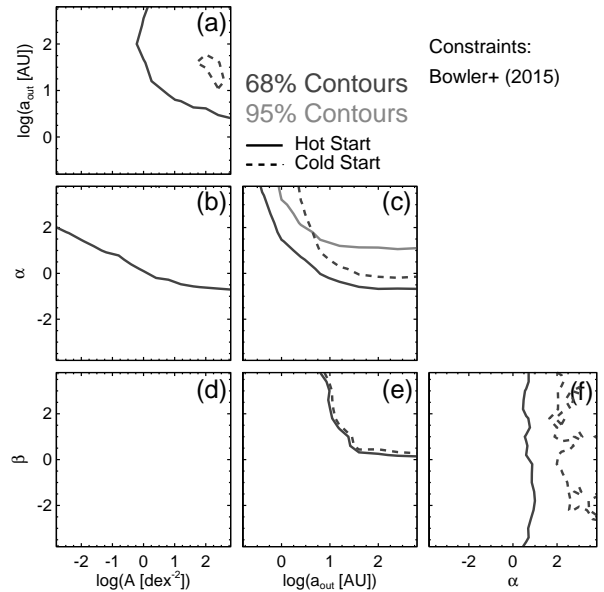}
	\caption{Likelihood contours as a function of pairs of parameters describing planet distribution functions that are found to be consistent with the non-detection of any planetary companions in the PALMS direct imaging survey \citep{Bowler2015}. Contours are drawn at levels of $68\%$ and $95\%$ of the peak likelihood and are marginalized over all other parameters except those being plotted.
		\label{fig:2d_constraints_bb15}}
\end{figure}

{\bf \citet{Bowler2015}:} We plot the likelihood contours derived from our comparison with the results of the PALMS survey in figure~\ref{fig:2d_constraints_bb15} (see \S~\ref{subsubsec:b15_methods_application} for methodology). These constraints are very similar, qualitatively, with those derived in our order of magnitude analysis of \S~\ref{subsubsec:oom_di}.

The PALMS survey can only weakly constrain combinations of $a_{\rm out}$ with $\alpha$, $\beta$, and $\mathcal{A}$. As shown in panels~(c) and (e), large values of $\beta$ and $\alpha$ are disallowed for large values of $a_{\rm out}$, since we would expect too many direct detections from such populations. In panel~(a), we see the same behavior with large values of $a_{\rm out}$ and $\mathcal{A}$, although here it is less pronounced since this set of contours marginalizes over all values of $\alpha$ and $\beta$. For sufficiently small values of $\alpha$ and/or $\beta$, most combinations of $a_{\rm out}$ and $\mathcal{A}$ are allowed. Our results are more sensitive to $\beta$, and especially more sensitive to $\alpha$, than they are to $\mathcal{A}$.

The major difference between the order of magnitude comparison presented in \S~\ref{subsubsec:oom_di} and the results of our detailed calculations we report here is this enhanced sensitivity to $\alpha$ shown in panels~(b), (c), and (f). In our order of magnitude comparison, we assumed that each star in the PALMS sample had the same detection limit (the median of the sample), whereas here, we employ the individual sensitivities for each star. This allows us to sample the planet distribution function at different contrasts (and thus, through the use of evolutionary models, different planet masses), providing more sensitivity to the value of $\alpha$. Not surprisingly, this is also where we see the largest difference between the ``hot-'' and ``cold-start'' models (see panel~c). However, as we will demonstrate in \S~\ref{subsec:results_syn_constraints}, there is not a significant difference in the constraints derived from the ``hot-'' and ``cold-start'' models over the regions of parameter space allowed by a joint comparison with all surveys.

\begin{figure}[!t]
	\epsscale{1.1}
	\plotone{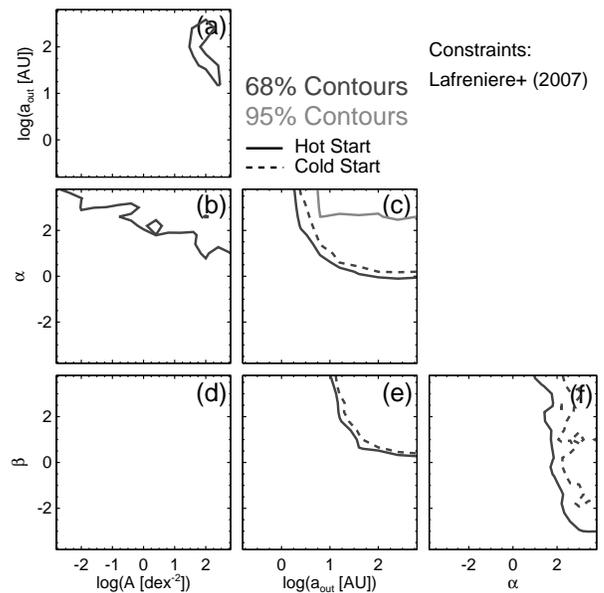}
	\caption{Likelihood contours as a function of pairs of parameters describing planet distribution functions that are found to be consistent with the non-detection of any planetary companions in the GDPS \citep{Lafreniere2007}. Contours are drawn at levels of $68\%$ and $95\%$ of the peak likelihood and are marginalized over all other parameters except those being plotted.
		\label{fig:2d_constraints_l07}}
\end{figure}

{\bf \citet{Lafreniere2007}:} Figure~\ref{fig:2d_constraints_l07} plots the likelihood contours on the parameters of populations consistent with the GDPS we derive following the procedure outlined in \S~\ref{subsubsec:l07_methods_application}. The interpretation of these constraints is identical to that discussed above for the PALMS survey. Although the GDPS is, in general, a deeper survey than PALMS, the large difference in sample sizes between the surveys means that the GDPS constraints are weaker than those derived from the PALMS survey.

\subsubsection{Long-Term RV Trends}
\label{subsubsec:results_rv_only}
\begin{figure}[!t]
	\epsscale{1.1}
	\plotone{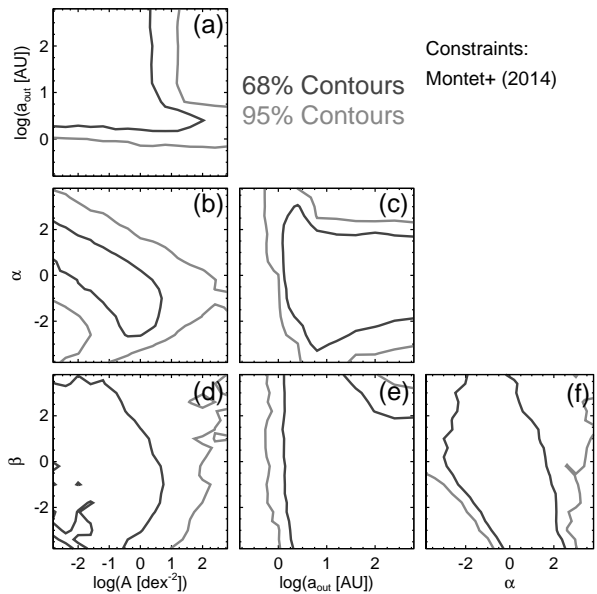}
	\caption{Likelihood contours as a function of pairs of parameters describing planet distribution functions that are found to be consistent with the detection of four significant, long-term RV trends in the CPS M dwarf sample \citep{Montet2014}. Contours are drawn at levels of $68\%$ and $95\%$ of the peak likelihood and are marginalized over all other parameters except those being plotted.
		\label{fig:2d_constraints_mb14}}
\end{figure}

{\bf \citet{Montet2014}:} We plot the likelihood contours on the parameters of our planet population model that are consistent with the CPS TRENDS survey in figure~\ref{fig:2d_constraints_mb14} (methodology for this comparison is presented in \S~\ref{subsubsec:m14_methods_application}). The constraints are qualitatively very similar to those of our order of magnitude comparison shown in \S~\ref{subsubsec:oom_rv}. As expected, the detection of RV trends places a lower bound on $a_{\rm out}$ shown in panels~(a), (c), and (e). We find an anticorrelation between $\alpha$ and $\mathcal{A}$ (panel~b) such that smaller values of $\alpha$, for which more planets are distributed to lower masses and thus induce smaller accelerations on their hosts, require larger normalizations in order to be consistent with the observed number of trends. The CPS TRENDS survey places a constraint on the combination of $\alpha$ and $\beta$ (panel~f); larger values of $\beta$ distribute more planets to larger separations where they might be detectable as trends, and thus smaller values of $\alpha$ are required to keep the number of planets massive enough to induce detectable accelerations consistent with the observations.

\subsection{Synthesized Constraints}
\label{subsec:results_syn_constraints}
\begin{figure*}[!ht]
	\epsscale{1.0}
	\plotone{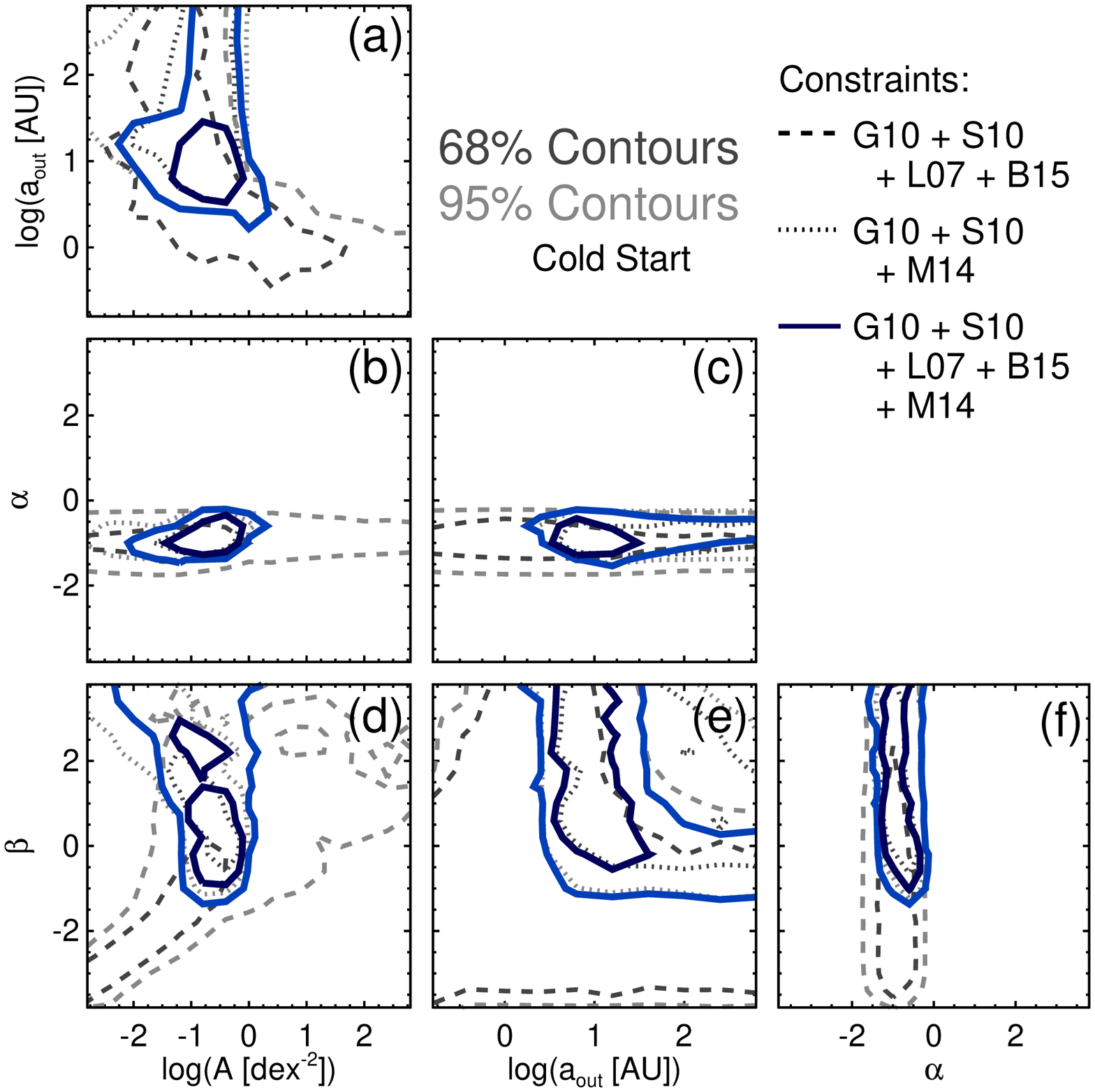}
	\caption{Likelihood contours as a function of pairs of parameters describing planet distribution functions that are found to be consistent with the results of microlensing (G10--\citealt{Gould2010}; S10--\citealt{Sumi2010}), direct imaging (L07--\citealt{Lafreniere2007}; B15--\citealt{Bowler2015}), and RV (M14--\citealt{Montet2014}) surveys. Here, we have utilized the predictions of the ``hot-start'' planet evolutionary models of \citet{Baraffe2003}. Contours are drawn at levels of $68\%$ and $95\%$ of the peak likelihood and are marginalized over all other parameters except those being plotted.
		\label{fig:2d_constraints_combined_hot}}
\end{figure*}

\begin{figure*}[!ht]
	\epsscale{1.0}
	\plotone{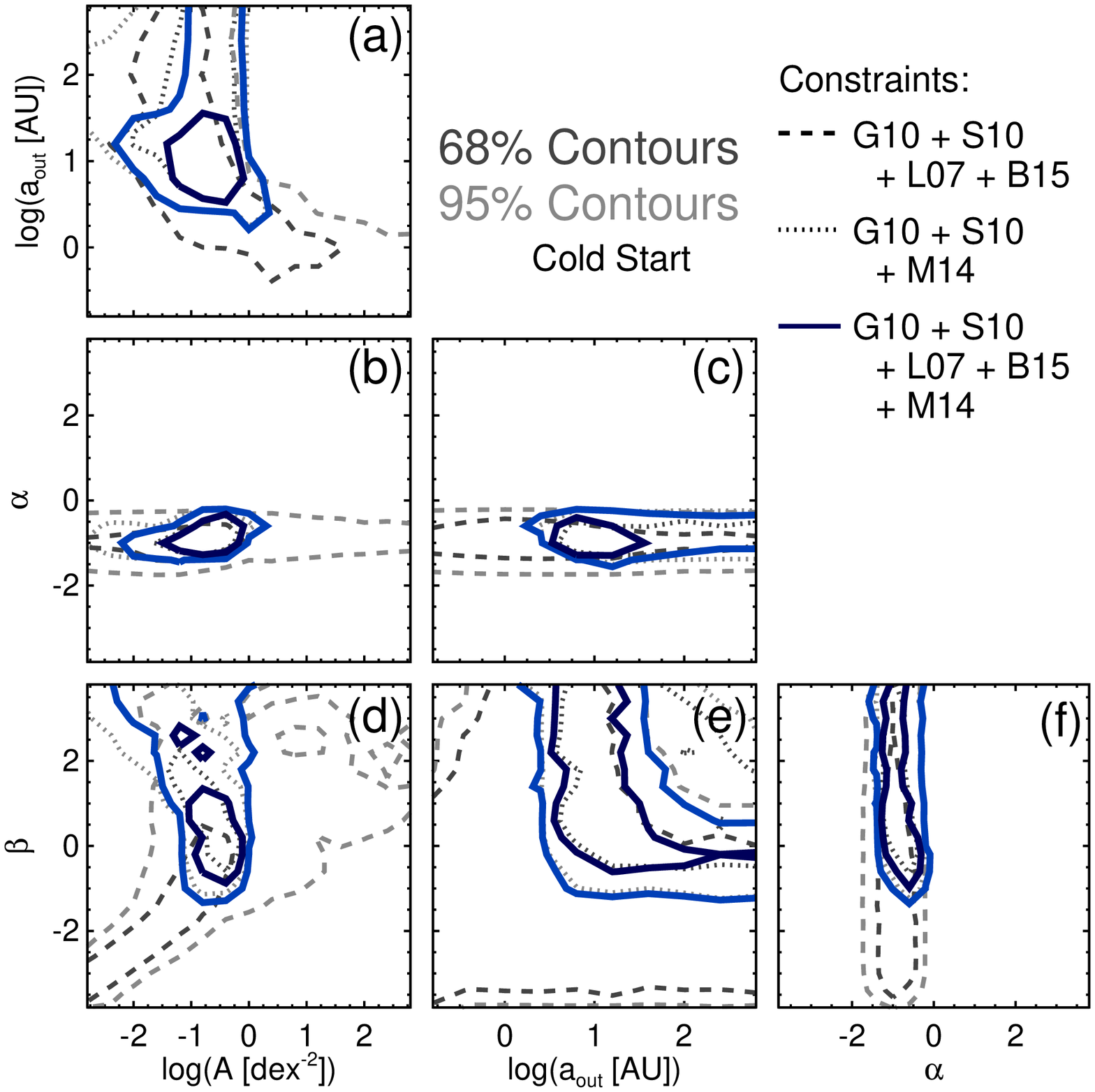}
	\caption{Identical to figure~\ref{fig:2d_constraints_combined_hot}, but utilizing the predictions of the ``cold-start'' planet evolutionary models of \citet{Fortney2008}.
		\label{fig:2d_constraints_combined_cold}}
\end{figure*}

We have shown that, individually, the results of any single exoplanet discovery survey can only constrain combinations of the parameters of our assumed planet distribution function. However, a joint comparison with the results of multiple detection methods enables us to derive meaningful constraints on the individual parameters of a planet population that is consistent with the results of all surveys we consider. We plot the likelihood contours on pairs of parameters describing a population that is consistent with the results of various combinations of surveys in figure~\ref{fig:2d_constraints_combined_hot} for our ``hot-start'' calculations and in figure~\ref{fig:2d_constraints_combined_cold} for our ``cold-start'' calculations.

Since the \citet{Gould2010} survey constrains a combination of $\alpha$ and $\mathcal{A}$ and the \citet{Sumi2010} survey constrains $\alpha$, the combination of these two surveys provides a relatively tight constraint on $\mathcal{A}$. With $\alpha$ and $\mathcal{A}$ more or less fixed, the CPS TRENDS survey places lower bound on $\beta$ and $a_{\rm out}$, below which there are not enough planets at large enough separations to be detected as long-term drifts. However, due to degeneracies between $\beta$ and $a_{\rm out}$, we are unable to place an upper bound on either of these parameters. Very large values of $\beta$ are allowed by the trends as long as $a_{\rm out}$ is relatively small. Inclusion of the direct imaging results does not improve the constraints much for similar reasons. Arbitrarily large values of $\beta$ are allowed by direct imaging surveys as long as $a_{\rm out}\lesssim 10~$AU. This is due to the fact that there is a sharp cutoff in detectability at projected separations approaching the inner working angle.

There is not much difference between the results of the ``hot-'' and ``cold-start'' calculations. The ``cold-start'' calculations are slightly less constraining, particularly on $a_{\rm out}$, but the difference is not much and well within the uncertainties (see table~\ref{tab:1d_medians}). One might assume that this is because the difference in expected planet luminosity between the two classes of evolutionary models is greatest at young ages, a region of parameter space occupied by just a fraction of the targets in direct imaging surveys. However, this is not actually the case, because it is these very targets (i.e. stars with ages $\lesssim 100~$Myr) that provide the best constraints from imaging surveys as overall; most planets are not expected to be detected around the older stars.

Instead, the lack of a significant difference between the ``hot-'' and ``cold-start'' predictions can be attributed to: 1) the difference in the predicted luminosities between the two classes of models decreases with planet mass at all ages, 2) our assumed power-law form for the planet distribution function, and 3) the most likely range of $a_{\rm out}$ we find lies within a region where direct imaging surveys are mostly insensitive to planets. The bottom-left panel in Figure~1 of \citet{Fortney2008} demonstrates that, at all ages, the predicted luminosities for lower-mass planets differ less between the ``hot-'' and ``cold-start'' models relative to the more massive planets. This is important, since we find that $\alpha$ is relatively well-constrained at negative values, and thus the mass function is heavily weighted towards lower-mass planets. Additionally, since our assumed planet distribution function has a power-law form, small changes in $\alpha$, $\beta$, and/or $a_{\rm out}$ can have a dramatic effect on the number of expected planet detections. As for the third reason, the detectability of exoplanets by direct imaging surveys dramatically decreases as the projected separation approaches the inner working angle (see e.g. figure~2 of \citealt{Lafreniere2007} and figure~3 of \citealt{Bowler2015}), which means that we do not predict any exoplanet detections with separations $a\lesssim 10~$AU for either survey, despite our choice of evolutionary model (see figures~\ref{fig:2d_constraints_bb15} and \ref{fig:2d_constraints_l07}).

\begin{figure*}[!t]
	\epsscale{0.8}
	\plotone{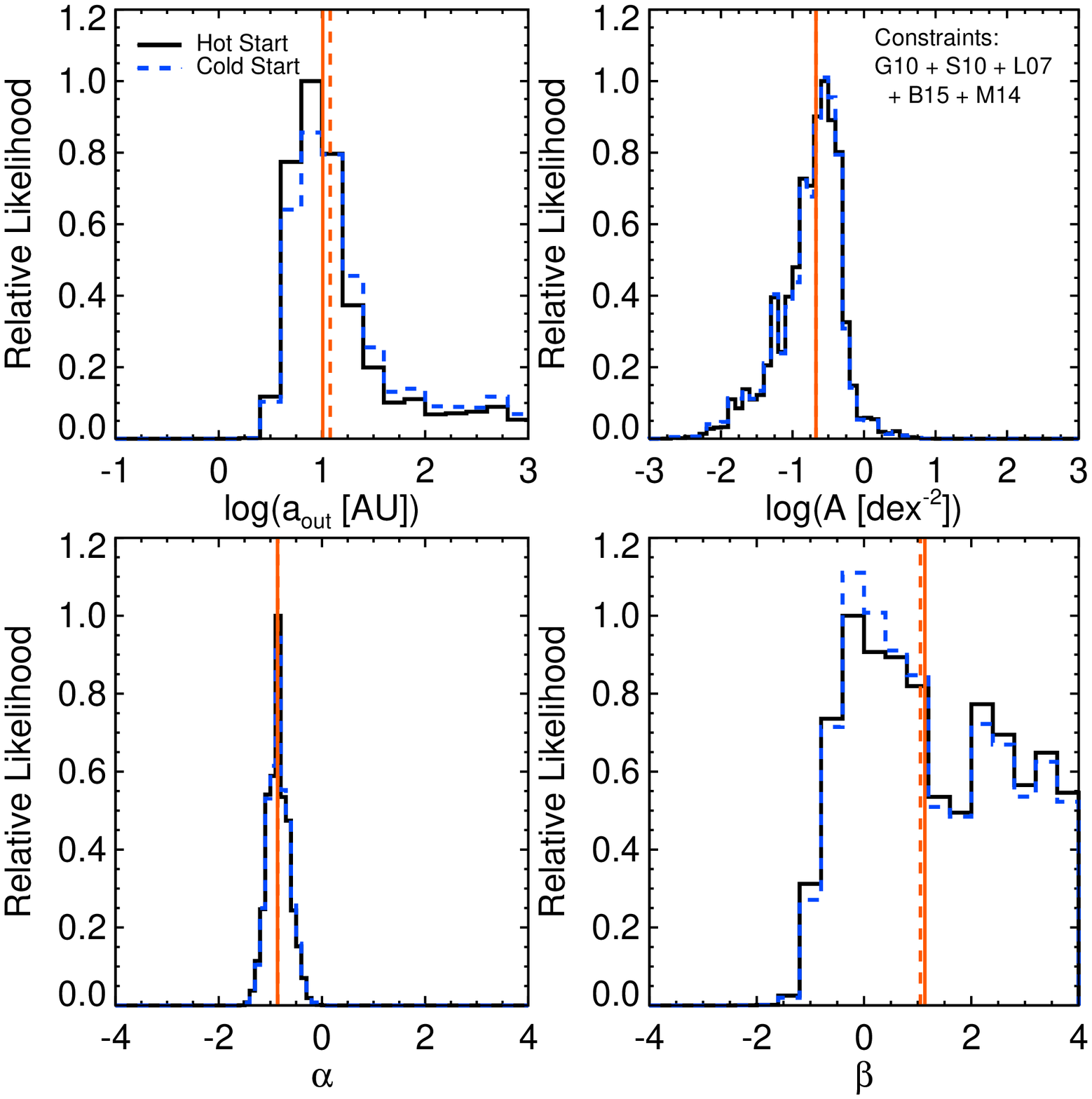}
	\caption{Relative likelihoods for each parameter, marginalized over the other three parameters, derived from our detailed comparison with microlensing (G10--\citealt{Gould2010}; S10--\citealt{Sumi2010}), direct imaging (L07--\citealt{Lafreniere2007}; B15--\citealt{Bowler2015}), and RV (M14--\citealt{Montet2014}) surveys. The vertical, red lines in each plot indicate the median values for the ``hot-start'' models (solid) and ``cold-start'' models (dashed). The median values and their 68\% confidence intervals are listed in table~\ref{tab:1d_medians}.
		\label{fig:1d_constraints_combined}}
\end{figure*}

We plot the corresponding one-dimensional likelihood functions on each parameter, marginalized over the other three parameters, in figure~\ref{fig:1d_constraints_combined}, and summarize the median values (and their 68\% confidence intervals) in table~\ref{tab:1d_medians} for both ``hot-'' and ``cold-start'' planet evolutionary models. The final parameter constraints we report in table~\ref{tab:1d_medians} do differ from those we found in our order of magnitude comparisons (presented in table~\ref{tab:1d_oom_medians}). While the median values are consistent between the two independent analyses, the results of our detailed calculations are more constraining, particularly on $\beta$ and $a_{\rm out}$ (compare figures~\ref{fig:oom_constraints_1d_main5} and \ref{fig:1d_constraints_combined}).

The likelihood distribution of $\alpha$ we find in our detailed analysis is slightly shifted towards smaller values, but has a 68\% confidence interval with no larger or smaller a ``spread'' than that derived from our order of magnitude analysis. The same is true of $\mathcal{A}$, although the shift is less pronounced. Although consistent between the two analyses, the resultant median value of $\beta$ of our detailed calculations is a factor $\approx 2$ larger than that of the order of magnitude comparison. Increased sensitivity to $\beta$ due to the inclusion of (real) variations in detection limits within the individual surveys, which thus sample our planet distribution function at different semimajor axes, means we also find that $\beta$ is better constrained relative to the results of our order of magnitude calculations (see \S~\ref{subsec:results_ind_constraints} for more discussion). In turn, due to the previously discussed degeneracy between $\beta$ and $a_{\rm out}$ (as well as variations in specific detection sensitivities), we find that $a_{\rm out}$ is also better constrained by our detailed comparison, although the median values between the two analyses are indistinguishable.

\begin{table*}[!t]
	\caption{\label{tab:1d_medians} Median values and 68\% uncertainties on the parameters of our planet distribution models for our detailed comparisons with the results of the microlensing surveys by \citet{Gould2010} and \citet{Sumi2010}, the GDPS \citep{Lafreniere2007} and PALMS \citep{Bowler2015} direct imaging surveys, and the CPS TRENDS survey \citep{Montet2014}.}
	\centering
	\begin{tabular}{c||c|c|c|c}
		\hline \hline
		Planet Evolutionary  &  \multicolumn{4}{c}{Median Values and 68\% Uncertainties} \\
		Models &  $\alpha$  & $\beta$  & $\mathcal{A}~[{\rm dex^{-2}}]$  & $a_{\rm out}~[{\rm AU}]$ \\
		\hline
		\begin{tabular}{@{}c@{}} ``Hot-Start'' \\ (\citealt{Baraffe2003})\end{tabular} & $-0.86^{+0.21}_{-0.19}$& $1.1^{+1.9}_{-1.4}$ & $0.21^{+0.20}_{-0.15}$ & $10^{+26}_{-4.7}$ \\
		\hline
		\begin{tabular}{@{}c@{}} ``Cold-Start'' \\ (\citealt{Fortney2008})\end{tabular} & $-0.85^{+0.21}_{-0.19}$& $1.1^{+1.9}_{-1.3}$ & $0.21^{+0.20}_{-0.15}$ & $12^{+50}_{-6.2}$ \\
		\hline\hline
	\end{tabular}
\end{table*}

We have demonstrated that there is overlap among the constraints we derive from a comparison with each of the individual surveys, which means that the results of all the surveys we consider can be explained by a single planet population described by a simple, joint power-law distribution function given by equation~(\ref{eqn:planet_dist_function}). As before in our order of magnitude comparison, we now determine how the likelihood of such a planet population, according to each survey individually, varies over the 68\% confidence intervals we derive on $\alpha$, $\beta$, $\mathcal{A}$, and $a_{\rm out}$.

\begin{figure}[!t]
	\epsscale{1.1}
	\plotone{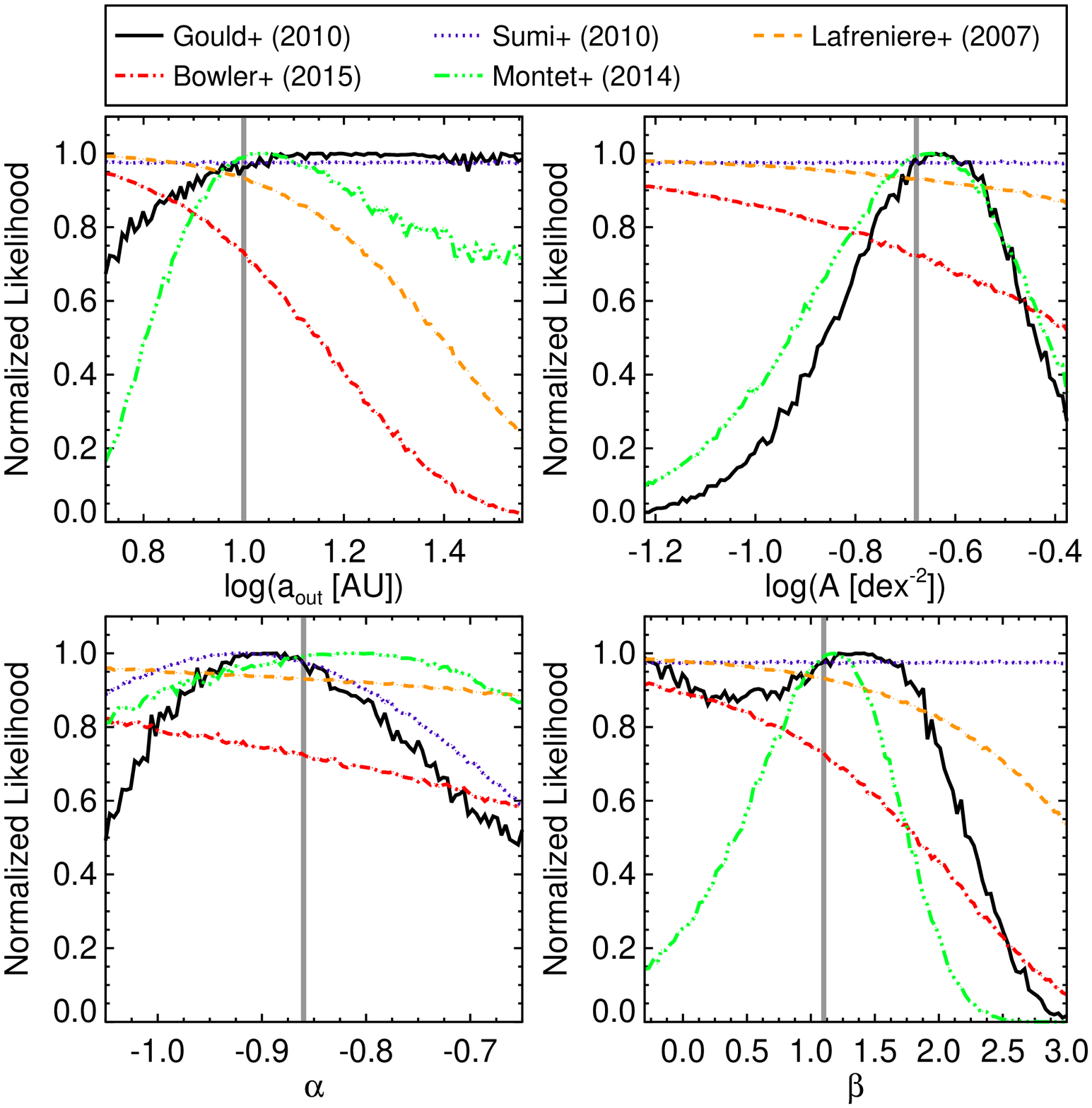}
	\caption{Likelihood functions of the parameters of our planet population model derived from individual exoplanet surveys across the 68\% confidence intervals we infer for each parameter from the combination of all surveys, assuming the remaining three parameters are at their inferred median values. The vertical, grey lines indicate the median values of each parameter listed in table~\ref{tab:1d_medians} for the ``Hot-Start'' results.
		\label{fig:range_lhoods_plot_hot}}
\end{figure}

\begin{figure}[!t]
	\epsscale{1.1}
	\plotone{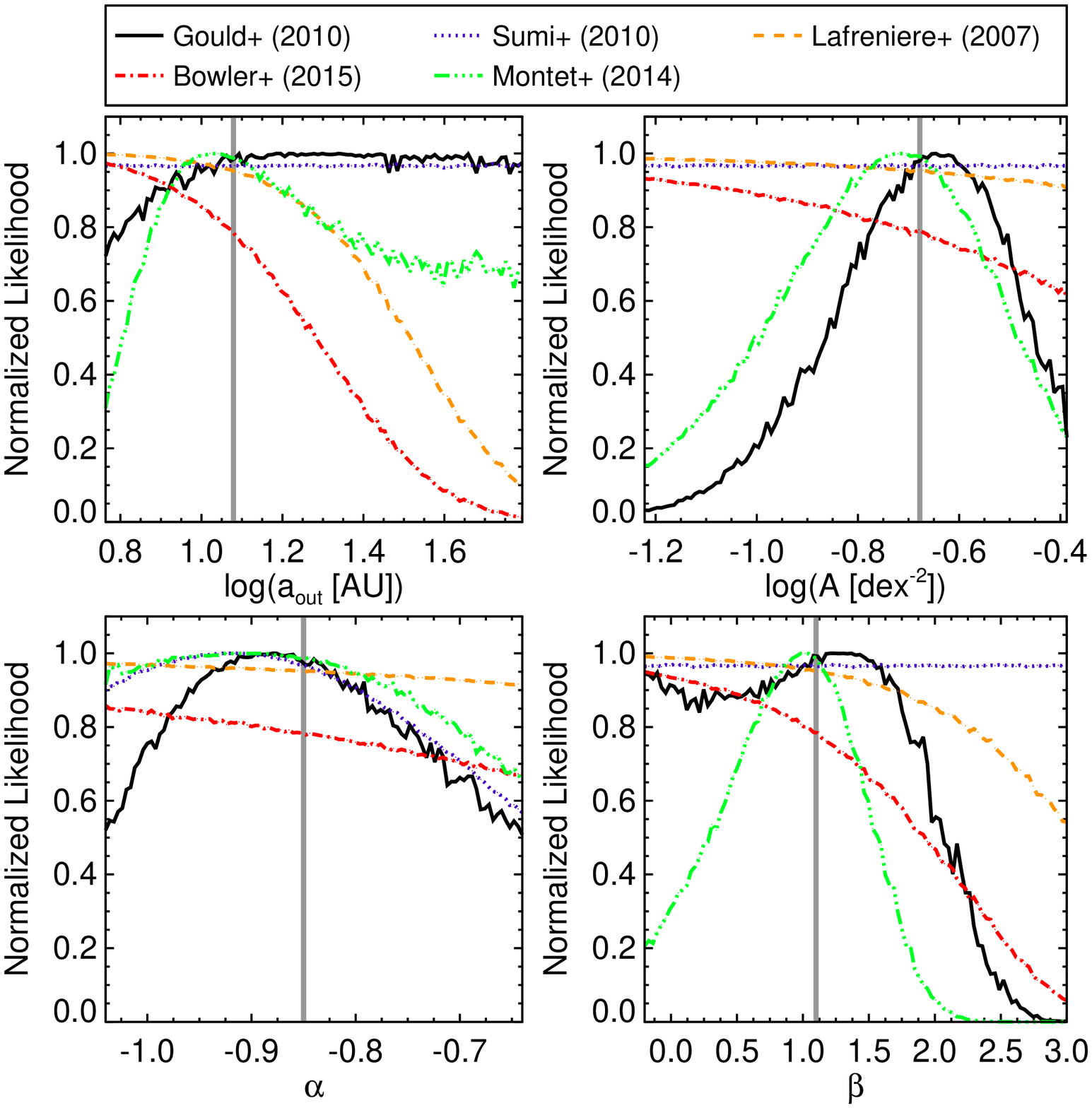}
	\caption{Identical to figure~\ref{fig:range_lhoods_plot_hot}, but utilizing the inferred median values and 68\% confidence intervals inferred for the ``Cold-Start'' comparison results listed in table~\ref{tab:1d_medians}.
		\label{fig:range_lhoods_plot_cold}}
\end{figure}

In figures~\ref{fig:range_lhoods_plot_hot} (``hot-start'') and \ref{fig:range_lhoods_plot_cold} (``cold-start''), we plot the normalized likelihood functions (i.e. the likelihoods normalized by the maximum likelihood value for each individual survey) of given parameters across their 68\% confidence intervals, assuming the remaining three parameters are at their median values (listed in table~\ref{tab:1d_medians}). These figures show that the likelihoods of these parameters from each individual survey are consistent to well within $1\sigma$ (i.e. their normalized likelihood is greater than $\approx 0.32$) across much of their inferred 68\% confidence intervals, and have normalized likelihoods $\gtrsim 0.7$ at the median value for all surveys (indicated by the vertical, grey lines). This demonstrates that this overlap in parameter space we identify is not improbable, and has a rather good chance of being real. Note that while in each of the panels there are some regions where the relative likelihood for a given survey seems low (e.g. in the upper left panel of figure~\ref{fig:range_lhoods_plot_hot}, the likelihood of $a_{\rm out}$ according to the PALMS survey of \citealt{Bowler2015} drops below 0.32 beyond $\log{a_{\rm out}}\approx 1.2$), this does not necessarily imply that a portion of these 68\% confidence intervals are relatively unlikely---keep in mind that the other three parameters are being held fixed at their median values in each panel. Although these likelihood curves are a little noisy, they are remarkably similar (qualitatively) to those we found in our order of magnitude analysis shown in figures~\ref{fig:range_lhoods_plot_main5_hot} and \ref{fig:range_lhoods_plot_main5_cold}.

\section{Uncertainties}
\label{sec:uncertainties}
The main sources of quantified uncertainty in this study are those of the measurements made by the six different surveys with which we jointly compare. By employing the full likelihood functions (or approximations thereof) of these measurements, rather than the maximum likelihood values, the uncertainties we derive on our results naturally include the contributions from each individual measurement. However, there are multiple sources of unquantifiable uncertainty in this study, in particular in the adopted models and priors necessary to derive the mapping between the properties of planets ($m_p$ and $a$) into the observables relevant to the different types of surveys. There are certainly differences in the properties of the samples of the individual surveys (e.g. distributions in stellar mass and metallicity), the magnitudes of which are unknown due to a lack of the appropriate observational data. We discuss the nature of these unquantified uncertainties below, arguing that at least some of them should be subdominant.

\subsection{Priors on Unknown Stellar, Galactic, and Planetary Orbital Properties}
\label{subsec:uncertainties_CFGKSR}
As we have demonstrated, a robust, joint comparison between the results of these different techniques is contingent on an ability to map basic planet properties into the observables relevant to each method. This requires knowledge of the physical properties of the hosts as well as the planetary orbital parameters, since each technique is sensitive to a unique combination thereof. While we attempt to account for uncertainties in these properties by marginalizing over prior distributions, imperfect knowledge of their actual distributions leaves some level of residual error.

The routine observables in a typical planetary microlensing event contain little information about the physical properties of the lens system beyond the planet-to-star mass ratio, $q$, and the projected separation in units of the Einstein radius, $s$. Measurement of the primary mass, and thus $m_p$, requires detection of higher order effects in the light curve (e.g. finite source effects, microlens parallax, xallarap) to break the degeneracy between lens mass and distance. The mapping necessary to compare our planet population models to the results of microlensing surveys, $(m_p,a)\rightarrow(q,s)$, thus depends on a Galactic model, priors on the lens mass and distance distribution functions, and priors on the planetary orbital parameters.

In this paper, we adopt the same Galactic model and prior on lens distances as in \citet{Clanton2014a} and \citet{Clanton2014b}. We demonstrated in those studies that we are able to reproduce the observed distributions of host star parameters ($R_E$, $t_E$, $\left|\boldsymbol{\mu}\right|$, $M_l$) of \citet{Gould2010} microlensing survey when weighting them by their expected event rates, with the possible exception of the distribution of lens distances. We found that events in our simulated sample typically have larger values of $D_l$ relative to the \citet{Gould2010} sample. However, given the small number of actual distance measurements in the microlensing samples, this ultimately may not be significant. Although we adopt a different lens mass function in this study than in \citet{Clanton2014a,Clanton2014b}, we still find agreement between our simulated events and the \citet{Gould2010} sample. We therefore expect uncertainties due to our choice of priors and Galactic model to be subdominant.

The priors over which we marginalize the planetary orbital parameters ($e$, $i$, $\omega$, $\Omega$, $M_0$) are those predicted for randomly-oriented orbits at random phases. We are effectively assuming there is no preference of orbital orientation to location in the Galaxy or host star properties, which would mean that any uncertainties introduced into our analysis by unknown orbital parameters should be naturally included in our analysis by virtue of marginalizing over these priors. However, we note that orbital eccentricity might actually depend on location in the Galaxy and/or on host star properties, resulting in a potential source of unquantified uncertainty in our analysis.

\subsection{Stellar Metallicities}
\label{subsec:uncertainties_HSMD}
The metallicities of the microlensing and direct imaging samples are unknown, however it is reasonable to expect that at least the microlensing sample should have a significantly different distribution than those of the RV and direct imaging surveys. The RV and direct imaging samples are limited to nearby M stars (within tens of parsecs), whereas microlensing probes stellar hosts several kpc into the Galaxy. Multiple studies have measured a Galactic metallicity gradient \citep[e.g.][]{Cheng2012,Hayden2014} with metallicity decreasing as distance from the Galactic center increases. The metallicities of the disk lenses (and at least a fraction of bulge lenses; see \citealt{Bensby2013}) are therefore expected to be enhanced relative to the local RV and direct imaging samples. Despite this, and the fact that there is a strong correlation between giant planet frequency and host star metallicity measured by RV surveys \citep{Gonzalez1997,Fischer2005,Johnson2010,Montet2014}, we found in \citet{Clanton2014a} that the measurements of giant planet frequency made by RV and microlensing surveys are consistent without a need to correct for any metallicity effects. This is discussed at length in \S~ 6.4 of \citet{Clanton2014a}. 

\subsection{Non-M Dwarf Lens Contamination in Microlensing Samples}
\label{subsec:uncertainties_CFGKSR}
The rate of microlensing events along a given sight line depends explicitly on the lens mass distribution function since $\Gamma\propto \theta_E \propto M_l^{1/2}$ \citep[see e.g.][]{Gaudi2012}. The slope of the mass function is such that there are greater number of low-mass stars relative to higher-mass stars, and therefore a majority of lenses are M dwarfs. However, there is some level of contamination of the microlensing samples by earlier spectral types (mostly GK), and stellar remnants. An estimated $\sim 20\%$ of microlensing events are due to remnants (primarily white dwarfs) and are essentially completely indistinguishable from other types of lenses by their timescale distribution \citep{Gould2000}. Nevertheless, we expect the uncertainties due to any such contamination to be small in comparison with the Poisson errors due to the small number of planet detections in the microlensing samples that are included in our analysis.

\subsection{Planet Evolutionary Models}
\label{subsec:uncertainties_PEM}
For the purposes of comparing our planet population models with the results of direct imaging surveys, we must derive the expected planet-to-star constrast, $\Delta {\rm mag}(m_p, {\rm age}, {\rm distance})$, for each system. Expected absolute $H$ and/or $K_s$ band magnitudes of each planet are estimated using the planet evolutionary models of \citet{Baraffe2003} and \citet{Fortney2008} given the planet's mass and age; the latter is assumed to be the same as the host's age. These are then converted into apparent magnitudes using the distance of the host. The contrast is then simply the difference between the apparent magnitude of the planet and that of the host in the appropriate band. While the uncertainties in the ages and distances of these systems are somewhat taken into account in our analysis by marginalizing over the appropriate probability density functions of these measurements (see below for additional discussion), uncertainties in the planet evolutionary models themselves are not.

There are two classes of these models, ``cold start'' \citep[e.g. ][]{Fortney2008} and ``hot start'' \citep[e.g.][]{Baraffe2003}, which are broadly distinguished by their assumptions about the initial conditions of a newly-formed protoplanet concomitant with different formation mechanisms. The ``cold-start'' models of \citet{Fortney2008} we adopt in this paper follow the prescription of giant planet formation by core accretion presented by \citet{Marley2007}. In their implementation, gas accreting onto the core during the phase of runaway gas accretion (occurring after the gaseous envelope has reached a mass roughly equal to that of the core and has hydrodynamically collapsed onto the core) is able to quickly radiate away most of its initial gravitational potential energy, lowering its specific entropy and causing the gas to reach thermal equilibrium with the local radiation field. This is the key reason for the significantly lower luminosity (at early ages) of the ``cold-start'' models relative to ``hot-start'' models. The ``hot-start'' models of \citet{Baraffe2003} require newly-formed giant planets to slowly radiate all of their initial gravitational potential energy, and thus have initially high entropy, as would be expected for planets formed directly by gravitational collapse \citep{Spiegel2012}. For a planet of a given luminosity, this amounts to a basic degeneracy in the models between its mass, initial conditions, and age, such that even if we knew the exact age, it would not be clear which model best describes the planet's evolutionary state since we do not know the exact formation mechanism.

Not only are the luminosities of planets of a given mass and age quite discrepant between the models of \citet{Fortney2008} and \citet{Baraffe2003} up to ages of some tens of Myr, or even $\sim 100~$Myr for more massive planets with $m_p \gtrsim 5~M_{\rm Jup}$, due to various assumptions about their initial conditions, but their spectra (and thus their apparent brightness in the various photometric bands) are also sensitive to the assumed atmospheric conditions (e.g. clouds, composition). Indeed, the treatment of atmospheric chemistry and condensation differs between the models of \citet{Fortney2008} and \citet{Baraffe2003}, resulting in non-convergence at late times when the initial conditions have been forgotten.

Our assumption that the ages of the planets are the same as that of their host could also be a source of uncertainty in our analysis. Giant planet formation by core accretion is believed to take place over the course of several Myr \citep[see e.g.][and references therein]{Kennedy2008} before dispersal of the gaseous disk, while formation by direct gravitational collapse can occur as quickly as a few orbital periods \citep[$\sim$ dynamical time;][]{Boss2000} and up to some $\sim 0.5~$Myr \citep{Helled2006} if captured planetesimals heat up the contracting protoplanet. Thus, without knowledge of the exact formation mechanism of each planet, we could be overestimating its age by up to several Myr when using the ``cold-start'' models. It could also be the case that there are two channels of planet formation, with some planets being formed by core accretion and others by fragmentation. Our hope is that by performing our analysis twice, once with each class of models, we can bracket the range of expected luminosities for a given planet mass and age.

We also note that these planet evolutionary models are poorly calibrated, and comparisons with the few existing benchmark brown dwarfs have revealed significant discrepancies between the model predicted luminosities and the actual measurements \citep{Dupuy2009,Dupuy2014,Crepp2012}. These studies have found that evolutionary models systematically underpredict luminosities by factors of up to $2-3$, corresponding to mass overestimates between $15-30$ percent. This caveat should be kept in mind when interpreting our conclusions.

Nevertheless, we do not find a significant difference between our results derived from the ``hot-'' and ``cold-start'' calculations (see \S~\ref{subsec:results_syn_constraints} for discussion), so we do not consider the use (or choice) of an evolutionary model a dominant source of uncertainty in our analysis (at least not over the region of parameter space we are constraining).

\subsection{Separation Distribution Function in Microlensing Surveys}
\label{subsec:uncertainties_SDFMS}
The results reported by the microlensing surveys of \citet{Gould2010} and \citet{Sumi2010} were obtained under the assumption that there are equal numbers of planets per logarithmic interval in projected separation, i.e. $dN_{\rm pl}/d\log{s}$ is a constant. The semimajor axis distribution function we adopt, $dN_{\rm pl}/d\log{a}\propto a^{\beta}$, is related, but not equivalent, to $dN_{\rm pl}/d\log{s}$ due to conflation with the distributions of lens masses and distances, and possibly the distribution of projection angles (a function of $e$, $i$, $\omega$, and $M_0$). Since $\beta$ is one of the parameters in our planet population model for which we are deriving constraints based on the results of \citet{Gould2010} and \citet{Sumi2010}, their assumption of a log-uniform distribution in projected separation potentially introduces uncertainties for which we do not account.

\begin{figure*}[!ht]
	\epsscale{0.8}
	\plotone{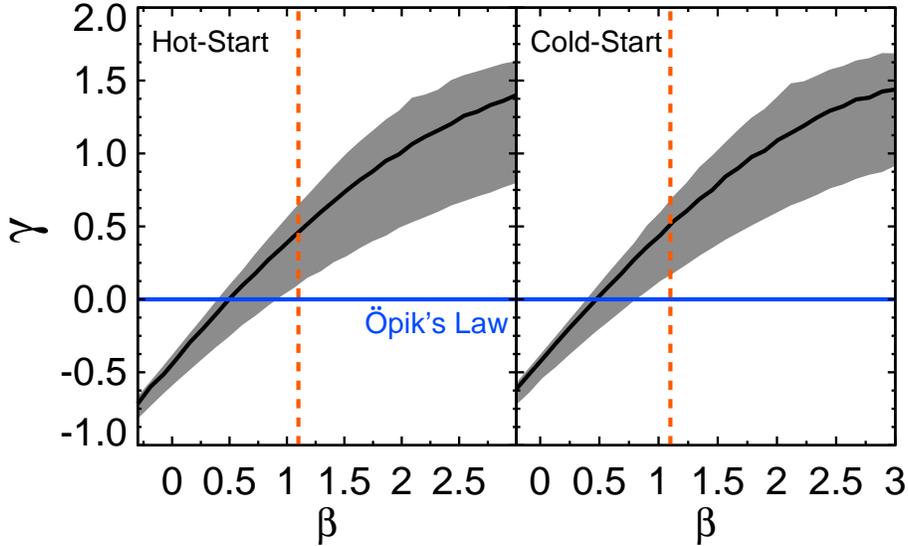}
	\caption{The slope of the distribution of $s$ measured over the range of $s$ to which the \citet{Gould2010} and \citet{Sumi2010} microlensing surveys are sensitive, $\gamma$, defined implicitly as $dN_{\rm pl}/d\log{s}\propto s^{\gamma}$, as a function of $\beta$. The range of $\beta$ shown in these panels is the $1\sigma$ uncertainty (i.e. the 68\% confidence interval) that we derive from our joint comparison with microlensing, RV, and direct imaging surveys. In each panel, the solid, black line shows the values of $\gamma$ we find when $\alpha$, $\mathcal{A}$, and $a_{\rm out}$ are held fixed at their median values, and the grey band shows the ``$1\sigma$ range'' that we find when $\alpha$, $\mathcal{A}$, and $a_{\rm out}$ are each allowed to vary (independently) within their 68\% confidence intervals (see table~\ref{tab:1d_medians}). The dotted, red, vertical lines represent the median values of $\beta$ we derive from our joint comparison with all surveys, and the solid, blue, horizontal line at $\gamma=0$ corresponds to a log-uniform distribution of $s$ (i.e. \"Opik's Law; \citealt{Opik1924}), as assumed by \citet{Gould2010} and \citet{Sumi2010}.
		\label{fig:logs_slopes_hot_cold}}
\end{figure*}

To investigate this issue, we model the distributions of $\log{s}$ for our planet populations over the range $|\log{s}|\leq (1/3)(\log{q_0}+4.3)$, where $q_0 = 5\times10^{-4}$ (the approximate range of $\log{s}$ to which the \citealt{Gould2010} and \citealt{Sumi2010} surveys are sensitive; see \S~\ref{subsec:mlens_constraints}), as $dN_{\rm pl}/d\log{s}\propto s^{\gamma}$ and measure the value of $\gamma$ with a least squares fit across the 68\% confidence interval we derive on $\beta$ for both the ``hot-'' and ``cold-start'' results (separately). We plot these best-fit values of $\gamma$ as a function of $\beta$ in figure~\ref{fig:logs_slopes_hot_cold}. We find in both the ``hot-'' and ``cold-start'' cases that our results are consistent with an \"Opik's Law distribution \citep[i.e. $\gamma=0$;][]{Opik1924} over some range of the 68\% confidence interval we derive on $\beta$ (the range of $\beta$ plotted in each panel), although $\gamma$ can differ significantly across the full confidence interval. Furthermore, the results of the \citet{Gould2010} and \citet{Sumi2010} surveys are only weakly dependent on their assumption of \"Opik's Law, since their surveys are sensitive to a relatively small range of $s$ (and thus, $a$), especially compared to the range we explore in this study. Thus, we argue that while this could be a non-trivial source of unquantified uncertainty in our analysis, is it probably not a dominant source.

\subsection{Comparisons with ``Intermediate Data Products''}
\label{subsec:uncertainties_CIDP}
The most correct and robust way to perform this study would be to compare our planet distribution model directly to the data collected by each respective survey. For example, to compare with long-term trends detected by the CPS TRENDS survey of \citet{Montet2014}, we would need to follow the procedure outlined below.
\begin{enumerate}
	\item For a given planet population, map each $(m_p,a)$ pair into the observables $(K,P)$, marginalizing over the appropriate stellar and orbital parameters.
	\item Using the derived $(K,P)$, generate simulated RV curves with the same number of observations and cadence for each star in the actual CPS sample, including uncertainties due to systematics and stellar jitter. Variations in orbital parameters (including phases) are included as part of the mapping $(m_p,a)\rightarrow (K,P)$.
	\item Pass these simulated RV curves through the detection pipeline of the CPS TRENDS survey to determine which planets produce detectable trends.
	\item Compute the probability that the detected trends have accelerations consistent with the actual trends reported by the CPS. For any trends we expect to be deteced by the CPS that are not actually detected, we would compute the associated Poisson probability.
	\item Multiplying together these probabilities, assign the planet population model under consideration a statistical weight. Repeat this procedure for all populations to constrain the parameters of our model, $\left\{\alpha, \beta, \mathcal{A}, a_{\rm out}\right\}$.
\end{enumerate}
Similar types of analyses would be performed to compare our planet models to microlensing and direct imaging surveys. Such a study would require all original data and detection software for each survey, and thus would be much more involved than the methods we have developed. However, given the levels of uncertainty due to the sources discussed previously in this section, we believe such a careful analysis is currently unwarranted.

We instead compare our planet population model to ``intermediate data products'' of the various surveys we consider. This results in sources of unquantified uncertainty that are only somewhat alleviated by the fact that we use the appropriate detection limits (or estimates thereof). Returning to the case of comparison with the trends detected by the CPS, our methodology is to use estimated detection limits (taking into account the actual number of observations, stellar jitter, measurement uncertainties, and time baseline of observations) for each star in their sample in the following manner. The derived observables, $(K,P)$, for each planet in a given population is determined to be either detectable or undetectable given the estimated sensitivites. We then assign a statistical weight to the population under consideration based on the total number of trends we expect to be detected by the CPS TRENDS survey and the likelihood function we approximate for their survey (see \S~\ref{sec:introduction}) on the total (actual) number of detected trends. The ``intermediate data product'' to which we compare for the CPS is thus the total number of detected trends.

\section{Summary and Conclusions}
\label{sec:summary_conclusions}
In this paper, we assume a planet distribution function described by a simple, power-law form with four free parameters: $d^2N_{\rm pl}/(d\log{m_p}~d\log{a})=\mathcal{A}(m_p/M_{\rm Sat})^{\alpha}(a/2.5~{\rm AU})^{\beta}$, with an outer cutoff radius of the separation distribution function of $a_{\rm out}$. We generate random planet populations from this distribution function, map their properties into the observables relevant to microlensing, RV, and direct imaging surveys, and compare the expected observations (determined using the actual detection limits) to the reported results of the five surveys we consider. We find that each survey individually can only constrain combinations of the  parameters $\{\alpha, \beta, \mathcal{A}, a_{\rm out}\}$, but by performing a simultaneous analysis with the results of all surveys, we demonstrate that there is a single planet population that is consistent with them all.

Each survey is sensitive to different, yet complementary, regions of parameter space with varying amounts of mutual overlap. This is important, as it allows each survey to sample the planet distribution function over different ranges of planet masses and semimajor axes, providing powerful constraints on the properties of planets at large separations. We find that the combination of the \citet{Gould2010} and \citet{Sumi2010} microlensing surveys tightly constrain $\alpha$ and $\mathcal{A}$. With these two parameters (essentially) fixed, including the results of the CPS TRENDS survey \citep{Montet2014} places a lower bound on the allowed values of $\beta$ and $a_{\rm out}$ in order to explain the four significant long-term RV drifts that \citet{Montet2014} (convincingly) argue must be due to giant planets. However, we find that the subsequent inclusion of the results of the GDPS \citep{Lafreniere2007} and PALMS \citep{Bowler2015} direct imaging surveys can only place relatively weak constraints on the values of $\beta$ and $a_{\rm out}$, and we are left with a large range of allowed values on these two parameters.

The median values and 68\% confidence intervals of the parameters of a planet population consistent with microlensing, RV, and direct imaging surveys that we derive when assuming ``hot-start'' planet evolutionary models are $\alpha=-0.86^{+0.21}_{-0.19}$, $\beta = 1.1^{+1.9}_{-1.4}$, $\mathcal{A} = 0.21^{+0.20}_{-0.15}~{\rm dex^{-2}}$, and $a_{\rm out} = 10^{+26}_{-4.7}~$AU, and when assuming ``cold-start'' evolutionary models, we find $\alpha=-0.85^{+0.21}_{-0.19}$, $\beta = 1.1^{+1.9}_{-1.3}$, $\mathcal{A} = 0.21^{+0.20}_{-0.15}~{\rm dex^{-2}}$, and $a_{\rm out} = 12^{+50}_{-6.2}~$AU.

We find that while the results from our analysis assuming ``hot-start'' evolutionary models are slightly more constraining than those we find when assuming ``cold-start'' models, the difference is not significant. The lack of a significant difference can be attributed to a few reasons: 1) the difference in the predicted luminosities between the two classes of models decreases with planet mass at all ages, 2) our assumed power-law form for the planet distribution function, and 3) the most likely range of $a_{\rm out}$ we find lies within a region where direct imaging surveys are mostly insensitive to planets. We find that $\alpha$ is relatively well-constrained to negative values, meaning the mass function is heavily weighted towards lower-mass planets. The predicted luminosities for lower-mass planets differ less between the ``hot-'' and ``cold-start'' models relative to the more massive planets (see figure~1 of \citealt{Fortney2008}). Furthermore, because our assumed planet distribution function has a power-law form, small changes in $\alpha$, $\beta$, and/or $a_{\rm out}$ can have a dramatic effect on the number of expected planet detections. Additionally, as the projected separation of a planet approaches the inner working angle, its detectability by direct imaging surveys dramatically decreases, meaning that we do not predict any detections of planets with projected separations $r_{\perp}\lesssim 10~$AU for either imaging survey, regardless of our choice of evolutionary model.

The degeneracy between $\beta$ and $a_{\rm out}$ suffered by our comparison with the results of direct imaging surveys, where arbitrarily large values of $\beta$ are allowed when $a_{\rm out}\lesssim 10~$AU (where these surveys are insensitive to planet detections) and large values of $a_{\rm out}$ are allowed for small values of $\beta$ (see figure~\ref{fig:2d_constraints_combined_hot}), limits our ability to derive tight constraints on these two parameters. If we knew the value of either $\beta$ or $a_{\rm out}$, we would immediately be able to break this degeneracy and derive much better constraints. Observational studies measuring (sub-)millimeter-wave continuum emission from protoplanetary disks around young M stars have inferred a wide range of characteristic radii, beyond which the dust surface density declines exponentially, between $\sim 20-200~$AU \citep{Andrews2007,Isella2009,Andrews2009,Andrews2010,Guilloteau2011}. The median value and 68\% confidence interval we derive on $a_{\rm out}$ are consistent with these observed disk sizes, however the long tail towards larger $a_{\rm out}$ where direct imaging surveys lose sensitivity due to a limited FOV thus seems less likely than we estimate for $a_{\rm out}\gtrsim 200$ (see figure~\ref{fig:1d_constraints_combined}). While planets formed in the disk could be placed into such very long-period orbits due to dynamical processes, such as planet-planet scattering, they are unlikely to end up on stable orbits \citep{DodsonRobinson2009}. Futhermore, in a cluster environment, the timescale for planet ``ionization'' by passing stars is $\sim 200~{\rm Myr}(n/{\rm pc^{-3}})^{-1}(a/200~{\rm AU})^{-1}(v/0.3~{\rm km~s^{-1}})^{-1}$, where $n$ is the number density of stars and $v$ is the relative velocity between the planetary system and the passing star \citep{Antognini2015}.

In this paper, we find that a single, continuous, joint-power law distribution function in both planet mass and semimajor axis can simultaneously explain existing observations from microlensing, RV, and direct imaging. However, we acknowledge the fact that the true planet distribution function is probably more complicated, with breaks in both mass and semimajor axis, and multiple power-law slopes. \citet{DodsonRobinson2009} demonstrate that giant planet formation by gravitational instability can succeed at large separations, but, even under the most favorable conditions, {\it in situ} giant planet formation by core accretion cannot occur beyond $a \sim 20~$AU around M stars. Furthermore, outward migration of giant planets from the inner disk to large separations seems implausible \citep{Peplinski2008}, and although planet-planet scattering can shuffle giant planets out to large separations, it is unlikely they end up on stable orbits \citep{DodsonRobinson2009}. Thus, if the mode of giant planet formation switches from core accretion to disk fragmentation as separation from the host star increases, it is reasonable to expect that the distribution of semimajor axes will have at least one break. The different formation mechanisms probably lead to distinct mass functions as well. However, since the existing observations can be explained by our simple distribution function, we argue that there is not yet a need to introduce more complicated distribution functions. As direct imaging surveys become more sensitive (e.g. {\it JWST}, {\it WFIRST-AFTA}), we may find evidence of such breaks in the planet distribution function.

Finally, while we have shown that wide-separation ($a\gtrsim 2~$AU) planets around M dwarfs can be explained by a single population, it would be interesting to see if such a population can explain the over-abundance of short-timescale microlensing events identified by \citet{Sumi2011}. These short-timescale events are consistent with planetary-mass objects that are either widely separated from their hosts, or are freely floating. In a future paper, we will investigate the question of whether or not our population of wide-separation planets could be responsible for these short-timescale events, and if not, determine what would be required to explain them.

\acknowledgments
This research has made use of NASA's Astrophysics Data System and was partially supported by NSF CAREER Grant AST-1056524. We thank Ben Montet and John Johnson for helpful conversations and for providing us with the necessary data to estimate the detection sensitivities for the CPS TRENDS sample. We thank Brendan Bowler for providing us the detection limits used in his statistical analysis of the PALMS survey. We thank Jonathan Fortney for providing clarification on the use of his planet evolutionary models. We also thank Ji Wang for helpful conversations.

\bibliographystyle{hapj}
\bibliography{myrefs}
\end{document}